\UseRawInputEncoding
\documentclass[preprint,linenumber,superscriptaddress,amsmath,amsfonts,nofootinbib]{revtex4-1}
\usepackage{graphicx}
\usepackage[width=0.8\textwidth,justification=justified]{caption}
\usepackage{bm}
\usepackage{hyperref}
\usepackage{color}
\usepackage[mathlines]{lineno}
\usepackage{float}
\usepackage[dvipsnames]{xcolor}

\begin{document}

\title{Analysis of the nonlinear dynamics of a chirping-frequency Alfv\'{e}n mode in a Tokamak equilibrium}

\author{X. Wang$^1$, S. Briguglio$^{2,*}$, C. Di Troia$^2$, M. Falessi$^2$, G. Fogaccia$^2$, V. Fusco$^2$, G. Vlad$^2$ and F. Zonca$^2$\\
\small {1. Max Planck Institute for Plasma Physics, 85748 Garching, Germany\\}
\small {2. ENEA, Fusion and Nuclear Safety Department, C. R. Frascati, Via Enrico Fermi 45, C. P. 65 - I-00044 - Frascati, Italy\\}
\small {*. sergio.briguglio@enea.it}}

\begin{abstract}

Chirping Alfv\'{e}n modes are considered as potentially harmful for the confinement of energetic particles in burning Tokamak plasmas because of their capability, by modifying their frequency, of extracting as much power as possible from such particles and, in turn, enhancing their transport. In this paper, the nonlinear evolution of a single-toroidal-number chirping mode is analysed by numerical particle simulation. This analysis can be simplified if the different resonant phase-space structures can be investigated as isolated ones; that is, if the phase space can be cut into slices that do not exchange particles and interact with the mode independently of each other. This can be done adopting a coordinate system that includes two global invariants of the system or, if it is not possible to identify these invariants (this is generally the case for chirping modes), two constants of motion. In the frame of a numerical simulation approach, we adopt as constants of motion, the magnetic momentum and 
a suited function of the initial particle coordinates.
The relevant resonant structures are then identified. The analysis is focused on the dynamics of two of them: namely, those yielding the largest drive during, respectively, the linear phase and the nonlinear one. 
It is shown that, for each resonant structure, a density-flattening region is formed around the respective resonance radius, with radial width that increases as the mode amplitude grows.
It is delimited by two large negative density gradients, drifting inward and outward.
If the mode frequency were constant, this density flattening would be responsible for the exhausting of the drive yielded by the resonant structure, 
which would occur as the large negative density gradients leave the resonance region. 
causes the resonance radius and the resonance region to drift inward. This drift, along with a relevant resonance broadening, delays the moment in which the inner density gradient reaches the inner boundary of the resonance region, leaving it. On the other side, 
the island reconstitutes around the new resonance radius; as a consequence, the large negative density gradient further moves inward. 
This process continues as long as it allows to keep the large gradient within the resonance region. When this is no longer possible, 
the resonant structure ceases to be effective in driving the mode. To further grow, the mode has to tap a different resonant structure, possibly making use of additional frequency variations.

\end{abstract}

\pacs{Valid PACS appear here}

\maketitle

\section{Introduction}
\label{sec:introduction}  

Alfv\'{e}n modes can be driven unstable, in Tokamak plasmas, by the resonant interaction with alpha particles produced by fusion reactions and/or energetic ions produced by auxiliary heating methods, characterised by speeds of the order of the Alfv\'{e}n velocity~\cite{chen07,fasoli07,heidbrink08,breizman11,Liu13,lauber13,gorelenkov14,pinches15,RMP16}. Mode-particle interaction can in turn deteriorate the confinement of these particles, preventing their thermalisation in the core plasma and eventually damaging the first wall. Some of the Alfv\'{e}n modes, like the Toroidal Alfv\'{e}n Eigenmodes~\cite{cheng85} (TAEs) have a MHD counterpart; that is, they exist as marginally stable or quasi marginally stable modes even in the absence of energetic-particle drive. Other modes, like the Energetic Particle Modes~\cite{chen94} (EPMs) have no such counterpart: they are oscillations of the Alfv\'{e}n continuum driven unstable only when the energetic-particle drive exceeds the continuum damping; in the absence of this drive, they are strongly damped. While the former modes are typically characterised by an approximately constant frequency, even in the nonlinear stage, essentially constrained to keep within the frequency gaps opened in the Alfv\'{e}n continuum by the coupling between different poloidal harmonics, the latter are able to vary their frequency (they have been named \textit{chirping modes}), during the nonlinear evolution, in order to extract as much power as possible from the energetic particles. Both because of this feature and the strong dependence on the energetic-particle drive they usually present above the instability threshold, such chirping modes are considered as potentially harmful for the confinement of energetic particles in burning Tokamak plasmas and have attracted much interest.

In this paper we want to investigate, by numerical particle simulation, the nonlinear dynamics of chirping modes. For the sake of simplicity, we consider the case of an Alfv\'{e}n spectrum characterised by a single toroidal mode number $n$. We have already studied the case of a single-$n$ mode, with constant frequency $\omega$ in Refs.~\cite{briguglio14,briguglio17}. Our approach consisted in adopting a coordinate system that includes two invariants: namely, the magnetic moment $M$ and the quantity $C\equiv \omega P_{\phi} - n E$, with $P_{\phi}$ and $E$ being, respectively, the toroidal angular momentum and the energy of the particle~\cite{white}. The phase space can then be seen as a set of slices characterised by given values of $M$ and $C$. Because of the invariance of these two coordinates, particles cannot cross their birth slice and the gradients of the distribution function orthogonal to the slices do not play any direct role in mode-particle power exchange: the slices can then be treated as isolated ones. Once the most relevant resonances (that is, the slices yielding the largest contribution to the mode drive or damping) are identified, saturation dynamics can be analysed by focusing on the evolution of each of them. 
To this aim, in Refs.~\cite{briguglio14,briguglio17} we have sampled the selected slices by a large number of test particles moving in the fields computed in a full-population, self-consistent simulation, and its interaction with the mode has been analysed in detail. 
We have shown that 
an island-like structure, enclosing the bounded orbits of particles 
instantaneously trapped in the potential well of the wave,
forms around the resonance radius in the plane $(\overline{\Theta},r_{\textrm{eq}})$; 
here, $\overline{\Theta}$ corresponds to the wave-phase seen by the particle and $r_{\textrm{eq}}$ is the ``equatorial radius" (see Sec.~\ref{sec:hamiltonian} and, respectively, Sec.~\ref{sec:useful_choices} for their definition). 
Mixing of trapped particles originating from the opposite sides of the resonance radius occurs. This gives rise to a flattened-density region delimited by large negative density gradients moving inward and outward as the island width grows with the increasing mode amplitude.
For fixed mode frequency saturation is reached when the density flattening completely covers the resonant-interaction region~\cite{briguglio14,briguglio17}. The width of this region is typically limited by the smallest of the resonance and mode widths: in the former case, the saturation mechanism has been named resonance detuning; in the latter, radial decoupling.

In the case of chirping modes, we can expect a more complex saturation dynamics. Indeed, the capability of the mode of changing its frequency
implies both that the resonant interaction of a given slice can be prolonged by counteracting the saturation mechanism just described, and that other phase-space regions can take over from the initially dominant one in destabilising the mode.
Unfortunately, a slavish application of the above procedure is not possible, as the quantity $C$ is no longer an exact invariant of the particle motion when $\omega$ is not a constant. The invariance of $C$ can be, of course, approximately satisfied if the rate of variation of $\omega$ is small enough. Moreover, the possibility of generalising the definition of $C$ in such a way to get an isolating invariant cannot be ruled out
~\cite{zonca2015,lichtemberg1983,lichtemberg2010}. In the present paper, however, we adopt a different choice: including in the phase-space coordinate system a constant of motion rather than a dynamical invariant; namely, a suited function of the initial values of the particle coordinates. Here, we use the words \textit{constant of motion} with the trivial meaning of a quantity which remains constant along the particle trajectory: the initial values of particle coordinates have, of course, this feature, then labelling each particle in a permanent way. 
It is worth observing that resorting to the initial values of particle coordinates (or to a suited function of them) is a viable choice only in the frame of numerical simulation, where the transformation that brings from the actual coordinates to the initial ones is readily available.

We will show that our approach allows us identifying the different resonant structures contributing to mode drive during the evolution of the system. In the case analysed in this paper, it is possible to take two of these structures as representative of the mode-particle interaction during the linear stage and, respectively, the nonlinear one. 
In both cases, the relatively small growth rate and the quite large mode structure would make the resonance detuning the dominant saturation mechanism in the constant-frequency case.
We will see, however, that the capability of the mode to change its frequency (a downward chirping, in the considered case) alters this mechanism. Indeed, it causes the resonance radius and resonance region to move inward. This drift allows to delay the moment in which the inner density gradient reaches the inner boundary of the resonance region and leaves it. On the other side, 
the island reconstitutes around the new resonance radius, drifting inward too. As a consequence, the large negative density gradient further moves inward. This process goes on as long as the frequency can decrease and the resonance region can move inward in such a way to keep the large gradient within the resonance region. When this is no longer possible, 
either because a further change in frequency is disadvantageous in terms of drive/damping balance, or because such a change does not result in a significant inward shift of the resonance radius and resonance region,
the gradient ceases to be effective in driving the mode. To further grow, the mode has to tap a different resonant structure, possibly making use of additional frequency variations. We will analyse the analogies and the differences between the behaviours of the two resonant structures, identified as the representative ones for the linear and, respectively, the nonlinear stages.

The paper is structured as follows: Sec.~\ref{sec:power} shows how to compute the power transfer in a coordinate system different from that adopted to push particles in the phase space. The only requirement is that the alternative coordinates of each particle are known in terms of the standard ones. Section~\ref{sec:useful_choices} presents some useful choices for such alternative coordinates. 
In particular, some advantages in adopting the so-called \textit{exact-invariant} coordinates in the constant-frequency single-toroidal-number case are discussed. For more general cases, the possibility of resorting to constants of motion instead of invariants is proposed, and it is observed that, in the view of a numerical simulation approach, a suited set of constants of motion can be immediately recognised in the initial coordinates of the particle. 
The numerical experiment performed to analyse the dynamics of a chirping mode is described in Sec.~\ref{sec:setting}. Section~\ref{sec:structures} presents the search for the most relevant phase-space resonant structures. In Sec.~\ref{sec:hamiltonian}, the Hamiltonian-mapping test-particle approach is adopted to investigate some aspect of the nonlinear dynamics of a resonant structure.
Section~\ref{sec:trapping_detrapping} analyses the phenomenon of particle trapping and de-trapping in detail.
 The relationship between mode-particle power transfer and fulfilment of the resonance condition is examined in Sec.~\ref{sec:resonance}. Section~\ref{sec:chirping} shows the evolution of density-flattening and resonance regions accompanying the mode chirping and describes the succession of different resonant structures in driving the mode during the nonlinear stage. Finally, a short summary of the paper and a discussion of the critical points of our approach are presented in Sec.~\ref{sec:conclusions}.

\section{Mode-particle power exchange and phase-space coordinates}
\label{sec:power}  

In this section we want to show how the calculation of the power exchange between mode and particles can be performed in the framework of a gyrokinetic simulation, and how it is possible, from a numerical point of view, to arbitrarily choose the set of phase space coordinates to represent this exchange.

Gyrokinetic simulations consist in solving the Vlasov equation for the particle distribution function for each relevant species, coupled with the equations for electromagnetic fields, in the low-frequency limit, that is for phenomena characterised by frequencies much smaller than the Larmor frequency of each species. The equations for the fields can be in the form of Poisson-Ampere equations (fully gyrokinetic codes~\cite{lee83,lee}) or MHD equations containing suited momenta of the particle distribution functions (hybrid MHD-gyrokinetic codes~\cite{park92pofb})\footnote{Hybrid MHD-gyrokinetic codes are yet a particular case of more general moment based kinetic approaches, including mixed gyrokinetic - fully kinetic codes~\cite{lin2005,lin2010,lin2012,chen2019}.}. 
Here, we consider for simplicity the case of a hybrid MHD-gyrokinetic code and concentrate on the energetic-particle population, though all the following treatment can be immediately generalised to any code and species.

The simulation is performed by discretising the phase space into microscopic volumes, representing each volume by a marker (macro-particle) that brings the whole electric charge contained in the volume, computing the electromagnetic fields in terms of the momenta of the marker distribution function and updating the phase-space coordinates of each marker according to the fields it experiences.

Let us adopt, for the phase space, the gyrocenter coordinate system $Z \equiv (r,\theta,\phi,M,U,\vartheta)$, where $r$ is the radial coordinate, $\theta$ and $\phi$ are the poloidal and toroidal angle, respectively (cf. Fig. ~\ref{fig:fig_toroid} below), $M$ is the conserved magnetic momentum, $U$  is the parallel (to the equilibrium magnetic field) velocity and $\vartheta$ is the gyrophase. Discretising the phase space allows us to represent the particle distribution function in terms of markers in the following way:
\begin{eqnarray}
\label{discretisation}
D_{z_c\rightarrow Z} F_H(Z,t) 
&\equiv& \int d^5Z' D_{z_c\rightarrow Z} F_H(Z',t)\delta^{(5)}(Z-Z') \\ \nonumber
&\simeq& \sum_l \overline{\Delta}_l F_{H}(Z_l)\delta^{(5)}(Z-Z_l).
\end{eqnarray} 
Here, $F_H$ is the energetic (``hot") particles, $D_{z_c\rightarrow Z}$ is the Jacobian of the transformation from canonical to gyrocenter coordinates,  $Z_l=Z_l(t)$ are the gyrocenter coordinates of the $l$-th macroparticle, which evolve according to the equations of motion. Moreover,
\begin{equation}
\delta^{(5)}(Z-Z_l)\equiv \delta(r-r_l)\delta(\theta-\theta_l)\delta(\phi-\phi_l)\delta(M-M_l)\delta(U-U_l)
\end{equation}
and
\begin{equation}
\overline{\Delta}_l\equiv \Delta_l(t) D_{z_c\rightarrow Z}(Z_l(t)),
\end{equation} 
with 
\begin{equation}
\Delta_l(t)\equiv [\Delta r \Delta \theta \Delta \phi \Delta M \Delta U]_l.
\end{equation}
Note that the integration in Eq.~\ref{discretisation} extends to a 5-D space, as $D_{z_c\rightarrow Z} F_H$ does not depend on the gyrophase $\vartheta$. Note also that  $\overline{\Delta}_l$, the phase-space volume element corresponding to the $l$-th macroparticle, is a constant of motion of the $l-$th macro-particle (Liouville's theorem).

The power transfer from energetic particles to the wave is given by $-d{\cal E}_H/dt,$ where 
\begin{equation}
\mathcal{E}_H=\frac{1}{m_H^3}\int d^6Z D_{z_c\rightarrow Z} F_H E
\end{equation}
is the total energy of fast particles, 
\begin{equation}
\label{eq:definizione_E}
E\simeq m_HU^2/2+M\Omega_H + e_H \langle \varphi \rangle - \frac{e_H U}{c} \langle\delta A_\| \rangle
\end{equation}
is the single particle energy, $m_H$ and $e_H$ are the energetic-particle mass and electric charge, respectively, $c$ is the speed of light, and $\langle\delta \varphi \rangle$ and $\langle\delta A_\| \rangle$ are the gyro-average of the fluctuating scalar potential and, respectively, the parallel (to the equilibrium magnetic field) component of the fluctuating vector potential. We can then write
\begin{eqnarray}
\label{e_h_dot}
-\frac{d \mathcal{E}_H}{dt} &=& -\frac{1}{m_H^3}\int d^6Z \left[\frac{\partial }{\partial t} (D_{z_c\rightarrow Z} F_H) E 
+ D_{z_c\rightarrow Z} F_H \frac{\partial E}{\partial t}\right]
= \\
&=& -\frac{1}{m_H^3}\int d^6Z \left[-\frac{\partial }{\partial Z^i} (D_{z_c\rightarrow Z} F_H \frac{dZ^i}{dt}) E 
+ D_{z_c\rightarrow Z} F_H \frac{\partial E}{\partial t}\right]= \nonumber \\ 
&=& -\frac{1}{m_H^3}\int d^6Z D_{z_c\rightarrow Z} F_H \left[\frac{\partial E}{\partial t} +
 \frac{dZ^i}{dt} \frac{\partial E}{\partial Z^i} 
\right]= \nonumber \\ 
&=& -\frac{1}{m_H^3}\int d^6Z  D_{z_c\rightarrow Z} F_H \frac{dE}{dt}.  \nonumber
\end{eqnarray}

After integrating over the gyrophase $\vartheta$ and replacing the quantity $D_{z_c\rightarrow Z} F_H$ by its discrete form, Eq.~\ref{discretisation}, we get
\begin{equation}
-\frac{d \mathcal{E}_H}{dt} =  -\frac{2 \pi}{m_H^3} \int d^5 Z  \sum_l \overline{\Delta}_l F_{H}(Z_l)\left(\frac{dE}{dt}\right)_l\delta^{(5)}(Z-Z_l).
\end{equation}
In the following, we will indicate by $Z$ the coordinates $(r,\theta,\phi,M,U)$, neglecting the gyrophase.

In order to identify the wave-particle resonances responsible for the destabilisation of the mode, it is worth reducing the dimensionality of the phase space by averaging the contribution of each marker over the poloidal and toroidal angles. In other words, we can define a power-transfer density in the 3-D space $(r,M,U)$ in the following way
\begin{equation}
-\frac{d \mathcal{E}_H}{dt} \equiv  \frac{1}{m_H^3} \int dr dM dU P(r,M,U,t),
\end{equation}
with $P(r,M,U,t)$ obtained from Eqs.~\ref{discretisation} and~\ref{e_h_dot} in the form
\begin{eqnarray}
P(r,M,U,t)&\equiv& -2\pi \int d\theta d\phi D_{z_c\rightarrow Z} F_H \frac{dE}{dt}= \nonumber \\
&\simeq&-2\pi \sum_l \overline{\Delta}_l F_{H}(Z_l)
\left(\frac{dE}{dt}\right)_l \delta(r-r_l)\delta(M-M_l)\delta(U-U_l).
\end{eqnarray}

Let us now consider a different coordinate system, $\tilde{Z}$. We can write the power transfer rate as
\begin{eqnarray}
\label{trasformazione_z_tilde}
-\frac{d \mathcal{E}_H}{dt} &=&  -\frac{2 \pi}{m_H^3} \int d^5 \tilde{Z} D_{Z \rightarrow \tilde{Z}}  \sum_l \overline{\Delta}_l F_{H}(Z_l)\left(\frac{dE}{dt}\right)_l\delta^{(5)}[Z(\tilde{Z})-Z_l] \\
&=& -\frac{2 \pi}{m_H^3} \int d^5 \tilde{Z}  \sum_l \overline{\Delta}_l F_{H}(Z_l)\left(\frac{dE}{dt}\right)_l\delta^{(5)}[\tilde{Z}-\tilde{Z}(Z_l)] \nonumber \\
&=& -\frac{2 \pi}{m_H^3} \int d^5 \tilde{Z}  \sum_l \overline{\Delta}_l F_{H}(Z_l)\left(\frac{dE}{dt}\right)_l\delta^{(5)}(\tilde{Z}-\tilde{Z}_l) \nonumber 
\end{eqnarray}
Here, we have used the $\delta$-function property
\begin{equation}
\int dx \delta(x-x_l) = \int dy D_{x \rightarrow y} \delta(x(y)-x_l) \equiv \int dy \delta(y-y(x_l)).
\end{equation}

Note that the transformation $Z \rightarrow \tilde{Z}$ could be very complicate and possibly a time dependent one. Equation~\ref{trasformazione_z_tilde} however shows that, if we choose these coordinates to perform the computation instead of the $Z$ ones, the only further quantities we need to know are the $\tilde{Z}$ coordinates of each marker, while we do not need to compute the Jacobian of the transformation (which could be, as we will see in the following, a difficult or even impossible task).

If the coordinate system $\tilde{Z}$ includes, as $Z$ does, the angles $\theta$ and $\phi$, we can define a power transfer density $\tilde{P}$ in the reduced space $(\tilde{r},\tilde{M},\tilde{U})$ as
\begin{equation}
-\frac{d \mathcal{E}_H}{dt} \equiv  \frac{1}{m_H^3} \int d\tilde{r} d\tilde{M} d\tilde{U} \tilde{P}(\tilde{r},\tilde{M},\tilde{U},t),
\end{equation}
or
\begin{eqnarray}
\label{eq:power_tilde}
\tilde{P}(\tilde{r},\tilde{M},\tilde{U},t)&\equiv& -2\pi \int d\theta d\phi D_{z_c\rightarrow Z} D_{Z \rightarrow \tilde{Z}} F_H \frac{dE}{dt}= \nonumber \\
&\simeq&-2\pi \sum_l \overline{\Delta}_l F_{H}(Z_l)
\left(\frac{dE}{dt}\right)_l \delta(\tilde{r}-\tilde{r}_l)\delta(\tilde{M}-\tilde{M}_l)\delta(\tilde{U}-\tilde{U}_l).
\end{eqnarray}
Here, we have used the notation $(\tilde{r},\tilde{M},\tilde{U})$ for the other $\tilde{Z}$ coordinates only conventionally, as there is no constraint on the meaning of any of these coordinates.

\section{Useful choices of phase space coordinates}
\label{sec:useful_choices}  
  
Among all the possible choices of phase-space coordinate system, some are particularly suited to yield pregnant information about particle dynamics. 
So, it can be worth replacing $r$ and $U$ by the angular momentum
\begin{equation}
\label{eq:definizione_p_phi}
P_{\phi} \simeq m_H R U + \frac{e_H R_0}{c} (\psi_{\textrm{eq}}-\psi_{\textrm{eq}0}) + \frac{e_H R_0}{c} \langle\delta A_\| \rangle,
\end{equation}
and the kinetic energy $E$; in Eq.~\ref{eq:definizione_p_phi} $R$ is the major-radius coordinate, $\psi_{\textrm{eq}}(r,\theta)$ is the poloidal flux of the equilibrium magnetic field ${\bf B}_{\textrm{eq}}\equiv R_0B_{\phi 0} \nabla \phi + R_0 \nabla \psi_{\textrm{eq}} \times \nabla \phi$ and the subscript $0$ indicates the quantities computed at the equilibrium magnetic axis. $P_{\phi}$ and $E$ are invariants of the unperturbed motion, and this allows to immediately obtain evidence, in the evolution of particle coordinates, of the effects of the mode-particle interaction.

An alternative and useful choice is that of the \textit{equatorial coordinates} $r_{\textrm{eq}}$ and $U_{\textrm{eq}}$, here defined as the value that the radial coordinate and, respectively, the parallel velocity of the particle with actual coordinates $(r,\theta,\phi,M,U)$ would assume at the next crossing of the equatorial plane at $\theta=0$ if its motion were unperturbed. In particular, for trapped\footnote{Note that here the word \textit{trapped} means ``trapped in the magnetic well". When the word means ``trapped in the potential well of the wave", it is explicitly indicated.} particles, we can convene to refer to the outermost equatorial crossing. These coordinates can be numerically computed, as functions of the $Z$ coordinates, from the conservation, in the unperturbed motion, of $P_{\phi}$, $M$ and $E$.
One of the merits of this choice is that the equatorial coordinates, along with the magnetic momentum $M$, are able to immediately identify unperturbed particle orbits, clearly separating passing particle contributions from trapped particle ones\footnote{Trapped-particle coordinates satisfy the following condition:
$$
\psi_{\textrm{eq}}\left(R_0\frac{m_H U_{\textrm{eq}}^2 -2M\Omega_{H0}\frac{r_{\textrm{eq}}}{R_0+r_{\textrm{eq}}}}
{m_H U_{\textrm{eq}}^2 +2M\Omega_{H0}\frac{R_0}{R_0+r_{\textrm{eq}}}},\pi
\right)
-\psi_{\textrm{eq}}(r_{\textrm{eq}},0)-\frac{m_Hc}{eR_0}(R_0+r_{\textrm{eq}})U_{\textrm{eq}} > 0.
$$
}.
This is only an approximation, of course, as particle motion is perturbed by fluctuating electromagnetic fields, but it is a satisfactory one until the perturbation is weak; that is, in the linear stage. During the nonlinear stage the description offered by this coordinate choice is still able to enlighten mode-particle dynamics, provided that the nonlinear perturbation of particle orbits keeps relatively small during a transit or bounce period.

A third possible choice resorts to exact invariants of the (perturbed) motion, if we are able to identify them. There is an immediate advantage in adopting coordinate systems that include such invariants: namely, the unambiguous identification of the most active particles in driving or damping the mode at each time. It can happen, during the nonlinear stage, that the mode is driven/stabilised by particles whose $Z$ coordinates are different from those of the particles driving/stabilising the mode during the linear stage. An interesting issue is distinguishing whether the nonlinearly driving/stabilising particles are actually the same as in the linear stage, with nonlinearly modified $Z$ coordinates, or different particles. In the former case, we could describe the mode-particle power exchange as a process in which the mode causes resonant particles to modify their orbits but, at the same time, adapts itself to further extract energy from them. In the latter case, the interaction between mode and resonant particles would apparently be described as a continuous search, by the mode, for the particles that can transfer power more efficiently.

Let us assume that the coordinates $(\tilde{r},\tilde{M},\tilde{U})$, or, more generally, some of them, are exact invariants of the perturbed motion. If we define $(\tilde{r}_{\textrm{max}}(t),\tilde{M}_{\textrm{max}}(t),\tilde{U}_{\textrm{max}}(t))$ the coordinates of the maximum power transfer at time $t$, any variation of the value corresponding to an exact invariant would indicate that new particles have replaced the previous ones in destabilising or stabilising the mode.

In practice, $M$ can be considered as a conserved quantity up to the required asymptotic expansion order in the collisionless gyrokinetic limit. So, a succession of different values of $M$ would readily be recognised as succession of different relevant particles taking part in the considered physical process. If, instead, the coordinates of the maximum power transfer corresponds to $r_{\textrm{max}}(t),M_{\textrm{max}},U_{\textrm{max}}(t)$, with constant $M_{\textrm{max}}$ (this is typically the case of modes driven unstable by transit resonances), the $Z$ coordinates do not allow us to distinguish between the two above situations (same particles versus succession of different particles). In order to succeed in this task, we have to identify more invariants, if they exist. In particular, if two more invariants can be selected, we get a complete identification, at any time, of the most driving or stabilising particles.

Another advantage in adopting a phase-space coordinate system including exact invariants of motion is that we can reduce the dimensionality of the free-energy source. This can be seen in the following way. The Vlasov equation for the energetic-particle distribution function can be written in the form
\begin{equation}
\frac{\partial \tilde{F}_H}{\partial t} + \frac{d \tilde{Z}^i}{dt} \frac{\partial \tilde{F}_H}{\partial \tilde{Z}^i} = 0,
\end{equation}
with $\tilde{F}_H(t,\tilde{Z})=F_H(t,Z(\tilde{Z})).$
If $\tilde{Z}^j$ is an invariant of the motion, we will have $d \tilde{Z}^j/dt =0$. Then, the gradient $\partial \tilde{F}_H / \partial \tilde{Z}^j$ will not take part in the mode-particle dynamics. We can express the same concept in a different way: if we cut the phase space into slices orthogonal to the invariant coordinate, there will be no particle flux from one slice to the other; that is, the nonlinear particle evolution will not yield any change of the distribution function along that coordinate. The dynamics of a given slice will then depend on other slice dynamics only through the fields, not because of mixing of the respective populations. This feature allows us to investigate separately each slice. Once the slices providing the most relevant resonances are identified, the details of mode-particle interaction can be analysed, e.g., by the so-called Hamiltonian mapping technique, which greatly enhance the resolution with which the resonance is examined by sampling the corresponding slice with a large number of test particles evolving in the fields self consistently computed by the considered simulation. The advantage of this approach is twofold: on one side, the dynamics of resonant particles is not obscured by the behaviour of non resonant ones; in particular, the local modifications of the distribution function can be highlighted even in the case of weak resonances, in spite of a general insensitivity of the overall population to the presence of the mode. On the other side, examining the evolution of a single-slice dynamics can make describing and identifying the saturation mechanisms easier.

In previous works~\cite{briguglio14,briguglio17}, we have investigated the nonlinear dynamics of Alfv\'{e}n modes characterised by a single toroidal number $n$ and a constant frequency $\omega$. In such cases, the quantity $C=\omega P_{\phi} - n E$ is an exact invariant of the perturbed motion. Indeed, the equations of motion in the Hamiltonian form read
\begin{equation}
\frac{dP_{\phi}}{dt}=-\frac{\partial H}{\partial \phi}
\end{equation}
 and 
 \begin{equation}
 \frac{dE}{dt}=\frac{\partial H}{\partial t},
 \end{equation} 
where $H$ is the single particle Hamiltonian. The conservation of $C$ immediately follows~\cite{white} from the dependence on time and toroidal angle of the Hamiltonian in the form $H=H(\omega t - n \phi)$. In those cases, we adopted a coordinate system $Z_C\equiv (r_{eq},\theta,\phi,M,C)$, performing the Hamiltonian mapping analysis of phase-space slices characterised by fixed values of $M$ and $C$. As explained in Ref.~\cite{briguglio14}, once a relevant slice has been fixed, the analysis proceeds storing the information concerning each test-particle orbit at its equatorial plane crossing; in particular, its coordinates $r_{eq}$ and $\phi$ ($\theta=0$ by definition) and the power transferred from/to the mode during the last poloidal orbit. In this way, the distribution function for particles belonging to the considered slice can be computed, along with its integral over $\phi$, which depends only on $r_{eq}$ and represents the free-energy source with which the slice can contribute to the destabilisation of the mode. Time evolution of such integral can then be related to that of the power transfer, getting insight in the saturation dynamics.

If multiple toroidal numbers are present and/or the frequency is not constant, C is no longer an exact invariant of the perturbed motion. This is the general case, of course, both because a large $n$ spectrum of Alfv\'{e}n modes can be destabilised in tokamak equilibria and chirping modes (that is, modes with time-varying frequency) are often observed in experiments. In these cases, 
slices orthogonal to the $C$ direction would not be isolated and the dynamics would be influenced also by the gradient of the distribution function along $C$ and the corresponding fluxes crossing each slice. In other words, the evolution of the free-energy source associated to a certain slice could not be adequately reproduced by simply looking at the gradient of the distribution function along $r_{eq}$.

It is however possible to envisage other, alternative, coordinate transformations, capable to yield the same advantages as the ``exact-invariant" one. In this paper, we propose to look at simple constants of motion (that is, quantities that are conserved during particle motion), without addressing the question whether an isolating integral or global invariant exists~\cite{lichtemberg1983,lichtemberg2010} besides $M$ (and can be considered instead of $C$) and which form it assumes. The reason is that any constant of motion could be used instead of the additional global invariant (isolating integral) of motion, 
as it will be shortly discussed at the end of this Section.
Identifying constants of motion is an elementary task, as the initial coordinates of a particle are, by definition, conserved along its motion. The transformation that links a certain coordinate system, $\tilde{Z}$, to the system represented by its initial values $\tilde{Z}_0$ can be obtained by the equations of motion
\begin{equation}
 \frac{d\tilde{Z}^i}{dt}=\tilde{V}^i(\tilde{Z},t)
 \end{equation}
in the following form
\begin{equation}
\label{eq:zeta_to_zeta0}
\tilde{Z}_0^i=\tilde{Z}^i - \int_0^t dt'  \tilde{V}^i(\tilde{Z},t').
\end{equation}
More generally, we can consider a coordinate transformation in which only some of the $\tilde{Z}^i$ are replaced by the corresponding $\tilde{Z}_0^i$, or by any function of them; in the following, any of these coordinate system will be named an \textit{initial-value coordinate system}.
Note that the choice of an initial-value coordinate system is, in general, impractical in the context of an analytical treatment, as it requires complicated backward transformations from the current particle coordinates to the initial ones. In the frame of a numerical approach, however, it is sufficient to memorise, in addition to the current coordinates of each marker, its initial coordinates.

On the basis of the arguments exposed in this Section, we find worth considering a three-constant system, $\tilde{Z}_3\equiv (r_{\textrm{eq0}},\theta,\phi,M,U_{\textrm{eq0}})$ for the identification of instantaneous maxima of the power transfer, and a two-constant system, $\tilde{Z}_2\equiv (r_{\textrm{eq}},\theta,\phi,M,U_{\textrm{eq0}})$, for investigating the nonlinear saturation dynamics. Here, $r_{\textrm{eq0}}$ and $U_{\textrm{eq0}}$ are the initial values of the equatorial coordinates $r_{\textrm{eq}}$ and $U_{\textrm{eq}}$ defined above.
Note that in the latter system the first coordinate is the current value of the equatorial radial coordinate: it is not a constant of motion. The corresponding gradient of the distribution function plays the role of a free-energy source for the mode destabilisation, and the associated fluxes contribute to mode saturation.

In concluding this section, we would like to emphasise the following points. For the purpose of a complete and unambiguous identification of instantaneous power transfer maxima, the use of any triad of invariant coordinates is equally effective, regardless of whether they are global invariants or not. 
As far as the goal of simplifying the description of nonlinear dynamics is concerned, the conclusion is similar, but requires some clarification. First of all, it is apparent how the choice of any invariant coordinate (besides $M$) allows a subdivision of the phase space into slices that do not exchange particles with each other. The fact, however, that these slices are treated as isolated is a mere artifice, consisting in treating the fluctuating field as an exogenous field: this is what is done in sampling each of the considered slices by a large number of test particles that passively evolve according to the field calculated, previously, in a self-consistent simulation. In the evolution of the real system, however, each slice communicates with the others, though not exchanging particles, precisely through the field. Moreover, one slice can take over from another in destabilising the mode, due to the modification of the frequency and/or the structure of the mode. One cannot therefore expect \textit{a priori} that the relevant dynamics will remain confined to a particular slice. 
In principle it is also possible that such a confinement occurs in correspondence with the choice of a given invariant, but, so far, no recipe is known to guide this choice in the general case: 
indeed, having such a recipe would be equivalent to a full ability to predict the nonlinear evolution of the system, even in very complex situations. 
And all this is irrespective of the particular invariant (global or not) included in the coordinate system. 
From this point of view there is, then, in general, no qualitative difference between the adoption of a coordinate system including a global invariant and that of a system containing any other quantity conserved during the motion of the particle: in both cases, only \textit{a posteriori} one can establish whether the choice made is effective for the purpose of a simplified description: namely, whether it allows to restrict the analysis to a small number of relevant slices.

\section{Setting of the numerical experiment}
\label{sec:setting} 

In this Section, the numerical experiment envisaged for analysing the dynamics of a chirping Alfv\'{e}n mode is presented.

\begin{figure}
\includegraphics[width=0.6\textwidth]{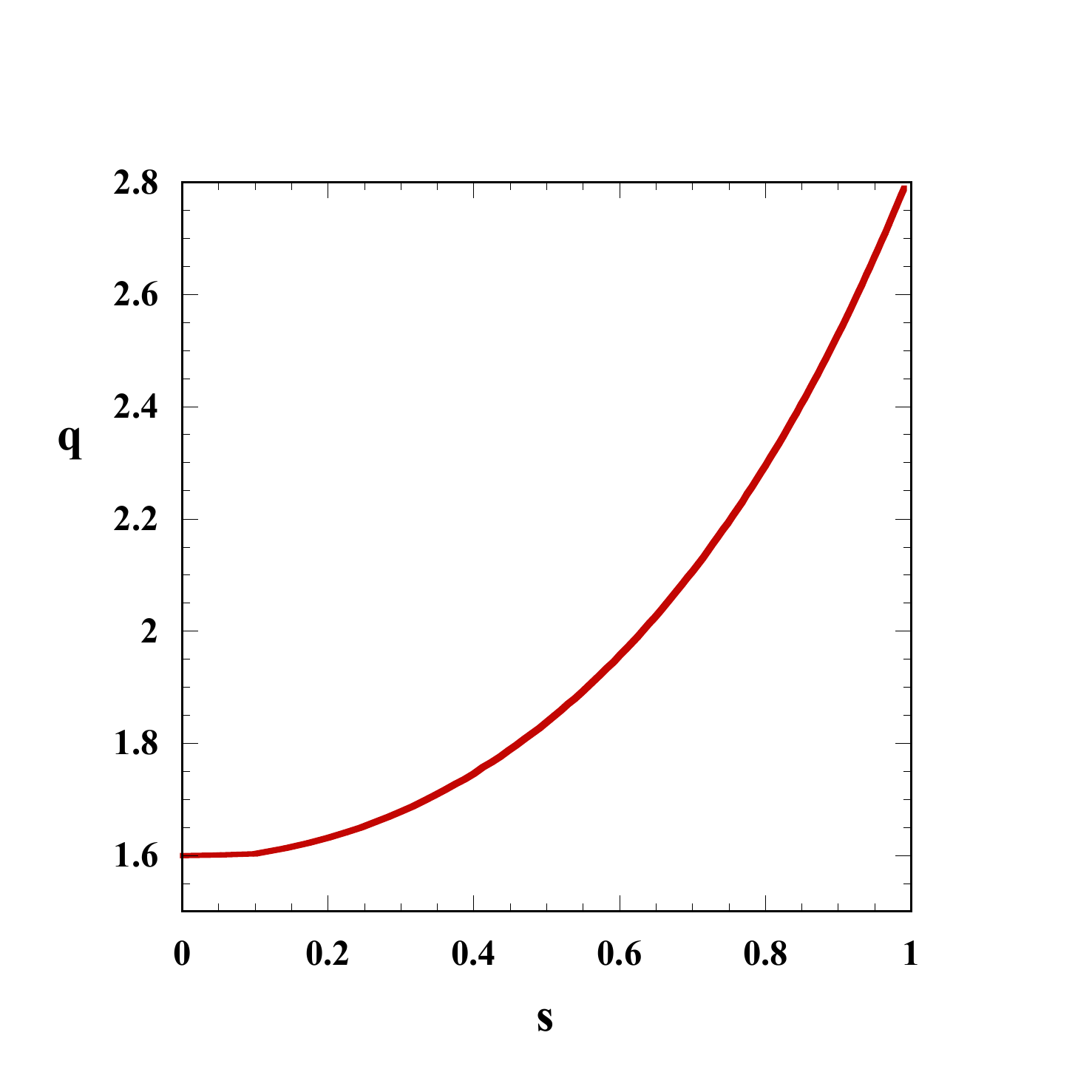}
    \caption{Radial profile of the safety factor $q$. Here $s\equiv \sqrt{1-\psi_{\textrm{eq}}/\psi_{\textrm{eq0}}}$.}
    \label{fig:q_profile} 
\end{figure}

\begin{figure}
\includegraphics[width=0.6\textwidth]{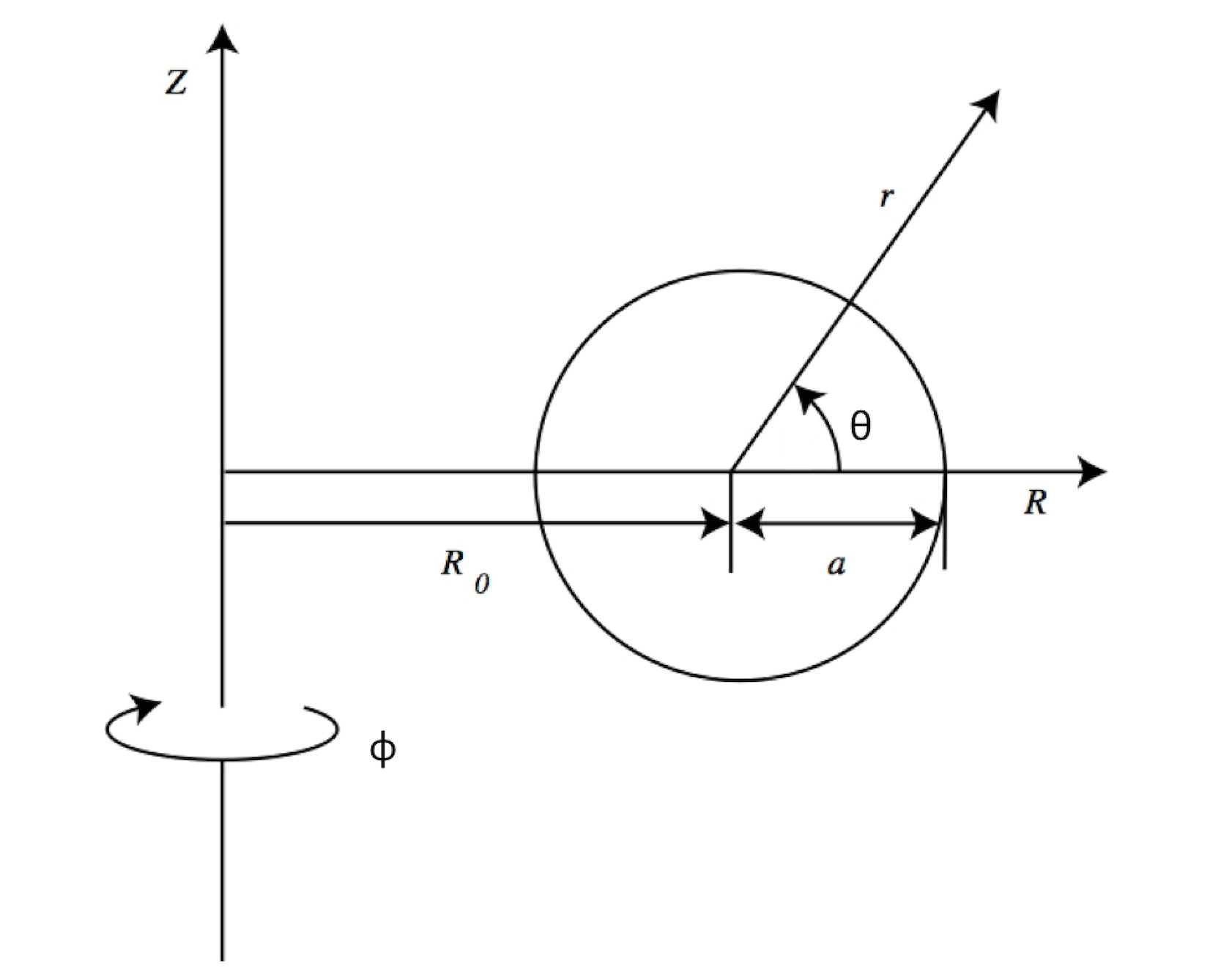}
    \caption{Toroidal coordinate system $(r,\theta,\phi)$ for a tokamak plasma equilibrium with major radius $R_0$ and minor radius $a$.}
    \label{fig:fig_toroid} 
\end{figure}

\begin{figure}
\includegraphics[width=0.6\textwidth]{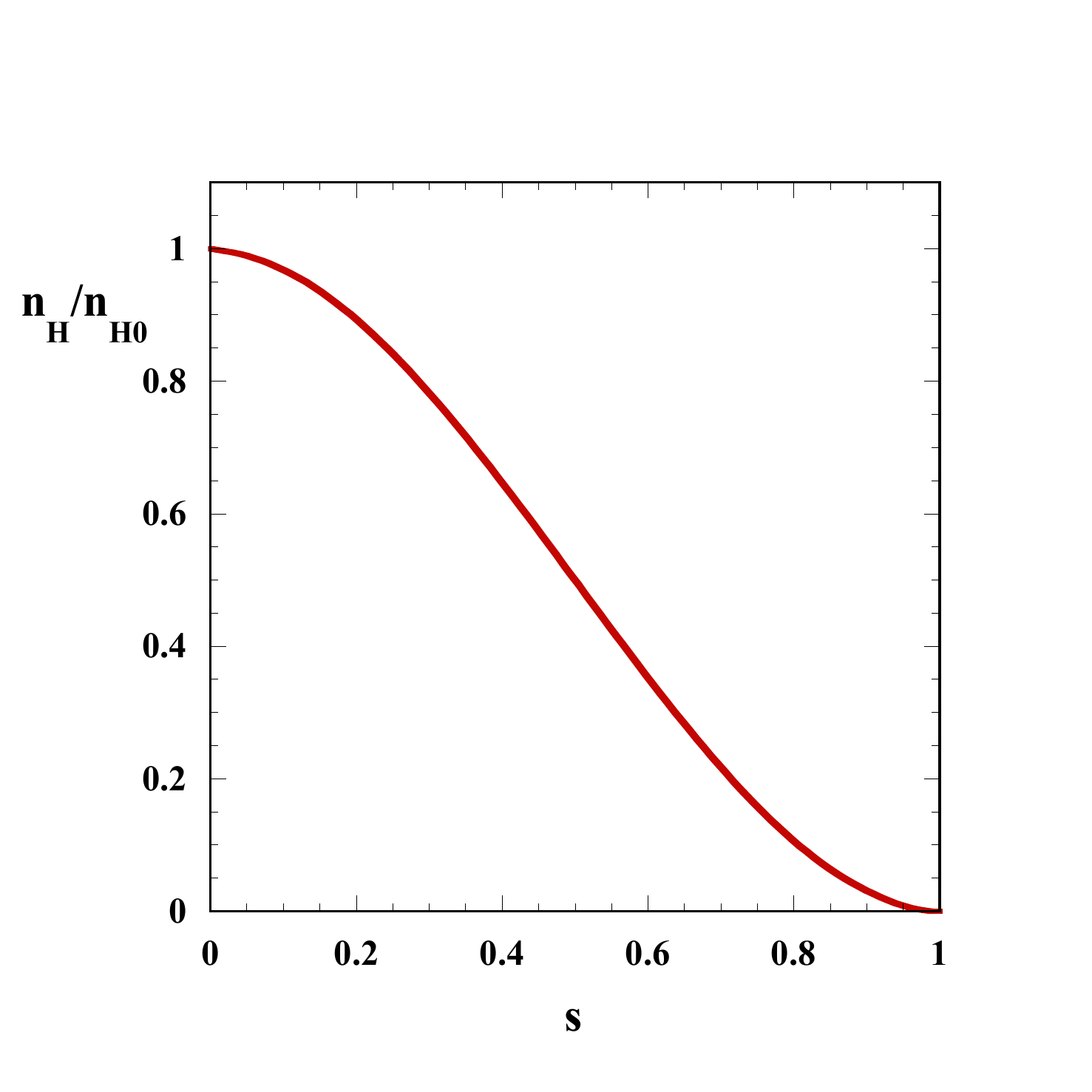}
    \caption{Energetic-particle density profile.}
    \label{fig:ep_density_profile} 
\end{figure}

We have simulated, by the hybrid MHD-gyrokinetic code XHMGC~\cite{briguglio95pop,briguglio98,wang11}, the evolution of an $n=3$ Energetic Particle Mode~\cite{chen94}, with poloidal harmonics $m=4\div9$, in a large aspect ratio Tokamak equilibrium ($R_0/a=10$, with $a$ being the minor radius of the torus), characterised by a safety factor $q$ monotonically increasing from $1.6$ to $2.8$ (Fig.~\ref{fig:q_profile}). The spatial coordinate system is shown in Fig. ~\ref{fig:fig_toroid}. A deuterium plasma is considered, along with an energetic deuteron population. Both bulk-ion and energetic-ion populations are treated kinetically\footnote{In the following, we focus on the energetic-ion population. Any reference to \textit{particles} or to related quantities must be understood, unless otherwise specified, as referring to that population.}. The relevant dimensionless parameters are the following: $n_{H0}/n_{i0}=0.01$, $v_{H0}/v_{A0}=0.3$, $\rho_{H0}/a=0.01$, $v_{i0}/v_{A0}=0.06$, $\rho_{i0}/a=0.002$. Here, $n_H$ and $n_i$ are the densities of the two species, $v_H$ and $v_i$ their thermal velocities, $\rho_H$ and $\rho_i$ their thermal Larmor radii, $v_A$ is the Alfv\'{e}n speed and subscripts $0$ denote quantities computed at the magnetic axis. The density profile of energetic particles is shown in Fig.~\ref{fig:ep_density_profile}, while their temperature profile, as well as bulk-ion density and temperature profiles are assumed flat. The initial distribution function is Maxwellian for both species. For the energetic particles, it is multiplied by an anisotropy factor $\Xi(\alpha;\alpha_{0}, \kappa)$, with
\begin{equation}
\Xi(\alpha;\alpha_{0}, \kappa) \equiv \frac{4}{\kappa \sqrt{\pi}} \frac{\exp \left[ - \left( \frac{\cos \alpha - \cos \alpha_{0}}{\kappa} \right)^{2} \right]}{ \mathrm{erf} \left( \frac{1 - \cos \alpha_{0}}{\kappa} \right) +\mathrm{erf} \left( \frac{1 + \cos \alpha_{0}}{\kappa} \right)},
\end{equation}
$\cos\alpha=U/(U^2+2M\Omega_H/m_H)$, $\alpha_{0}=0$ and $\kappa=0.1$.
The small initial relaxation due to the fact that such functions are not equilibrium ones is inhibited.

\begin{figure}
\includegraphics[width=0.5\textwidth]{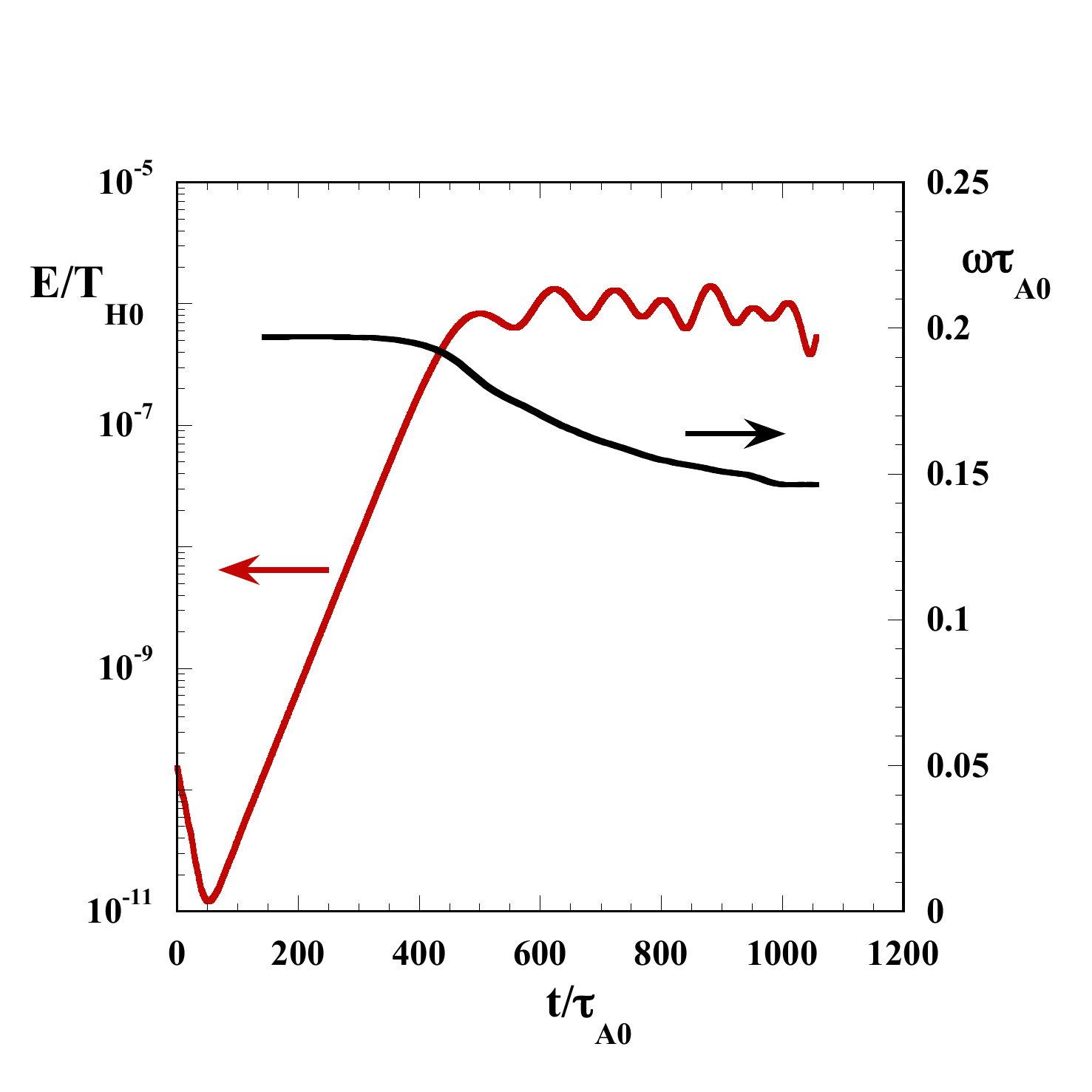}
    \caption{Time evolution of the mode energy (red) and frequency (black).}
    \label{fig:energy_vs_t} 
\end{figure}

\begin{figure}
\includegraphics[width=0.4\textwidth]{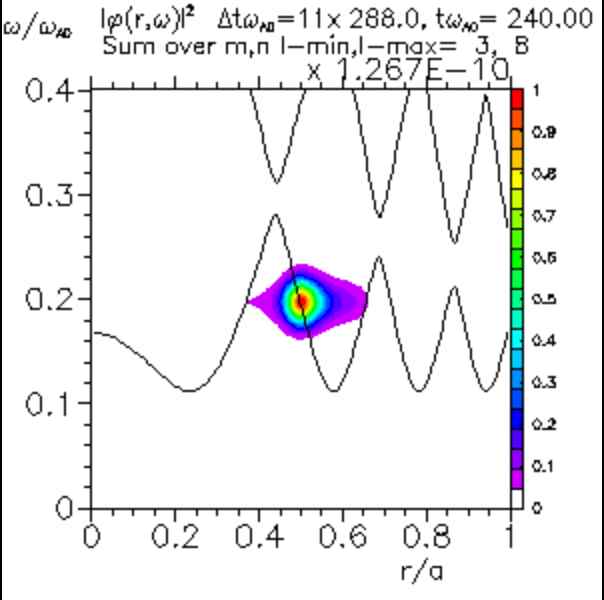}
\includegraphics[width=0.4\textwidth]{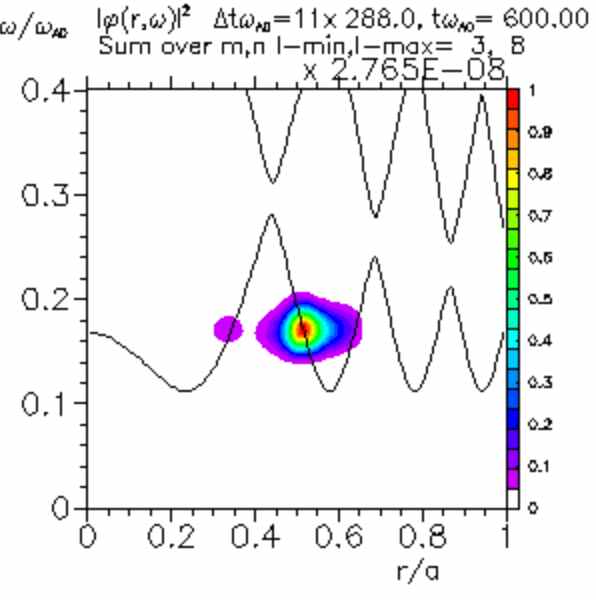}
 \caption{Spectrum of the scalar potential in the space $(r,\omega)$ at two times: $t=240 \tau_{A0}$, during the linear stage (left), and $t=600 \tau_{A0}$,  the nonlinear one (right). The Alfv\'{e}n continuum is also represented}
    \label{fig:spectrum_r_omega} 
\end{figure}

\begin{figure}
\includegraphics[width=0.4\textwidth]{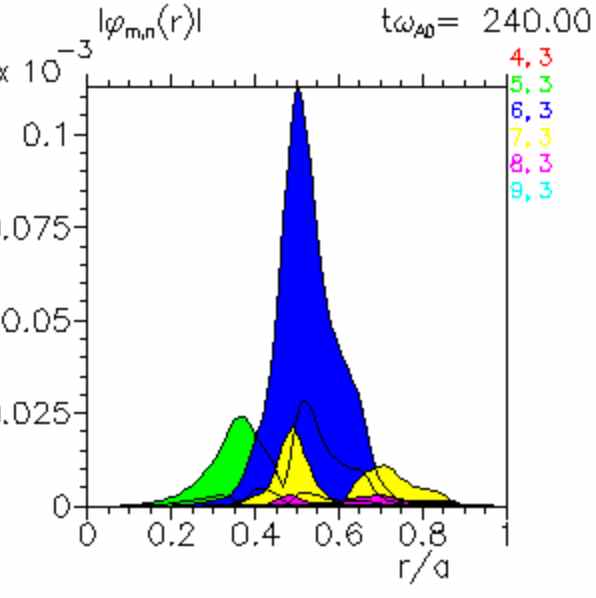}
\includegraphics[width=0.4\textwidth]{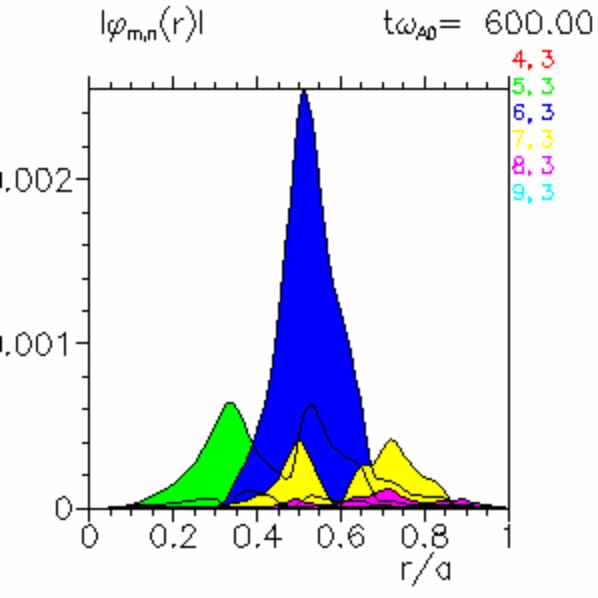}
    \caption{Radial structure of the poloidal harmonics of the mode at two times: $t=240 \tau_{A0}$ (left) and $t=600 \tau_{A0}$ (right).}
    \label{fig:poloidal_harmonics} 
\end{figure}

Figure~\ref{fig:energy_vs_t} shows the time evolution of mode energy and frequency. The first saturation of the mode is followed by a stage in which the mode gets further drive and saturates again, several times. A significant decrease of frequency is observed in the nonlinear phase (see also Fig.~\ref{fig:spectrum_r_omega}, which shows the power spectrum of the scalar potential in the space $(r,\omega)$ at two times: $t=240 \tau_{A0}$ (with $\tau_{A0}\equiv R_0/v_{A0}$), during the linear phase, and $t=600 R_0/v_{A0}$, in the nonlinear one). In spite of the significant nonlinear variation of the frequency, the radial structure of the mode is scarcely modified, as it can be seen from Fig.~\ref{fig:poloidal_harmonics}.

\section{Search for the most relevant phase-space structures}
\label{sec:structures}

Our aim is yielding a detailed description of the nonlinear evolution of the mode. 
As stated before, the frequency variation causes $C$ to lose its exact invariance. Although, in the presente case, the rate of variation of $C$ keeps quite small, instead of controlling the amount of non conservation and evaluating its relevance, we prefer to include in the coordinate system different conserved quantities, as proposed in Sec.~\ref{sec:useful_choices}:
the initial values of minor radius and/or parallel velocity; more precisely, those of the equatorial minor radius and/or the equatorial parallel velocity. Then, while pushing particles in the usual gyrocenter coordinates $Z$, we adopt two alternative systems: a three-constants system, $\tilde{Z}_3\equiv (r_{\textrm{eq0}},\theta,\phi,M,U_{\textrm{eq0}})$ for the identification of instantaneous maxima of the power transfer, and a two-constant one, $\tilde{Z}_2\equiv (r_{\textrm{eq}},\theta,\phi,M,U_{\textrm{eq0}})$, for investigating the nonlinear saturation dynamics, with $r_{\textrm{eq0}}$ and $U_{\textrm{eq0}}$ being the initial values of the equatorial coordinates $r_{\textrm{eq}}$ and $U_{\textrm{eq}}$.

The first issue we want to face is the identification of the most relevant resonant structures; that is, the phase-space regions playing the most important role in destabilising or stabilising the mode. To this aim, we consider the power transfer from energetic particles to the mode averaged over the poloidal and toroidal angles, computed, after discretising the energetic-particle distribution function, in the form given by Eq.~\ref{eq:power_tilde}.
As discussed in Sec.~\ref{sec:useful_choices}, different coordinate systems $\tilde{Z}$ can be adopted to describe the phase space.

The usual choice made in investigations based on gyrokinetic simulation~\cite{novikau} consists in referring the power transfer to the actual gyrocenter coordinates ($\tilde{Z}\equiv Z$). As stated above, it can be more interesting referring the power transfer to an equatorial-coordinate system, $\tilde{Z}\equiv (r_{\textrm{eq}},\theta,\phi,M,U_{\textrm{eq}})$. As the triad $(r_{\textrm{eq}},M,U_{\textrm{eq}})$ fully identify a poloidal particle orbit, this choice is justified by the interest in identifying the contribution of each particle along its whole orbit, rather than the instantaneous one. To state this in a slightly different way, until a certain particle maintains its unperturbed orbit, it will contribute to the power transfer to the same point in the space 
$(r_{\textrm{eq}},M,U_{\textrm{eq}})$, while it would spread its contribution on a curve 
$U(\theta)=U(r(\theta),M)$ in the space $(r,M,U)$.

\begin{figure}
\includegraphics[width=0.3\textwidth]{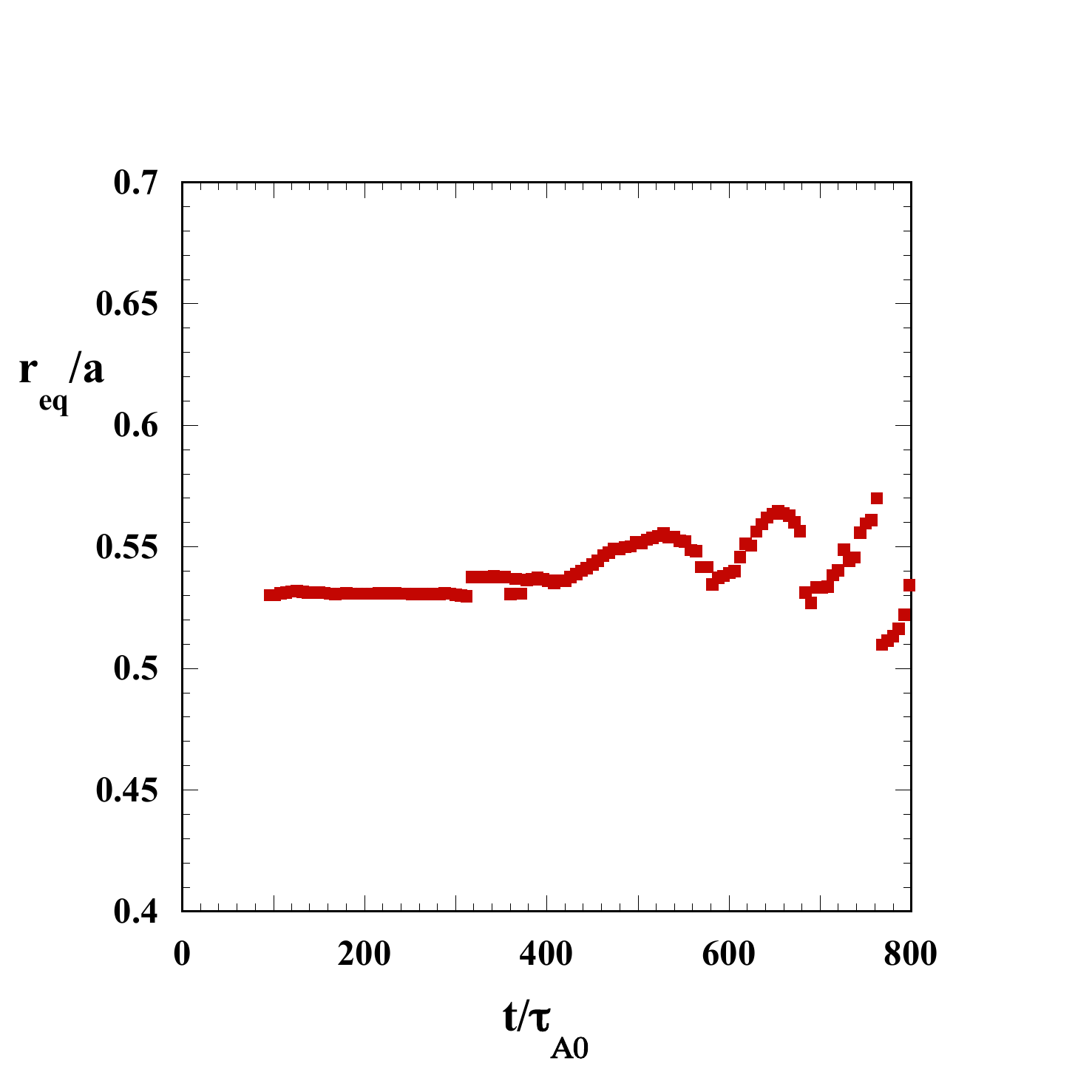}
\includegraphics[width=0.3\textwidth]{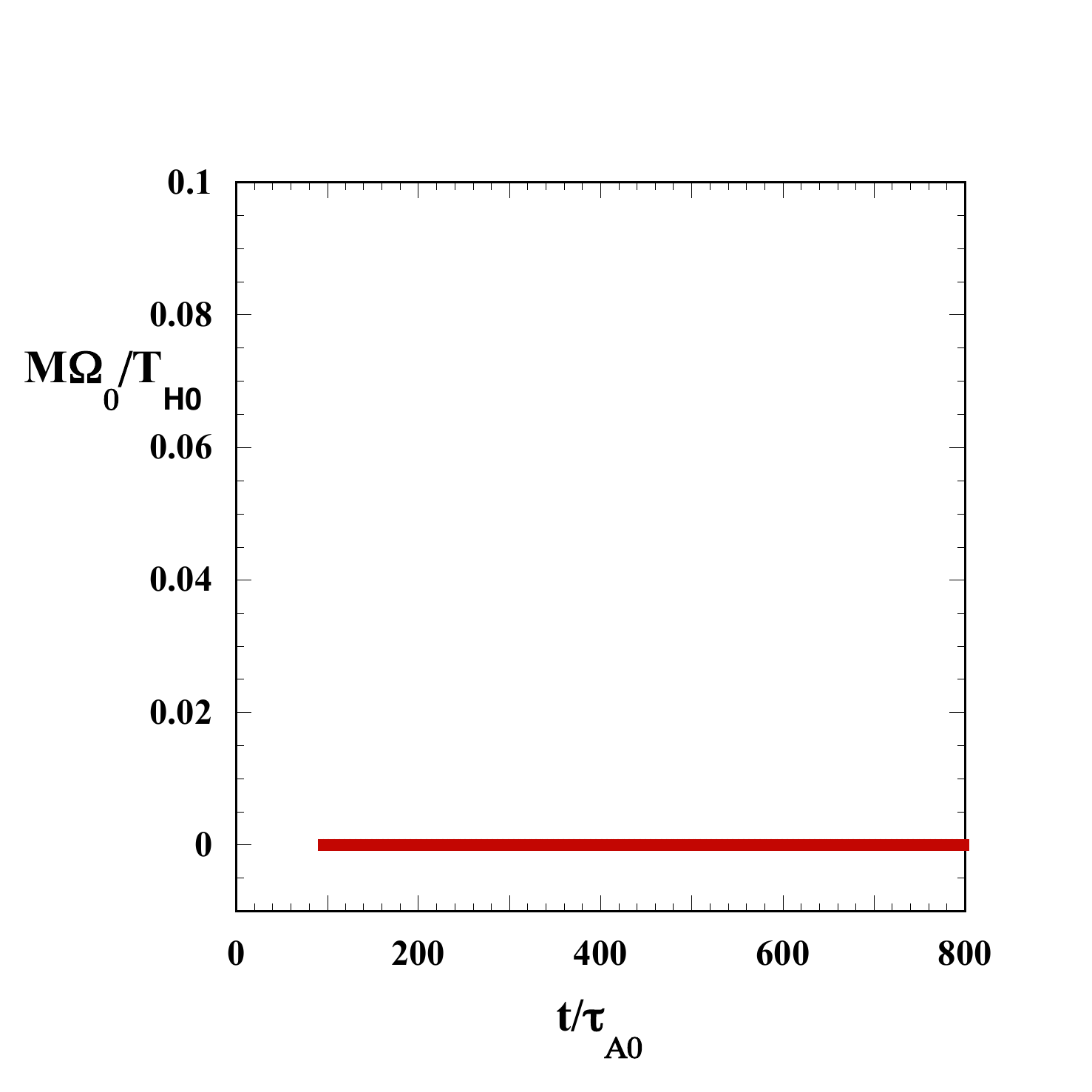}
\includegraphics[width=0.3\textwidth]{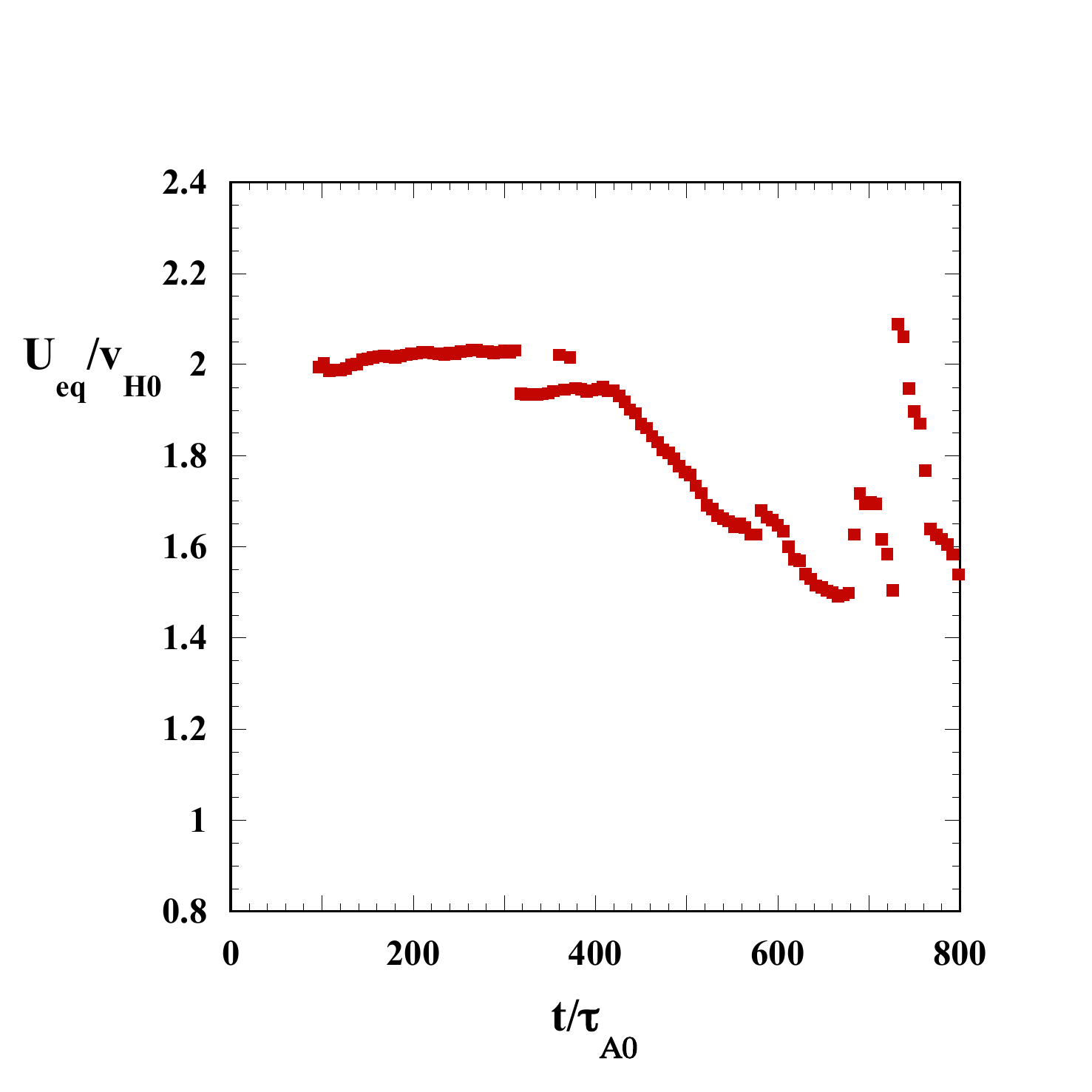}
    \caption{Coordinates, in the reduced phase space $(r_{\textrm{eq}},M,U_{\textrm{eq}})$, of the maximum of the power transfer density averaged over $\theta$ and $\phi$. 
    }
    \label{fig:max_actual} 
\end{figure}

\begin{figure}
\includegraphics[width=0.3\textwidth]{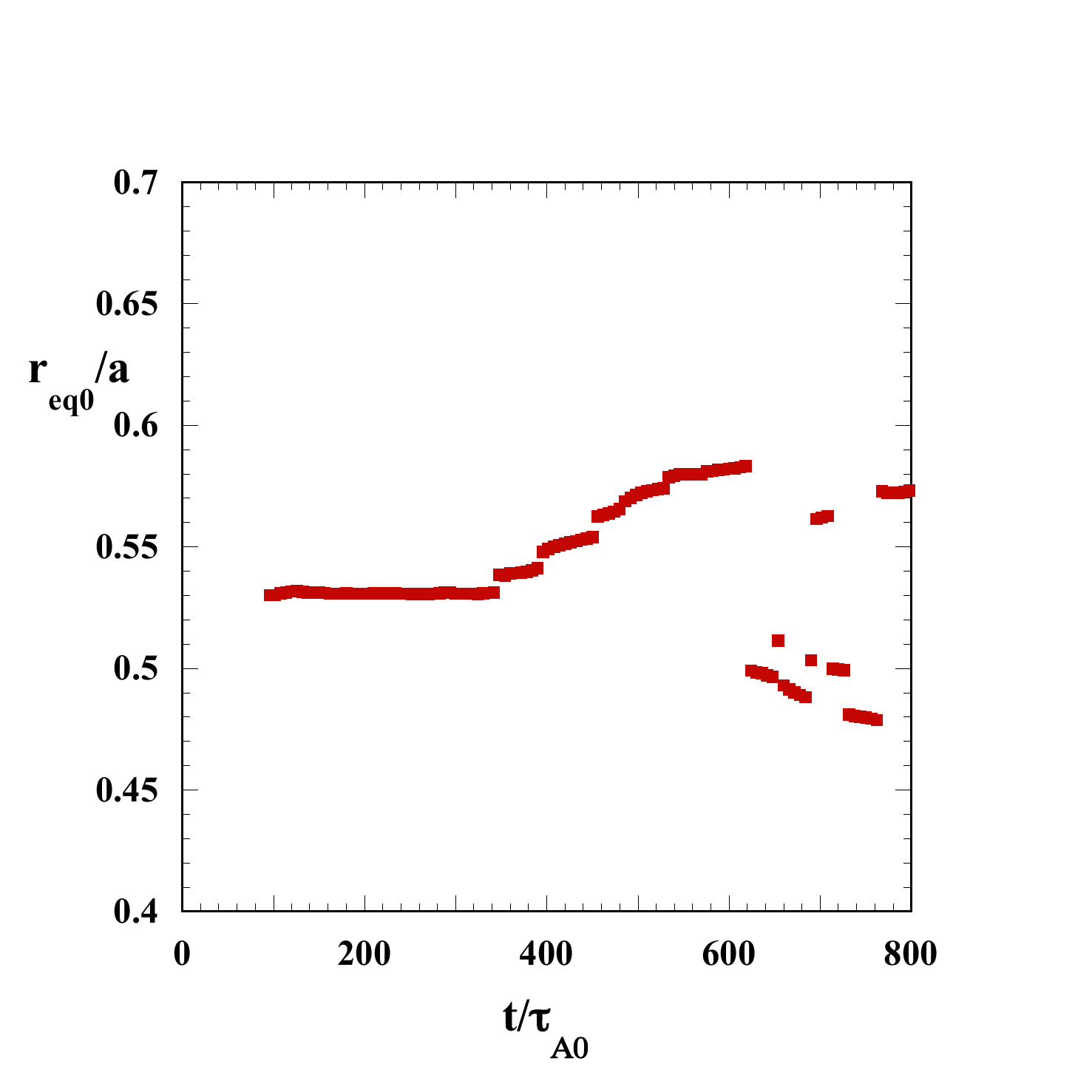}
\includegraphics[width=0.3\textwidth]{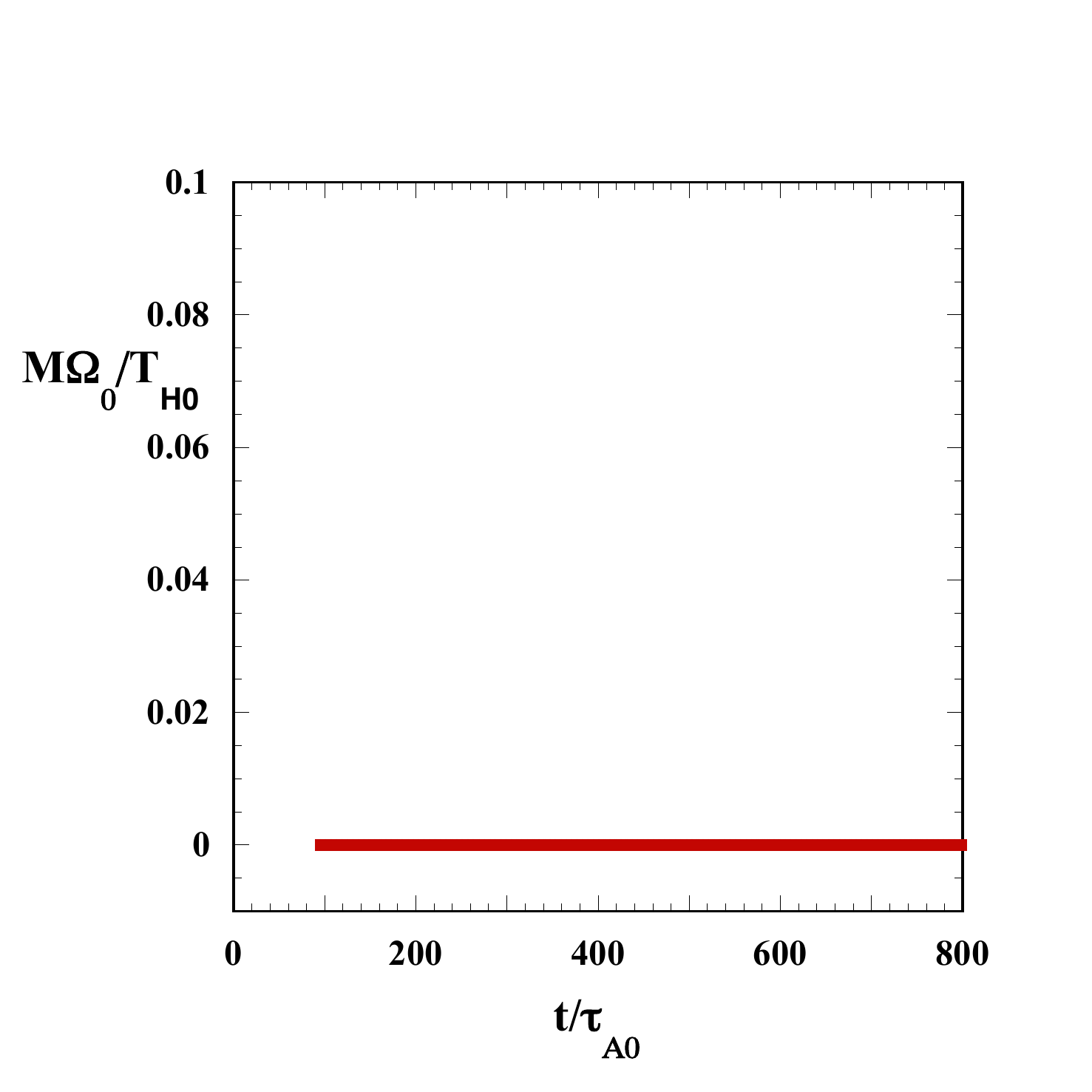}
\includegraphics[width=0.3\textwidth]{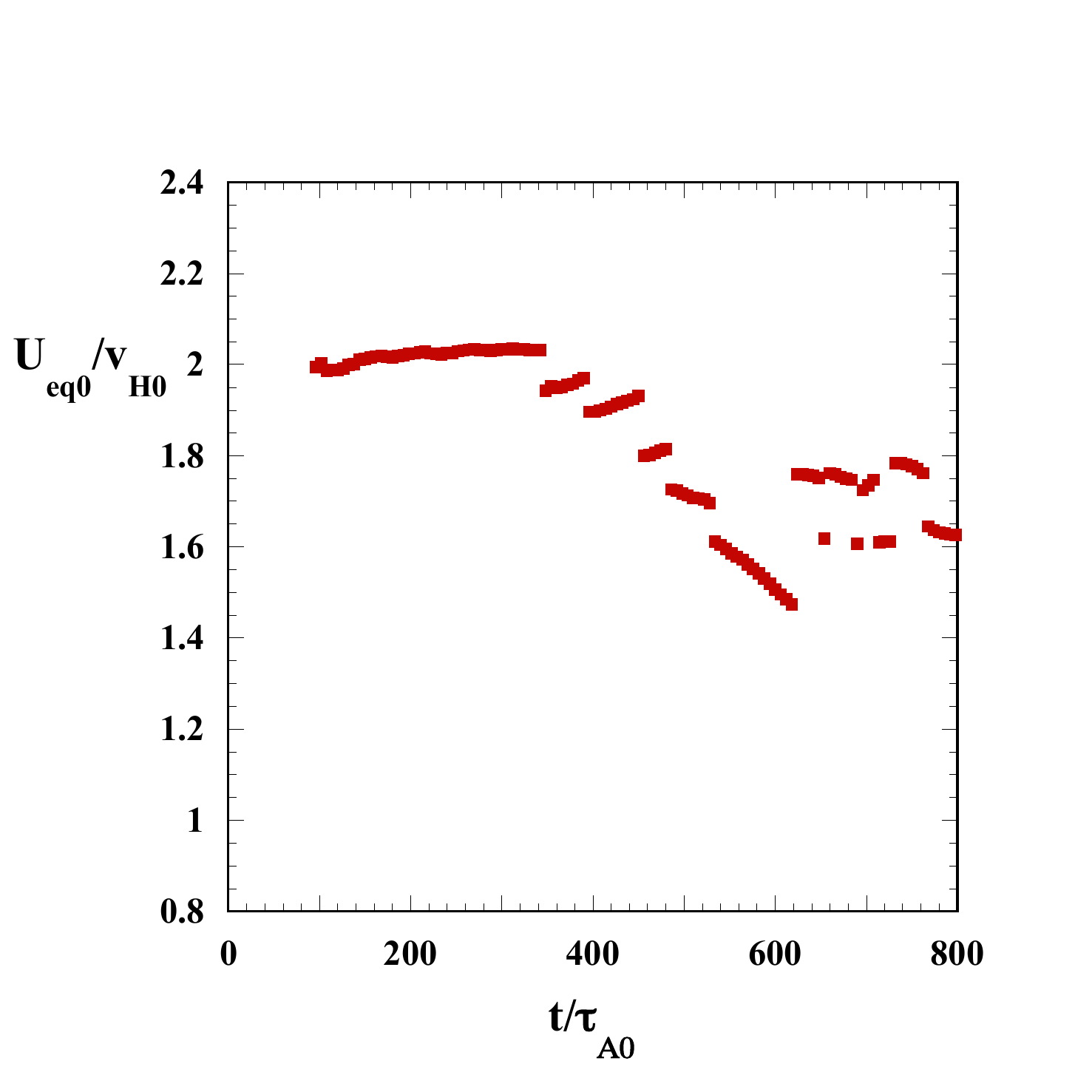}
    \caption{Same as Fig.~\ref{fig:max_actual}, but for coordinates $(r_{\textrm{eq0}},M,U_{\textrm{eq0}})$.}
    \label{fig:max_initial} 
\end{figure}

Figure~\ref{fig:max_actual} shows the time evolution of the coordinates of the maximum of $\tilde{P}(r_{\textrm{eq}},M,U_{\textrm{eq}},t)$ for the considered case. 
We see that the drive is always related to co-passing particles ($M=0$, $U_{\textrm{eq}}>0$). During the nonlinear stage the maximum power transfer coordinates $r_{\textrm{eq max}}(t)$ and $U_{\textrm{eq max}}(t)$ change, while $M_{\textrm{max}}(t)$ keeps constant. 
This time dependence is consistent with both a nonlinear drive yielded by the same particles responsible for the linear destabilisation, with coordinates progressively modified by the interaction with the mode; and with a drive yielded by a succession of different particles.
As discussed in Sec.~\ref{sec:useful_choices}, the fact that there is no time dependence of $M_{\textrm{max}}$ (with $M$ being the only invariant coordinate in the considered system) makes it impossible to readily show which of these two interpretations is correct. To do so unambiguously, we adopt the $\tilde{Z}_3$ coordinate system and, after averaging the power transfer density over the toroidal and poloidal angles, look at its maxima in the reduced space $(r_{\textrm{eq0}},M,U_{\textrm{eq0}})$ (Fig.~\ref{fig:max_initial}). In this case, the time variation of $r_{\textrm{eq0 max}}(t)$ and $U_{\textrm{eq0 max}}(t)$ clearly indicates that the maximum drive is generally due, at different times, to different particles (characterised by different initial coordinates).

\begin{figure}
\includegraphics[width=0.7\textwidth]{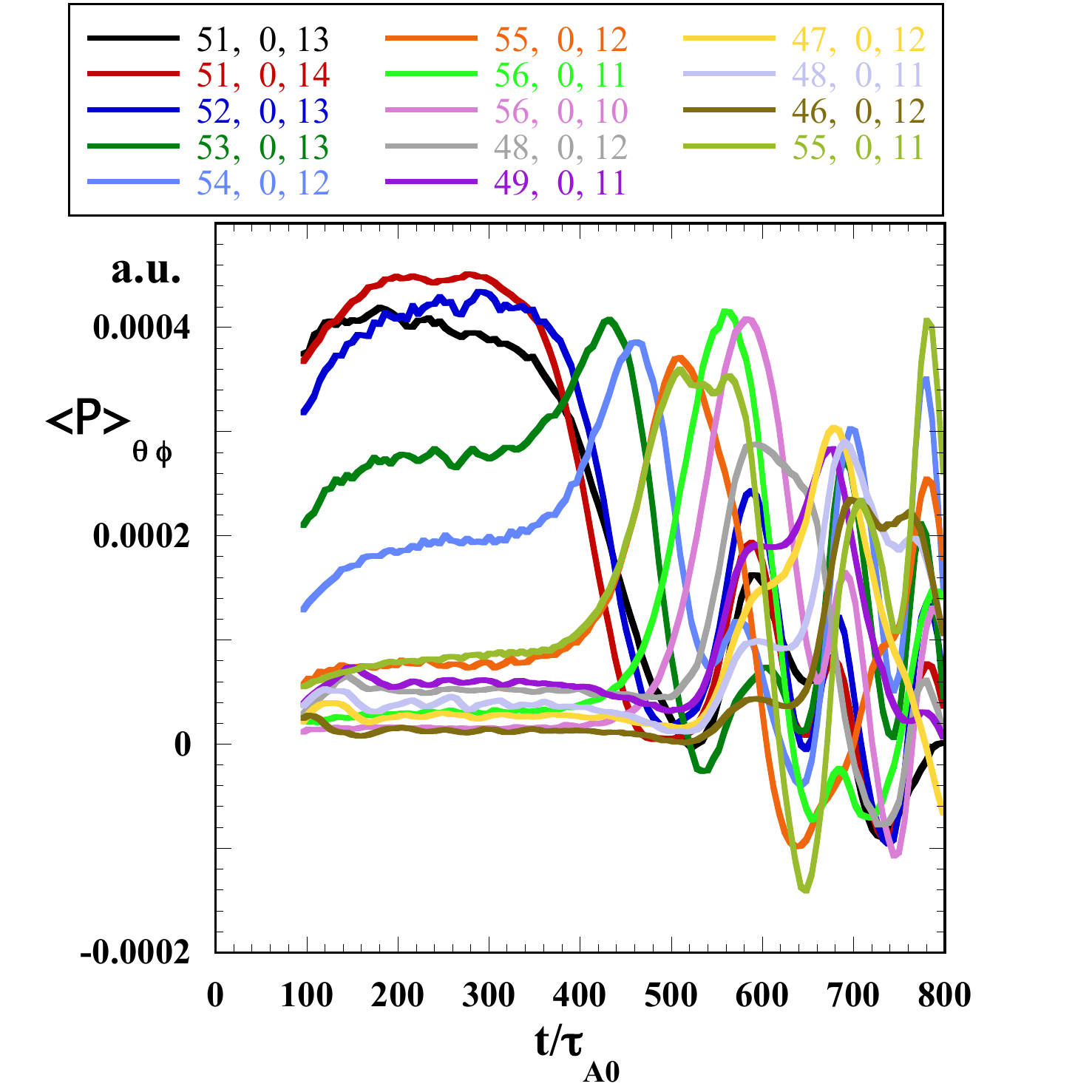}
    \caption{Time evolution of the power transfer (averaged over $\theta$ and $\phi$) at the most relevant grid points in the discretised space $(r_{\textrm{eq0}},M,U_{\textrm{eq0}})$. Only those grid points
    that yield a maximum of the power transfer at some time in the interval $[0,800 \tau_{A0}]$ have been considered. Grid points are labelled by three indices. The first one, relative to the radial coordinate, ranges from $46$ ($r_{\textrm{eq0}}\simeq 0.48 a$) to $56$ ($r_{\textrm{eq0}}\simeq 0.58 a$). The spacing between adjacent points is $\Delta r_{\textrm{eq0}}\simeq 0.01 a$. The second one, relative to $M$, is $0$ for all grid points ($M=0$). The third one, relative to the parallel velocity, ranges from $10$ ($U_{\textrm{eq0}}\simeq 1.48 v_{H0}$) to $14$ ($U_{\textrm{eq0}}\simeq 2.07 v_{H0}$), with a grid spacing $\Delta U_{\textrm{eq0}}\simeq 0.1475 v_{H0}$.}
    \label{fig:maxpower_3d_vs_t} 
\end{figure}

Figure~\ref{fig:maxpower_3d_vs_t} shows the quantity $\tilde{P}$ versus time at the grid points, in the discretised space $(r_{\textrm{eq0}},M,U_{\textrm{eq0}})$, that result to host the maximum value for some time, in the time interval $[0,800 \tau_{A0}]$\footnote{Note that the maxima shown in Fig.~\ref{fig:max_initial} are computed, at each time, by looking at the appropriate grid point considered in Fig.~\ref{fig:maxpower_3d_vs_t}, approximating $\tilde{P}$ near this point by a 2nd-order polynomial and looking for its closest maximum.}. We see that grid points dominating the drive during the linear stage, progressively lose importance, being replaced by grid points less and less contributing to linear destabilisation.

Computing power transfer in the coordinate system $\tilde{Z}_3$ is worth to label particles providing the maximum-drive at different times, but it is not suited for investigating the saturation dynamics. To address this issue, we have to adopt a coordinate system with only two constants of motion, in such a way that particle fluxes, responsible for the relaxation of the free-energy source and the related saturation mechanism can be observed. Then, we resort to the $\tilde{Z}_2=(r_{\textrm{eq}},\theta,\phi,M,U_{\textrm{eq0}})$ coordinate system, containing the two constants of motion $M$ and $U_{\textrm{eq0}}$. In this way, the gradient of the distribution function along the equatorial radial coordinate $r_{\textrm{eq}}$ represents the free-energy source, and the corresponding flow contributes to mode saturation.

In principle, we can expect that in the present case, differently from the cases characterised by single toroidal number and constant frequency~\cite{briguglio14,briguglio17}, in addition to the linear-stage resonance, further resonances may contribute the mode drive during the nonlinear stage. 
Then, we have first to identify the most relevant structures in the 2-D reduced phase space $(M,U_{\textrm{eq0}})$. Once these structures are identified, we can analyse each of them independently of the others. Indeed, the phase space can be cut into distinct slices, orthogonally to the subspace $(M,U_{\textrm{eq0}})$, with flows entering or leaving the slices being forbidden by the conservation of $M$ and $U_{\textrm{eq0}}$.

\begin{figure}
\includegraphics[width=0.5\textwidth]{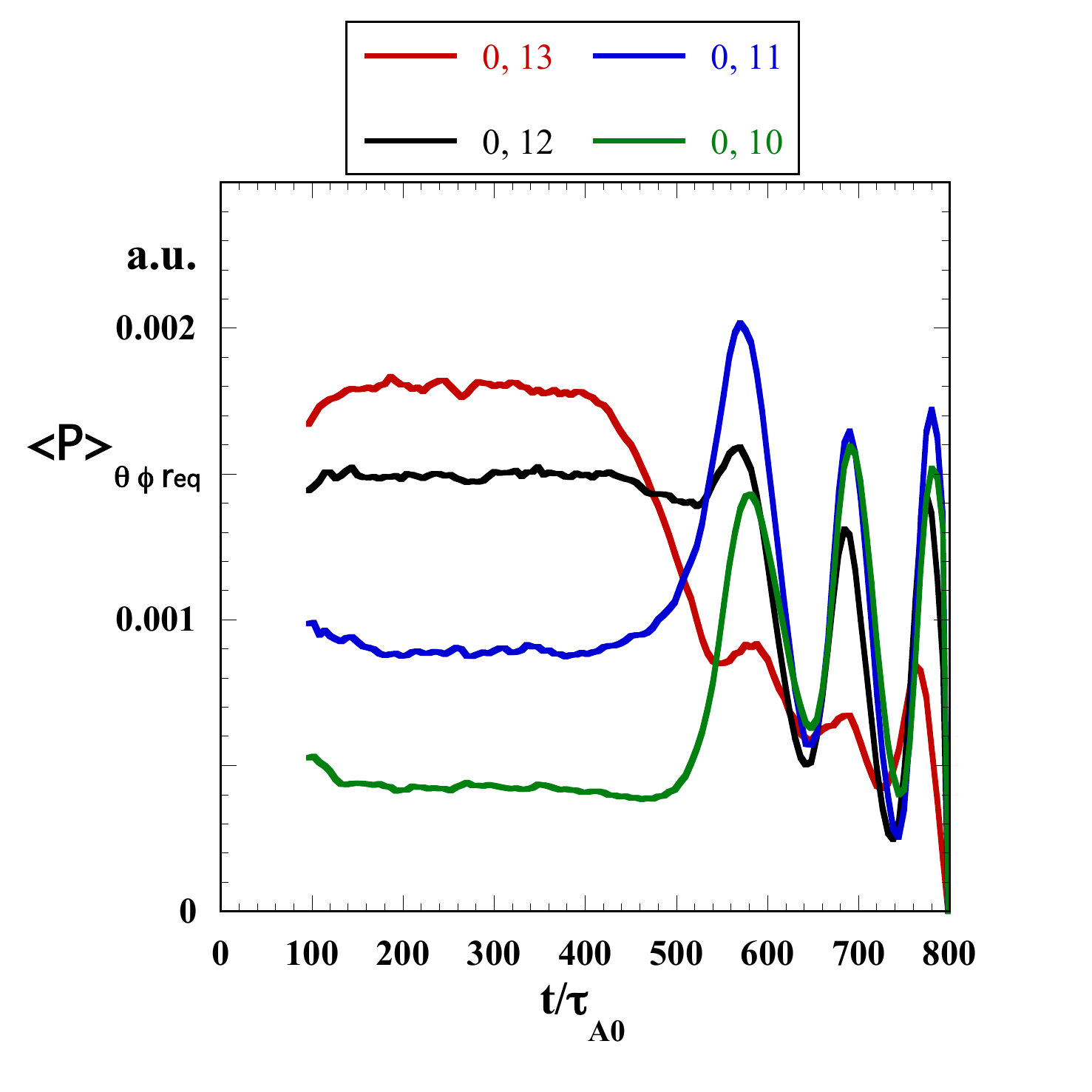}
    \caption{Time behaviour of the power (integrated over $r_{\textrm{eq}}$, $\theta$ and $\phi$) yielded by the most relevant phase-space slices. Each slice corresponds to a grid point in $(M,U_{\textrm{eq0}})$. As in Fig.~\ref{fig:maxpower_3d_vs_t}, only grid points yielding a maximum power transfer at some time in the considered interval are plotted. Grid point labels, relative to $M$ and $U_{\textrm{eq0}}$, have the same meaning as in Fig.~\ref{fig:maxpower_3d_vs_t}.}
    \label{fig:relevant_slices} 
\end{figure}

\begin{figure}
\includegraphics[width=0.3\textwidth]{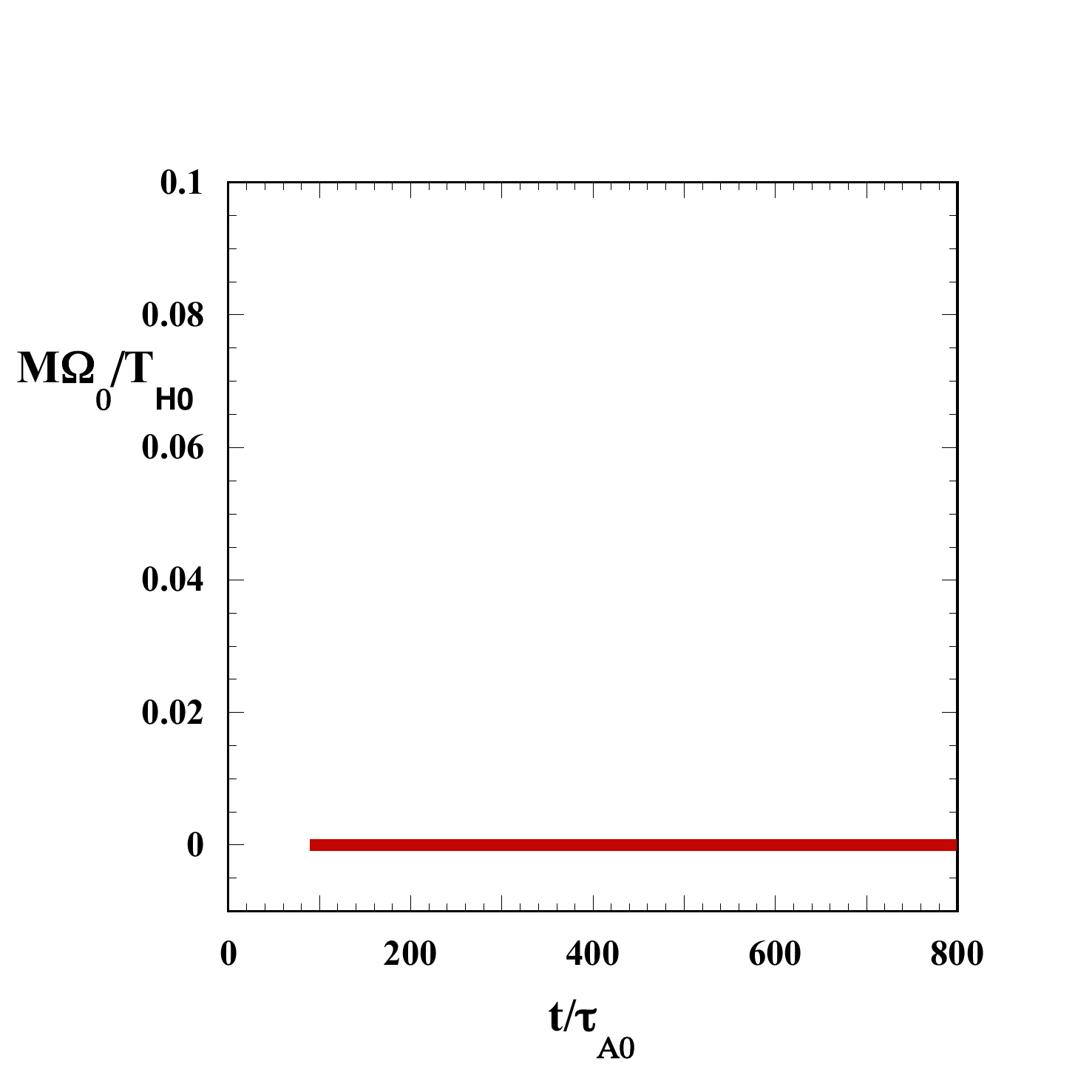}
\includegraphics[width=0.3\textwidth]{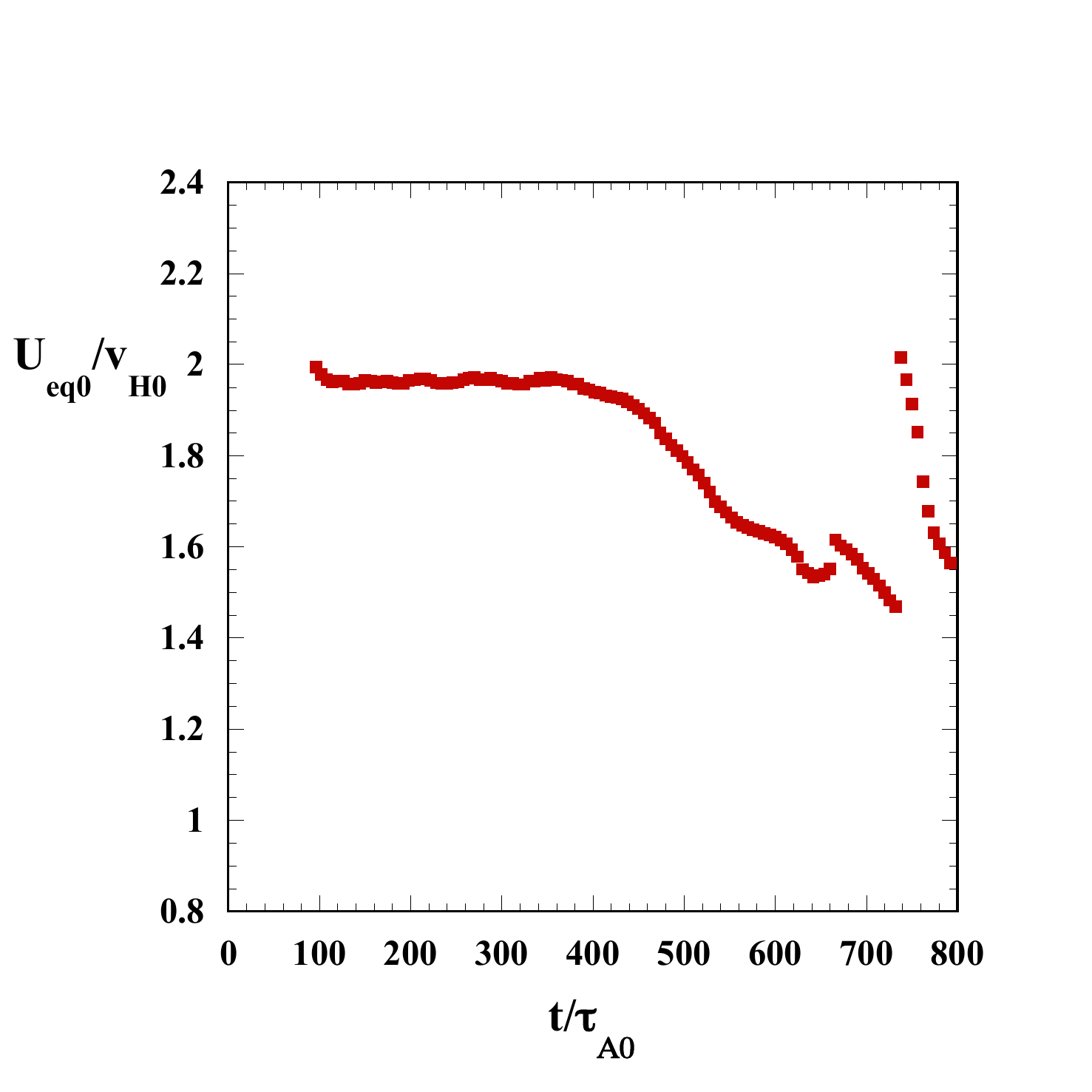}
    \caption{Coordinates, in the reduced phase space $(M,U_{\textrm{eq0}})$, of the maximum of the power transfer density averaged over $\theta$, $\phi$ and the radial coordinate. 
    }
    \label{fig:max_initial_2d} 
\end{figure}

\begin{figure}
\includegraphics[width=0.65\textwidth]{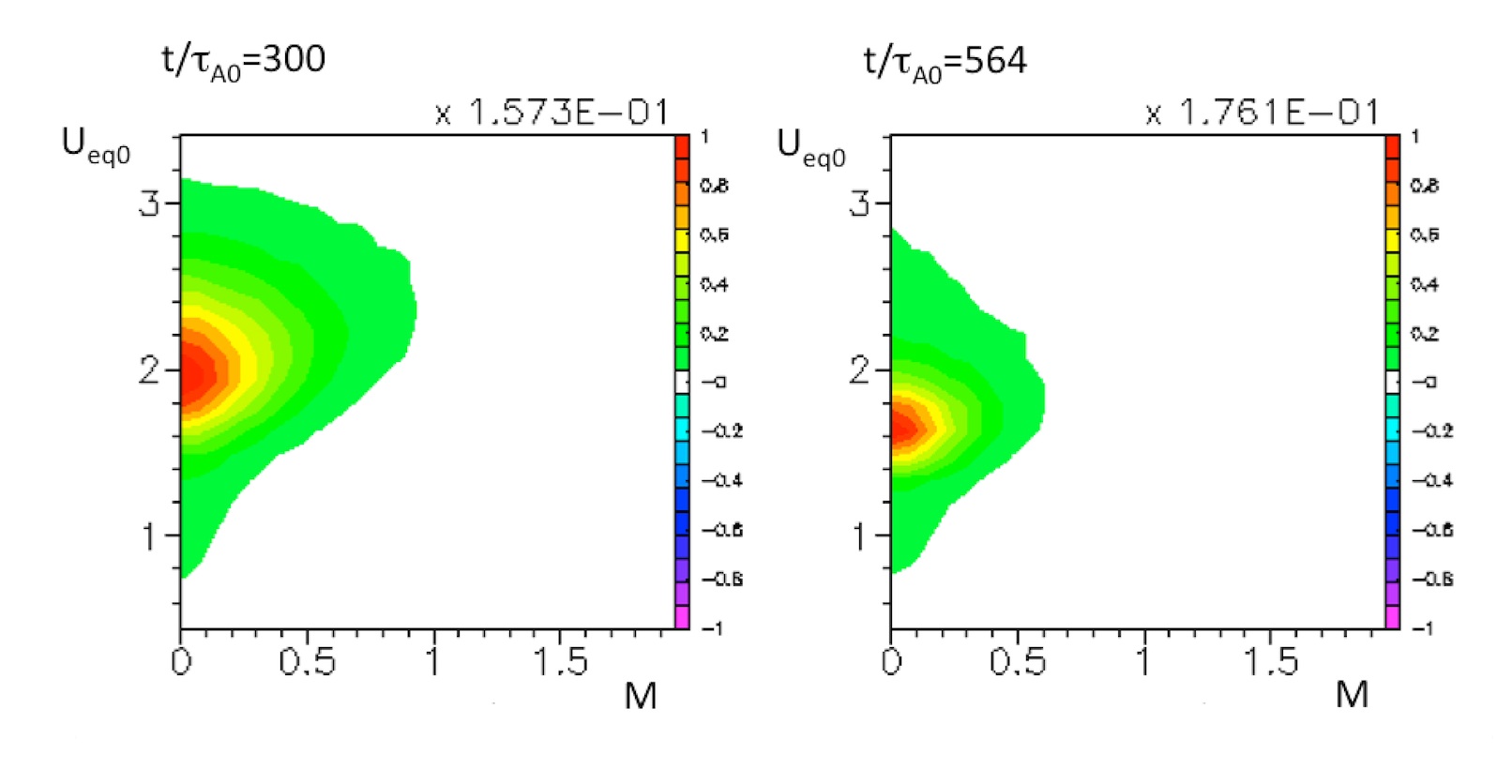}
    \caption{Contour plots, in the plane $(M,U_{\textrm{eq0}})$, of the power transfer density averaged over $\theta$, $\phi$ and the radial coordinate, at two times: $t=300 \tau_{A0}$ (left) and $t=564 \tau_{A0}$ (right).}
    \label{fig:an_power} 
\end{figure}

Figure~\ref{fig:relevant_slices} show the time behaviour of the power yielded by the most relevant slices. As in Fig.~\ref{fig:maxpower_3d_vs_t}, only those grid points that yield a maximum power transfer at some time in the considered interval are plotted. Consistently with the previous observation, we find that all the relevant slices are characterised by $M=0$.
Figure~\ref{fig:max_initial_2d} shows the time evolution of the 2-D maximum coordinates obtained by approximating, as done for Fig.~\ref{fig:max_initial}, the average power transfer near the grid points by a 2nd-order polynomial and looking for the closest maximum.
The succession of different slices in playing the role of dominant drive can also be seen from Fig.~\ref{fig:an_power}, showing the contour plots of the power transfer in the plane $(M,U_{\textrm{eq0}})$ (integrated over all the other coordinates) at $t=300 \tau_{A0}$ and $t=564 \tau_{A0}$.
We observe that, although qualitatively similar, Fig.~\ref{fig:relevant_slices} presents a simpler situation than Fig.~\ref{fig:maxpower_3d_vs_t}. The reason is, of course, that the former is obtained by integration over the radial coordinate, and it is insensitive to redistribution of the power transfer density within the same $(M,U_{\textrm{eq0}})$ slice. In the following, we will take advantage from this simplification, concentrating our analysis on two slices only: namely, the slice dominating during the linear stage ($M=0$, $U_{\textrm{eq0}}\simeq1.92$) and that dominating around $t=564 \tau_{A0}$ ($M=0$, $U_{\textrm{eq0}}\simeq1.63$), after the contribution of the former one has strongly decreased. We will conventionally refer to these two slices as the \textit{linear} slice and the \textit{nonlinear} one.

\section{Hamiltonian mapping analysis}
\label{sec:hamiltonian}

In order to analyse the behaviour of each of the two phase-space structures identified in the previous Section, we apply the Hamiltonian mapping approach, described in previous papers~\cite{briguglio14,briguglio17}. The slice is sampled by a large number of test particles, initialised with the following coordinates: $r=r_{\mathrm{min}}\div r_{\mathrm{max}}$, with $[r_{\mathrm{min}},r_{\mathrm{max}}]$ being a radial interval covering the whole radial domain of interest (here, $r_{\mathrm{min}}=0.4 a$, $r_{\mathrm{max}}=0.65 a$), $\theta=0$, $\phi=0\div2\pi$, $M=0$, $U=\bar{U}_{\textrm{eq0}}$, with $\bar{U}_{\textrm{eq0}}$ being the value characterising the slice. As all particles are initialised at $\theta=0$, the initial values of $r$ and $U$ assume the meaning of the equatorial coordinates $r_{\textrm{eq}}$ and $U_{\textrm{eq}}$. Test particles are pushed in the fields computed self-consistently, at each time step, in the considered simulation. Their coordinates are detected at each crossing of the equatorial plane, and the wave-particle phase $\Theta\equiv \int_0^t dt' \omega(t') + m\theta -n\phi$ is calculated. Note that we have defined the phase taking into account that the mode frequency is not a constant. The wave-phase at the $j-$th crossing will then be given by $\Theta_j= \int_0^{t_j} dt' \omega(t') + 2\pi j m\sigma -n\phi_j$, where $\sigma\equiv\mathrm{sign}(U)$, $t_j$ is the time at which the crossing occurs and $\phi_j$ the value assumed by the toroidal angle at that time.
The resonance condition can be written as
\begin{equation}
\Delta\Theta_j\equiv \Theta_{j}-\Theta_{j-1} = \int_{t_{j-1}}^{t_{j}} dt' \omega(t') + 2\pi m\sigma -n\Delta\phi_j=2\pi k,
\label{eq:resonance_condition}
\end{equation}
with $k=0,\pm1,\pm2,...$ being the \textit{bounce harmonic}.
Here, $\Delta \phi_j\equiv \phi_j -\phi_{j-1}$.
The average power exchanged with the mode along the poloidal orbit is also computed.

\begin{figure}
\includegraphics[width=1.\textwidth]{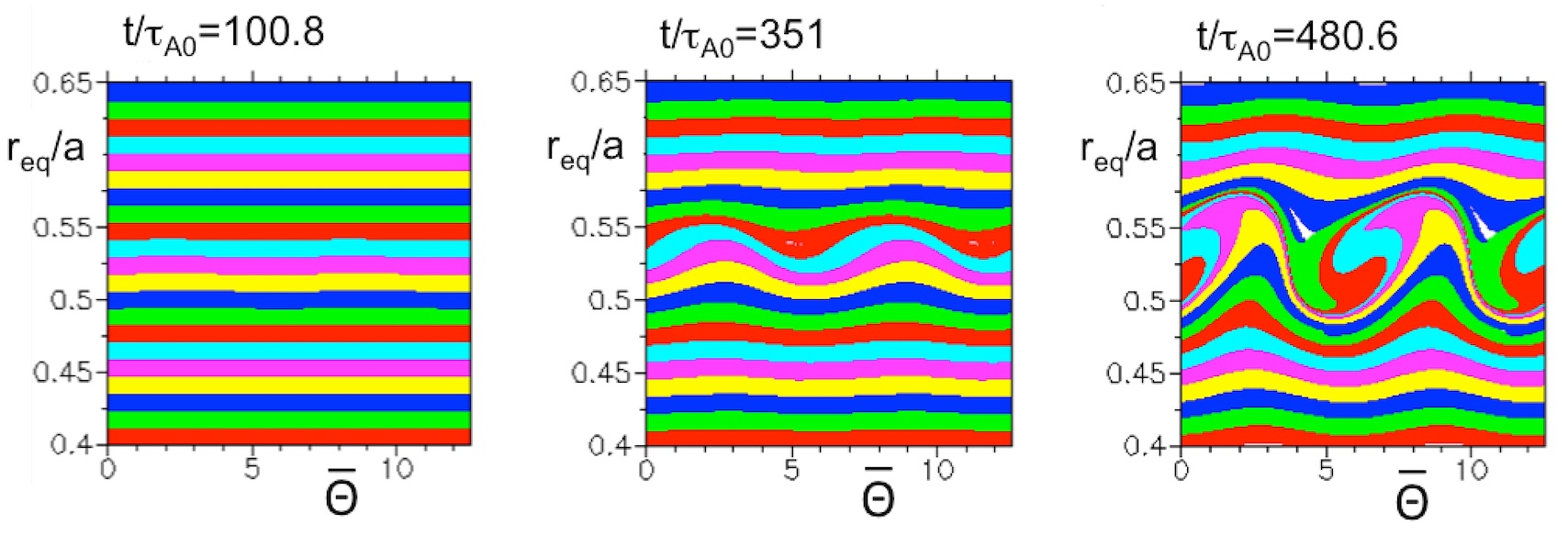}
    \caption{From left to right: test-particle markers in the $(\overline{\Theta},r_{\textrm{eq}})$ plane at three successive times of the considered simulation: $t=100.8 \tau_{A0}$ (left), $t=351.0 \tau_{A0}$ (center) and $t=480.6 \tau_{A0}$ (right). Each marker is coloured according to the birth $r_{\textrm{eq}}$ value of the particle.}
    \label{fig:an_er_phase_0} 
\end{figure}

Figure~\ref{fig:an_er_phase_0} shows, for each test particle, a coloured marker in the plane $(\overline{\Theta},r_{\textrm{eq}})$, where $\overline{\Theta}\equiv \Theta \,\,\textrm{module}\,\, 2\pi$. From Eq.~\ref{eq:resonance_condition}, we see that the resonance conditions reads, in terms of $\overline{\Theta}$, 
\begin{equation}
\Delta\overline{\Theta}_j=0.
\label{eq:resonance_condition_2}
\end{equation}
To get a clearer view of particle evolution, a companion marker is also drawn at $\overline{\Theta} + 2\pi$. Marker colour is chosen according to the birth $r_{\textrm{eq}}$ value of the particle. Three times are considered, relative, respectively, to linear ($t=100.8 \tau_{A0}$), early-nonlinear ($t=351.0 \tau_{A0}$) and fully-nonlinear stage of mode evolution ($t=480. \tau_{A0}$).
In the unperturbed motion, $r_{\textrm{eq}}$ is constant. Then, during the linear phase of the mode evolution (Fig.~\ref{fig:an_er_phase_0}-left), particle trajectories in the $(\overline{\Theta},r_{\textrm{eq}})$ plane essentially reduce to fixed points for $r_{\textrm{eq}}=r_{\mathrm{eq\,res}}$ (in this case, $r_{\textrm{eq\,res}}\simeq 0.54 a$), while they correspond to drift along the $\overline{\Theta}$ axis in the negative (positive) direction, for $r_{\textrm{eq}}$ greater (less) than $r_{\mathrm{eq\,res}}$.
As the mode amplitude grows (Fig.~\ref{fig:an_er_phase_0} center), $r_{\textrm{eq}}$ varies because of the mode-particle interaction (e.g., radial $\bf{E}\times \bf{B}$ drift). Even particles that  were initially resonant are brought out of resonance, drifting in phase and eventually undergoing an inversion of the drift in $r_{\textrm{eq}}$. 
Particles that cross the $r_{\textrm{eq}}=r_{\mathrm{eq\,res}}$ line revert the phase drift as well. Thus, their trajectories are bounded, giving rise to the formation of an island-like structure in the $(\overline{\Theta},r_{\textrm{eq}})$ plane (Fig.~\ref{fig:an_er_phase_0} right). Its radial extension grows with mode amplitude, 
consistent with equations of motions accounting for radial $\mathbf{E}\times \mathbf{B}$ drift and particle motion due to radial magnetic field perturbation.
Particles outside the island undergo a secular drift in phase, as the $\bf{E}\times \bf{B}$ drift is not able to cause them crossing the resonance radius.

\begin{figure}
\includegraphics[width=0.50\textwidth]{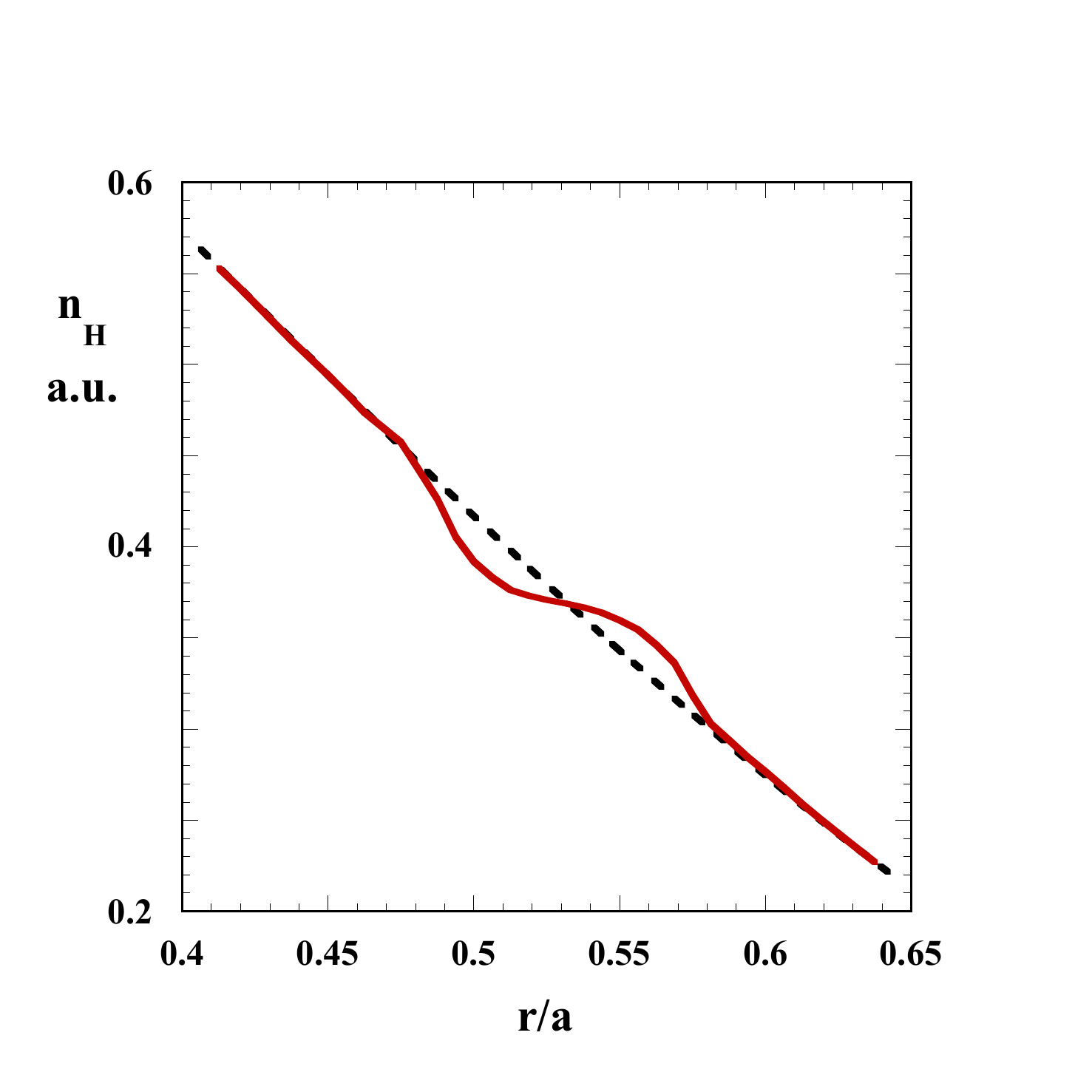}
    \caption{Normalised density profile, for the considered phase-space slice, at $t=480.6 \tau_{A0}$ (red), compared with the initial one (black).}
    \label{fig:flattening_480p6} 
\end{figure}

The formation of the island mixes particles born on both sides of the resonance radius, causing a density flattening around that radius. This can be seen from Fig.~\ref{fig:flattening_480p6}, where the initial density profile and the flattened one at $t=480.6 \tau_{A0}$ are compared. At the same time, large negative density gradients emerge at the boundaries of this flattening region\footnote{Using the expression ``density flattening" we only mean that the negative density gradient is reduced; not necessarily that it vanishes. In certain cases, however, it can vanish or even invert its sign.}.
This flattening process is the basis for mode saturation, or, to be more precise, for the exhausting of the destabilising contribution of the considered phase-space slice. If the mode were constrained to keep constant its frequency and radial structure, saturation would be reached as soon as the flattened-density region extends over the whole resonant interaction region. This, in turn, would be limited by the intersection between the resonance region (defined as the region, around the resonance, in which the rate of phase drift is smaller than a certain amount proportional to the linear growth rate) and the radial extension of the mode. 
Typically, this intersection coincides with the narrowest of these two regions.
If it is the resonance region, the saturation mechanism has been dubbed \textit{resonance detuning}; if it is the radial extension of the mode, \textit{radial decoupling}~\cite{briguglio14,briguglio17,zonca2015,RMP16}. If the equilibrium allows the mode to modify its frequency and/or structure, several elements can make it able to further extract energy from particles, prolonging its growth. First, modification of the radial structure can prevent or delay radial decoupling, provided it is the relevant mechanism. Second, frequency modifications cause the resonance radius and the resonance region to move radially, thus managing to reach an area with a density profile not yet flattened. Third, different sensitivities of different resonances to the former two processes can bring to a succession of dominating resonances in driving the mode in the nonlinear stage, such that the exhaustion of the contribution of one resonance does not imply a definitive saturation of the mode. We will see in Sec.~\ref{sec:chirping} that, in the present case, the latter two elements are essential in determining the nonlinear evolution of the mode.

\section{Power transfer rate and resonance condition}
\label{sec:resonance}

Before addressing the problem of mode saturation in greater detail, we want to investigate the behaviour of the most destabilising particles, within the same slice, and its connection with their capability of fulfilling the resonance condition.

\begin{figure}
\includegraphics[width=0.8\textwidth]{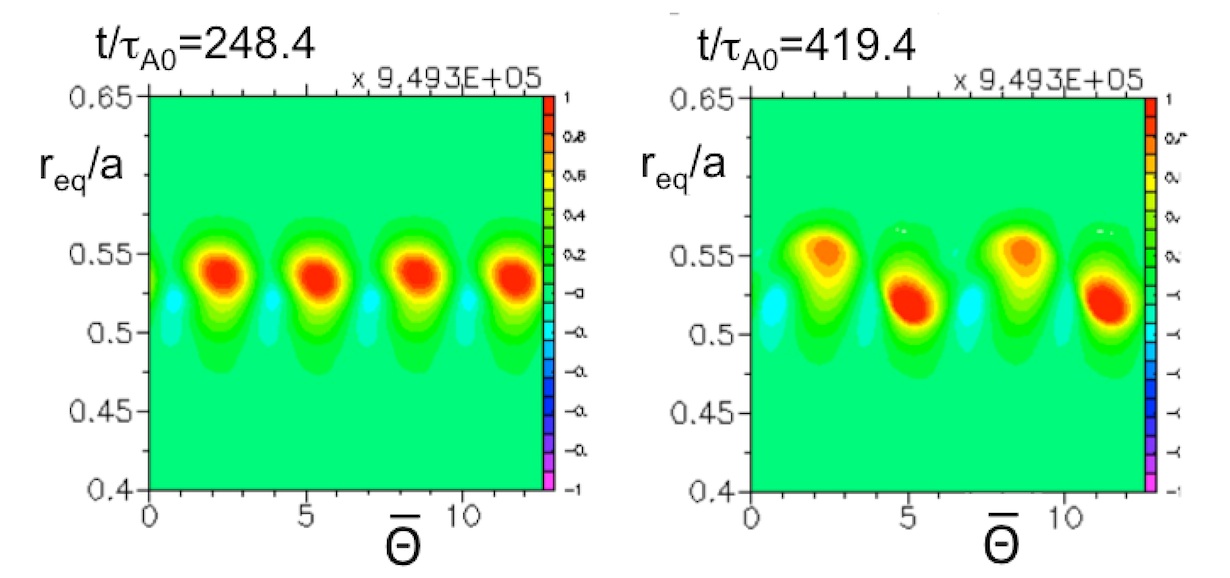}
    \caption{Power transfer structures in the plane $(\overline{\Theta},r_{\textrm{eq}})$. Each marker is coloured according to its own power-transfer rate (averaged over the last poloidal orbit). Two times are considered: $t=248.4 \tau_{A0}$ and $t=419.4 \tau_{A0}$.}
    \label{fig:an_er_phase_1} 
\end{figure}

\begin{figure}
\includegraphics[width=0.8\textwidth]{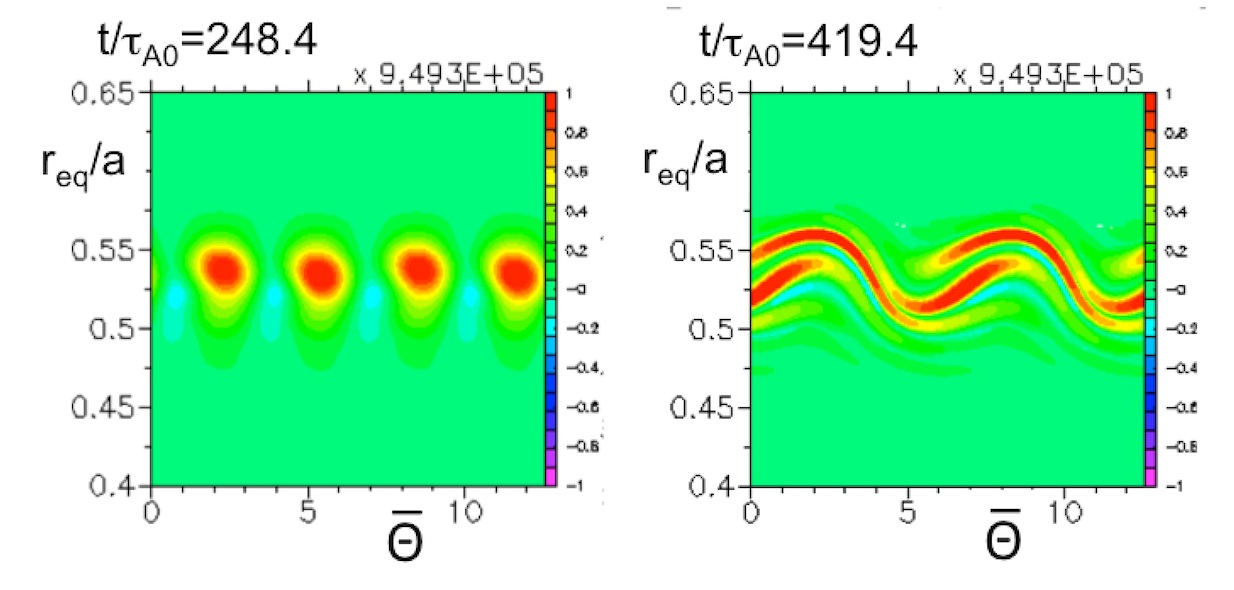}
    \caption{Same as Fig.~\ref{fig:an_er_phase_1}, but with markers coloured according to their power-transfer rate at $t=248.4 \tau_{A0}$. Two times are considered: $t=248.4 \tau_{A0}$ (left) and $t=419.4 \tau_{A0}$ (right). The left frame is, by definition, coincident with the left frame of Fig.~\ref{fig:an_er_phase_1}. Comparing the right frame with that of the latter Figure shows that the mode is driven, at different times, by not the same particle clusters.}
    \label{fig:an_er_phase_2} 
\end{figure}

Figure~\ref{fig:an_er_phase_1} shows analogous plots to those presented in Fig.~\ref{fig:an_er_phase_0}. In this case, however, markers are coloured according their own power-transfer rate (averaged over the last poloidal orbit). Two times are considered: $t=248.4 \tau_{A0}$ and $t=419.4 \tau_{A0}$. We see that, consistently with island formation, density flattening and large negative gradients appearance, the power transfer maxima, originally located around the resonance radius, separate from each other, following, respectively, the inner and the outer gradient.
We wonder whether the most destabilising particles at $t=248.4 \tau_{A0}$ (represented by the red markers) are the same that drive the mode at $t=419.4 \tau_{A0}$\footnote{Note that here we are analysing the dynamics of particles belonging to the same phase-space slice, different from what presented in Figs. ~\ref{fig:max_initial} and ~\ref{fig:max_initial_2d}.}. To answer this question, we plot, in Fig.~\ref{fig:an_er_phase_2} the same markers shown in Fig.~\ref{fig:an_er_phase_1}, colouring them according to the power-tranfer rate they have at $t=248.4 \tau_{A0}$. The left frame is, by definition, coincident with the left frame of Fig.~\ref{fig:an_er_phase_1}. The right frame makes apparent the position, at $t=419.4 \tau_{A0}$, of the $t=248.4 \tau_{A0}$ most destabilising particles. Comparing it with the right frame of Fig.~\ref{fig:an_er_phase_1}, it is evident that the mode is driven by not the same particle clusters at the two considered times.

\begin{figure}
\includegraphics[width=0.90\textwidth]{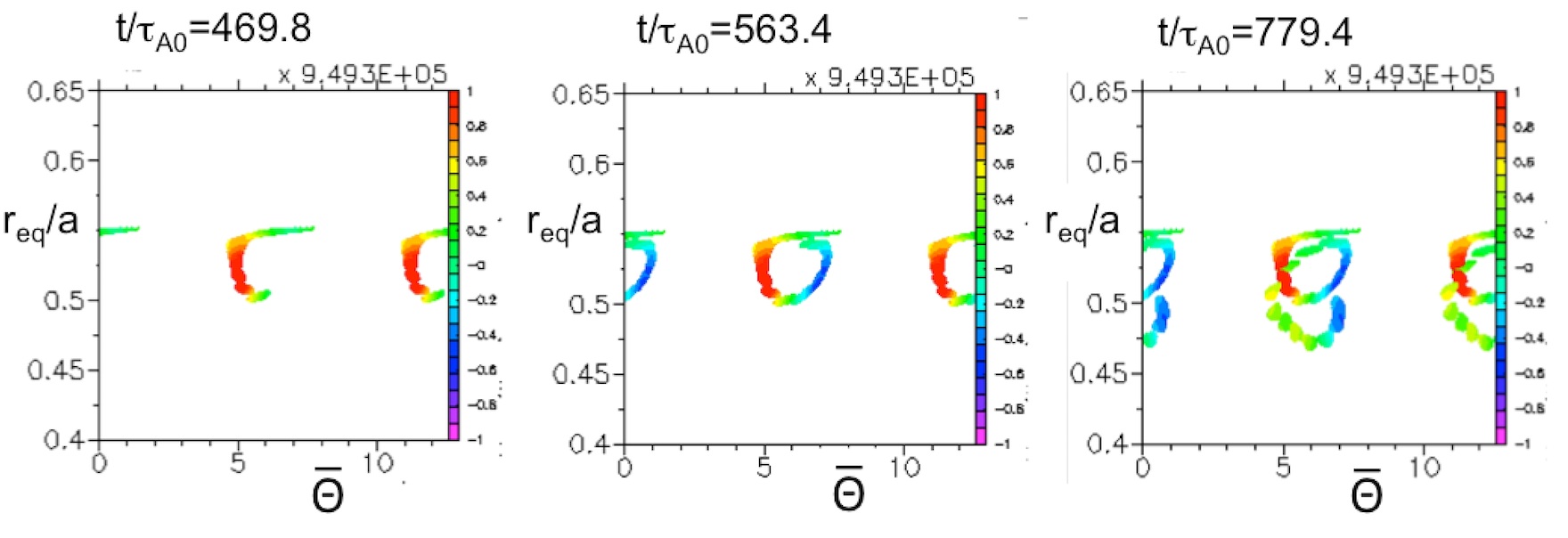}
    \caption{Trajectories of a small group of test particles in the plane $(\overline{\Theta},r_{\textrm{eq}})$. Only particles yielding, at a certain time ($t=415.8 \tau_{A0}$), more than 95\% of the maximum power transfer at the same time are considered. Markers are coloured according to their instantaneous power transfer level. Different frames refer to three different times: $t=469.8 \tau_{A0}$ (left), $t=563.4 \tau_{A0}$ (center) and $t=779.4 \tau_{A0}$ (right). All the previous-time positions are plotted, in order to get information about the whole trajectories.}
    \label{fig:traces} 
\end{figure}

\begin{figure}
\includegraphics[width=0.40\textwidth]{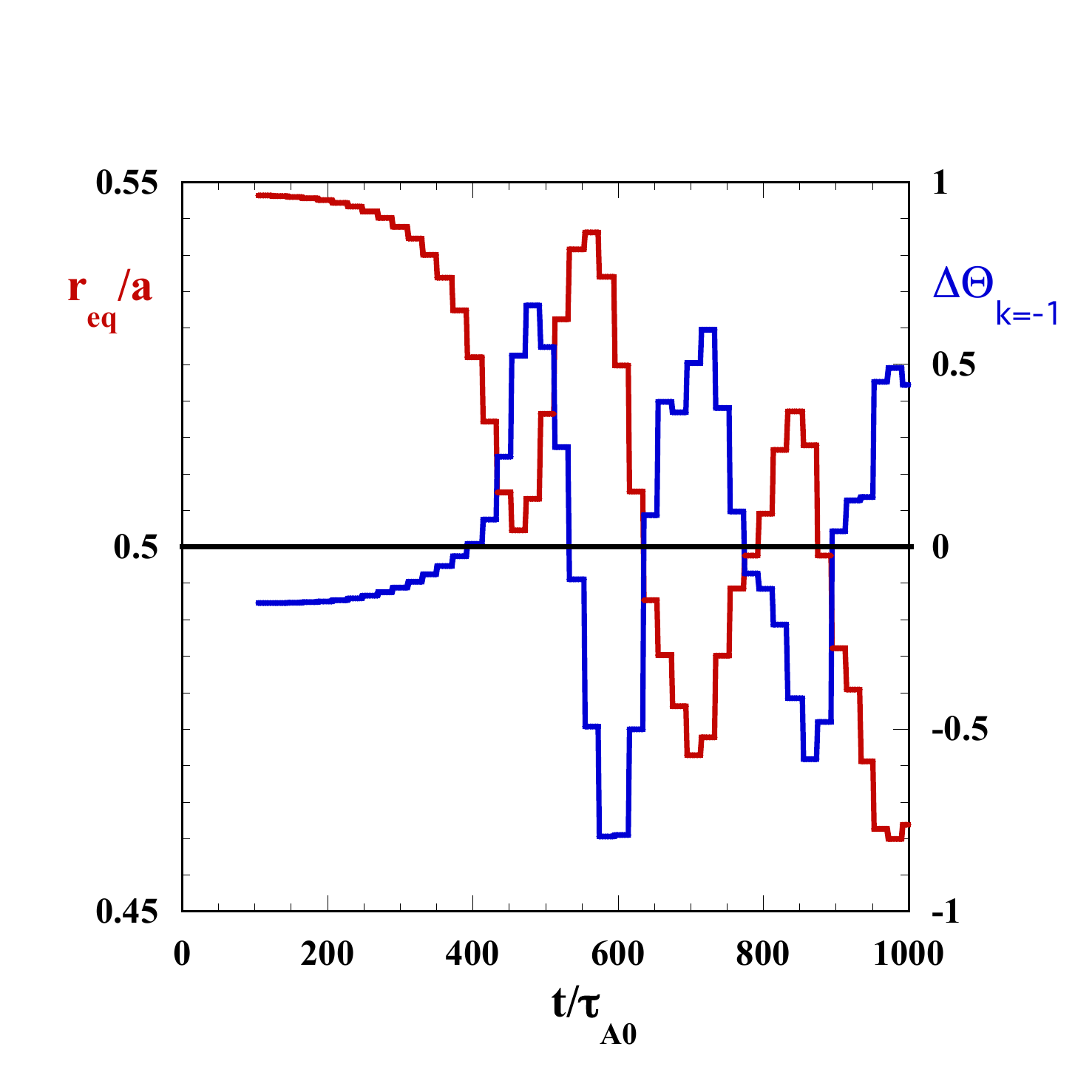}
\includegraphics[width=0.40\textwidth]{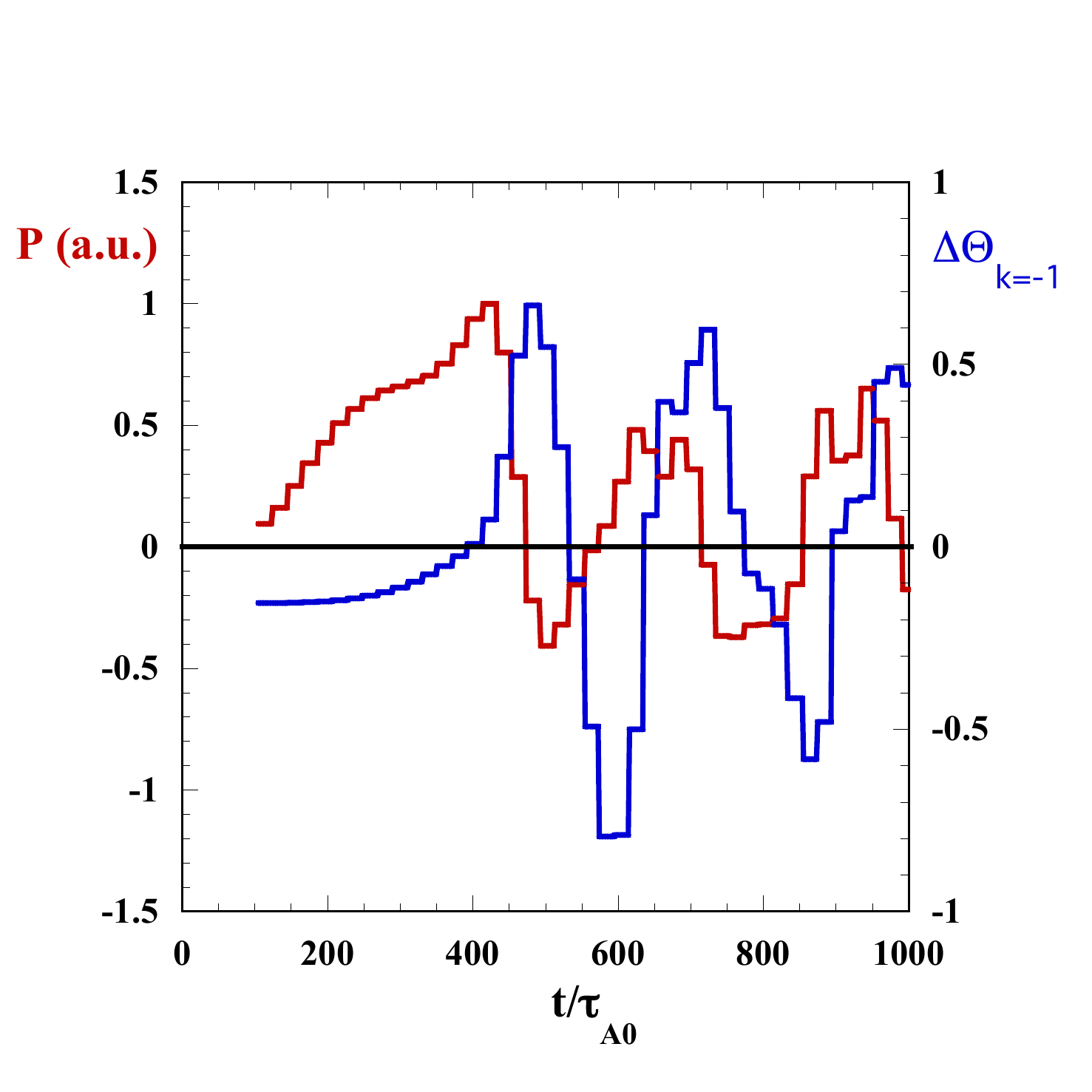}
    \caption{Time evolution of equatorial radial position (left) and power transfer (right), compared with that of $\Delta \Theta_{k=-1}$, for the  most destabilising of the particles considered in Fig.~\ref{fig:traces} at $t=415.8 \tau_{A0}$. Here, $\Delta \Theta_{k=-1}$ is the wave-phase variation given by Eq.~\ref{eq:resonance_condition} for $k=-1$.}
    \label{fig:er_power_delta_theta} 
\end{figure}

\begin{figure}
\includegraphics[width=0.80\textwidth]{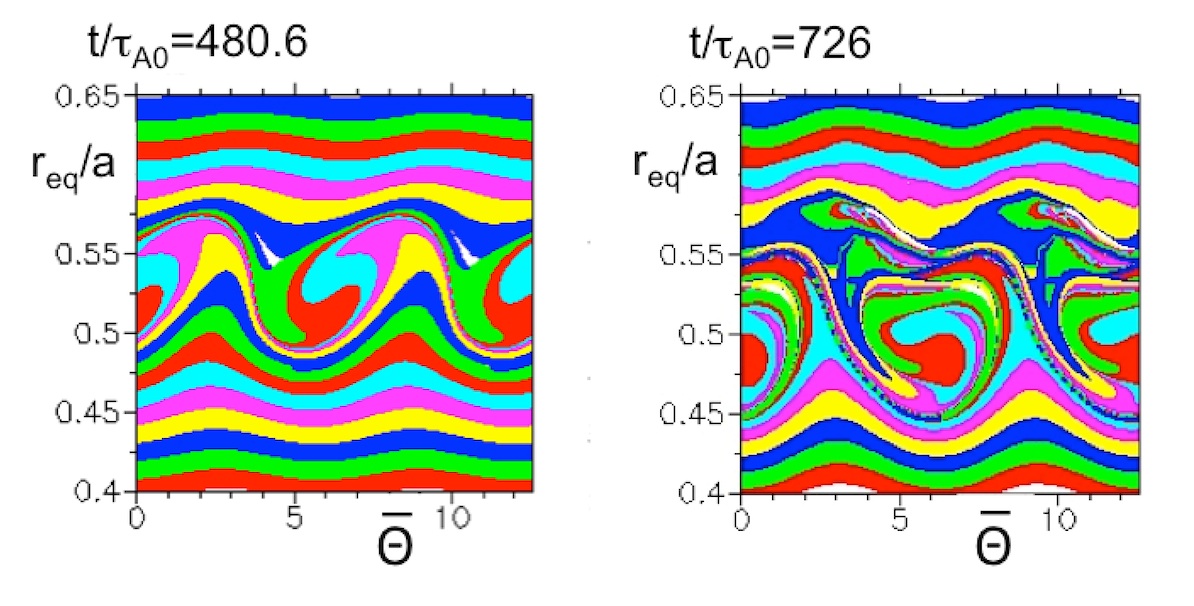}
    \caption{Same as Fig.~\ref{fig:an_er_phase_0}, at two different times in the nonlinear stage: $t=480.6 \tau_{A0}$ (left) and $t=726 \tau_{A0}$ (right). An inward shift of the island center is observed.}
    \label{fig:an_er_phase_0_2} 
\end{figure}

Given that a particle does not permanently belong to the group of the most destabilising particles, we want to check now whether the power exchange between the mode and a certain particle is limited to only a short time interval or it can recur later. To this aim, we follow (Fig.~\ref{fig:traces}) a small group of test particles in the plane $(\overline{\Theta},r_{\textrm{eq}})$. Particles are chosen selecting the most destabilising ones at a given time. Markers are coloured according to the same prescription as in Fig.~\ref{fig:an_er_phase_1}. All the different-time positions are plotted, in order to get information about their whole trajectories in that plane. 
Several facts deserve to be observed. First, the same particles can assume, at different times, the role of driving or damping particles. In the former case, particles move radially inward; in the latter case, outward. In both situations, particles satisfy the resonance condition, Eq.~\ref{eq:resonance_condition_2}, quite well and, indeed, they are wave-trapped particles (Fig.~\ref{fig:traces} center). Second, the time interval in which a particle is in resonance with the mode is of order of few poloidal orbits. Third, particle radial excursion is of order of the island width. Fourth, trajectories show a overall inward radial drift, corresponding to an analogous drift of the resonance radius (Fig.~\ref{fig:traces} right); this will be discussed in the next Sections.
These facts are better shown in Fig.~\ref{fig:er_power_delta_theta}, where the radial position and the power transfer are reported, for the most destabilising of the particles considered in Fig.~\ref{fig:traces} at $t=415.8 \tau_{A0}$, along with the wave-phase variation, versus time. It is easy to see that maximum power transfer (alternatively, positive and negative) and radial excursion (respectively, inward and outward) occur when the particle is in phase with the mode ($\Delta \overline{\Theta}=\Delta \Theta_{k=-1}$ close to zero). The overall inward drift of the radial motion is also evident; it can be appreciated looking at the inward drift of the island, shown in Fig.~\ref{fig:an_er_phase_0_2}.

\section{Trapping/de-trapping process}
\label{sec:trapping_detrapping}

\begin{figure}
\includegraphics[width=0.50\textwidth]{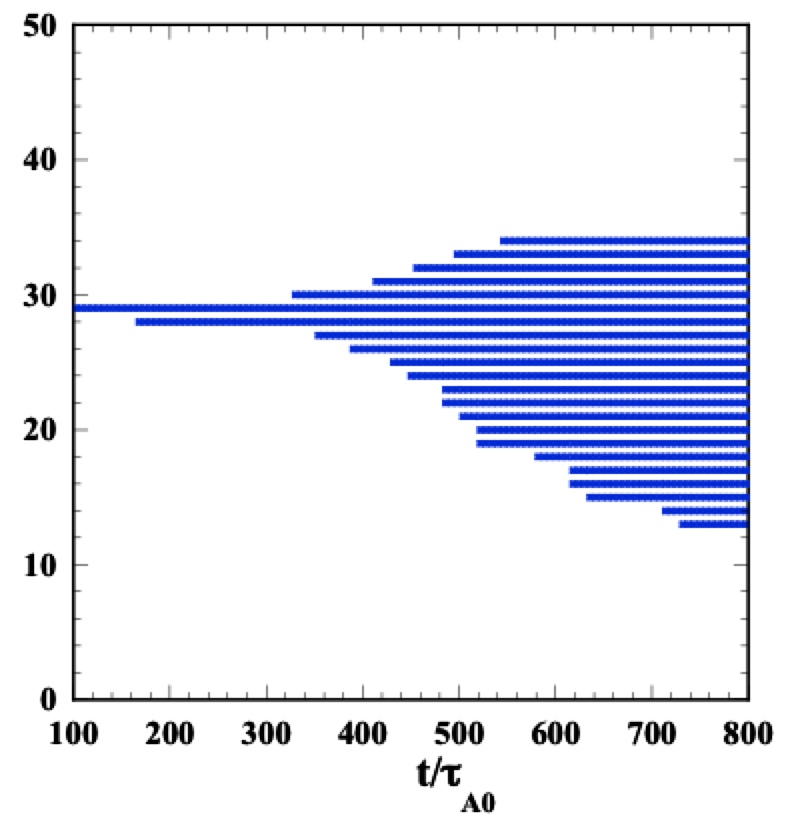}
    \caption{At each time, the indices of the radial bands of test particles owning at least one particle able to satisfy fairly well the resonance condition (last phase variation less than $0.015$). A progressive asymmetric (mainly inward) trapping of new bands is observed, while no bands get fully de-trapped in this case.}
    \label{fig:trapping} 
\end{figure}

\begin{figure}
\includegraphics[width=0.80\textwidth]{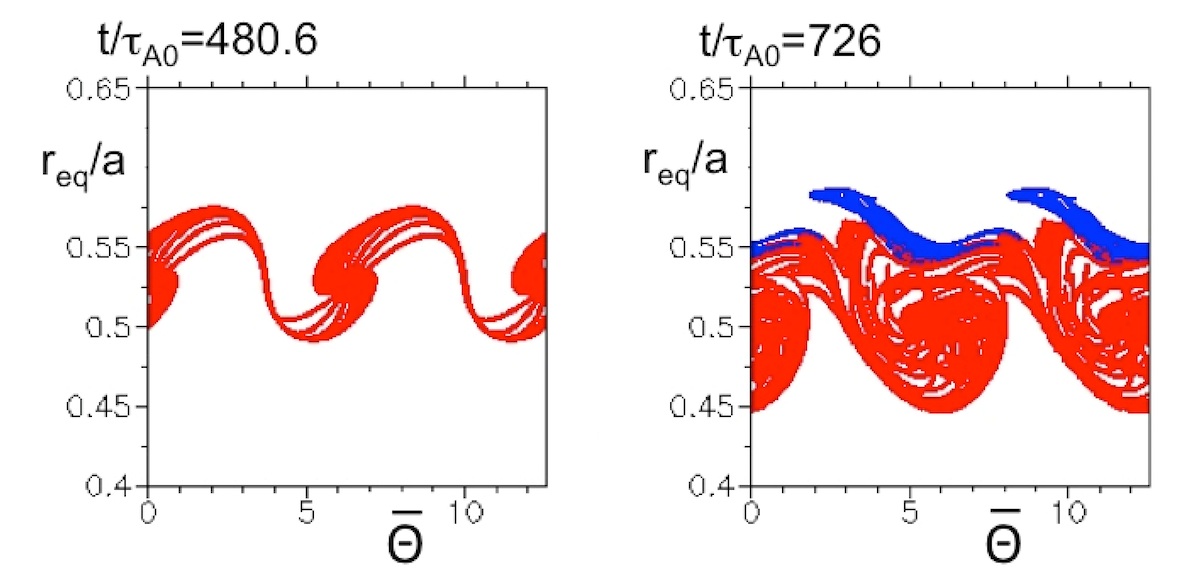}
    \caption{Only markers corresponding to particles that have already fairly satisfied the resonance condition are plotted (``already-trapped" particles). Particles that appear to be still trapped in the wave (``actually-trapped" particles) are coloured in red, while those that have cumulated a phase drift greater than a conventional threshold value ($2.5\pi$) are coloured in blu. Two different times are considered, as in Fig.~\ref{fig:an_er_phase_0_2}: $t=480.6 \tau_{A0}$ (left) and $t=726 \tau_{A0}$ (right.). A significant amount of particle de-trapping is observed, on the outer side of the island, at $t=726 \tau_{A0}$.}
    \label{fig:trapping_detrapping} 
\end{figure}

The evolution of the island structure shown in Fig.~\ref{fig:an_er_phase_0_2} can be analysed as follows: after dividing test particle into bands according to their birth $r_{\mathrm{eq}}$ values, as done in assigning colours to markers in Fig.~\ref{fig:an_er_phase_0_2}, we conventionally consider a band as ``wave trapped" if at least one of its particles has its last phase variation smaller than a certain value ($0.015$, in this case), corresponding to the resonance condition fairly well satisfied by at least one particle. Fig.~\ref{fig:trapping} reports, at each time, the indices of actually wave-trapped bands. We observe a progressive, asymmetric (mainly inward) trapping of bands. From this analysis, no de-trapping clearly emerges.
The reason is that a band is considered ``wave trapped" if at least one of the particles it includes is fairly resonant.
A different picture can be obtained by looking at the individual behaviour of test particles. In Fig.~\ref{fig:trapping_detrapping} only markers corresponding to particles that have already fairly satisfied the resonance condition are plotted (``already-trapped" particles). Colour is red for particles that appear still trapped in the wave (``actually-trapped" particles); it switches to blue as the particle cumulates a phase drift greater than a certain conventional threshold ($2.5\pi$, in this case).
The same two times considered in Fig.~\ref{fig:an_er_phase_0_2} are examined. We see that, at $t=726 \tau_{A0}$, a significant amount of particle de-trapping takes place on the outer side of the island.
Here, the de-trapping phenomenon becomes apparent because each particle is considered according to its specific dynamics, which can be different from that of closely born particles.

\begin{figure}
\includegraphics[width=0.90\textwidth]{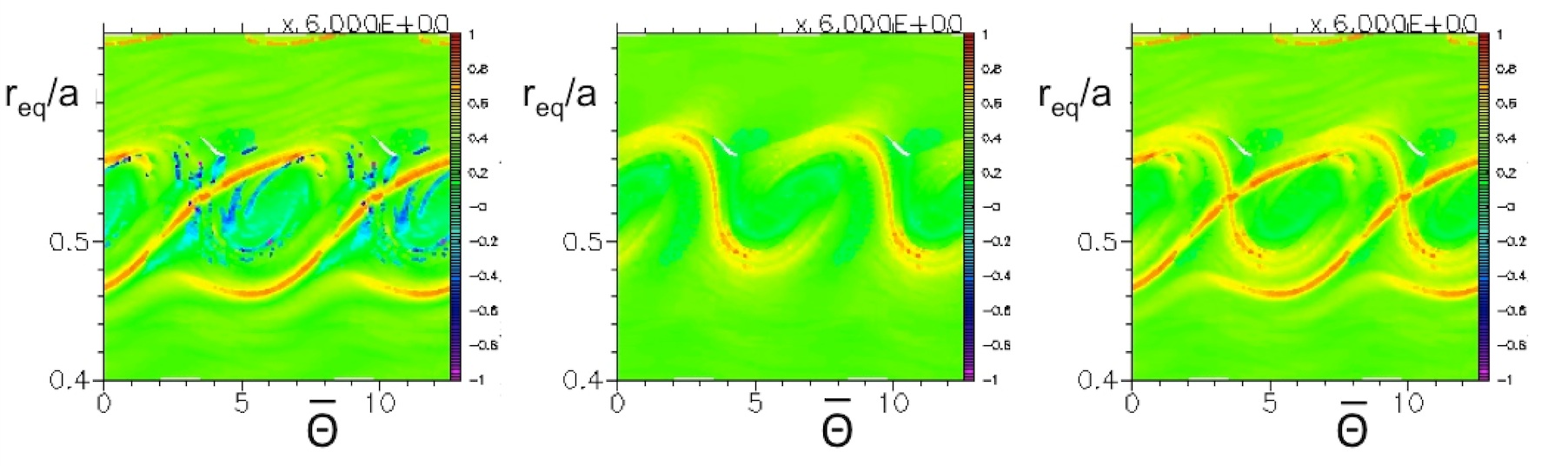}
    \caption{Lagrangian Coherent Structure in the plane $(\overline{\Theta},r_{\mathrm{eq}})$. Markers are coloured according to the values of their forward Lyapunov exponent $\lambda_{f}$ (left), those of their backward Lyapunov exponent $\lambda_{b}$ (centre) or the largest of the two (right), where $\lambda_f\equiv \log(d_{t+\Delta t}/d_t)$ and $\lambda_d\equiv \log(d_{t-\Delta t}/d_t)$. Here, $d_{t+\tau}$ is the distance in the plane, at time $t+\tau$ ($\tau = 0,\pm \Delta t$, between a marker and its nearest neighbour at time $t$. In this Figure, $t=504 \tau_{A0}$ and $\Delta t=200 \tau_{A0}$. In the left frame, the reddest lines are the repulsive lines; in the central frame, the reddest ones are the attractive lines. In the right frame, the two kind of lines are superimposed.}
    \label{fig:lyapunov} 
\end{figure}

\begin{figure}
\includegraphics[width=0.60\textwidth]{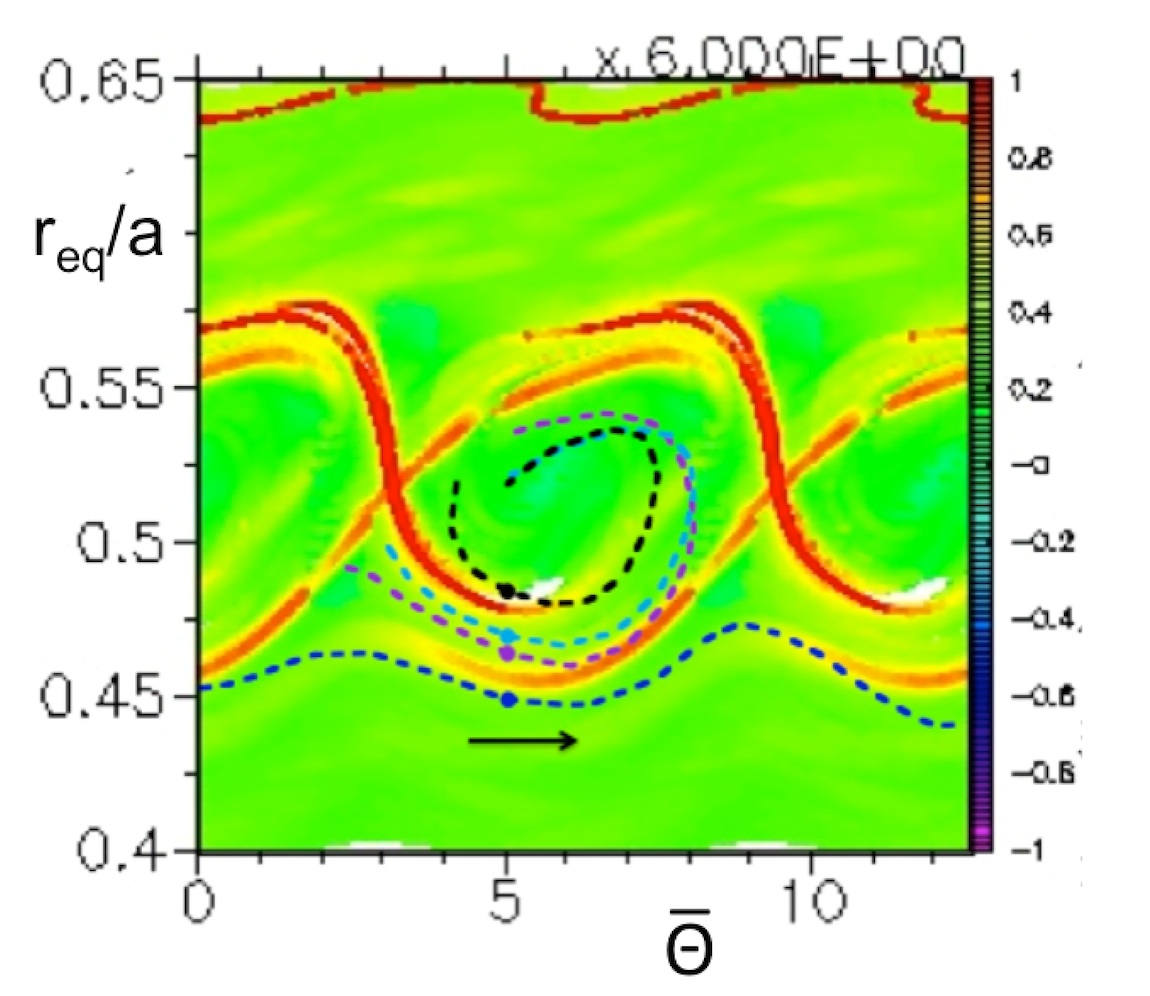}
    \caption{Lagrangian Coherent Structures (repulsive and attractive lines superimposed) for $t=558 \tau_{A0}$ and $\Delta t=200$, as defined in in Fig.~\ref{fig:lyapunov}. The trajectories of some particles are also shown. Particles have been chosen in such a way that they are located, at $t=558 \tau_{A0}$, just below the lower channel (the blue big dot), within the channel (violet and light blue big dots) and just above the channel (the black big dot). The arrow indicates the verse of motion.
     Particles that enter the lower channel become trapped.}
    \label{fig:lyapunov_trapping} 
\end{figure}

\begin{figure}
\includegraphics[width=0.60\textwidth]{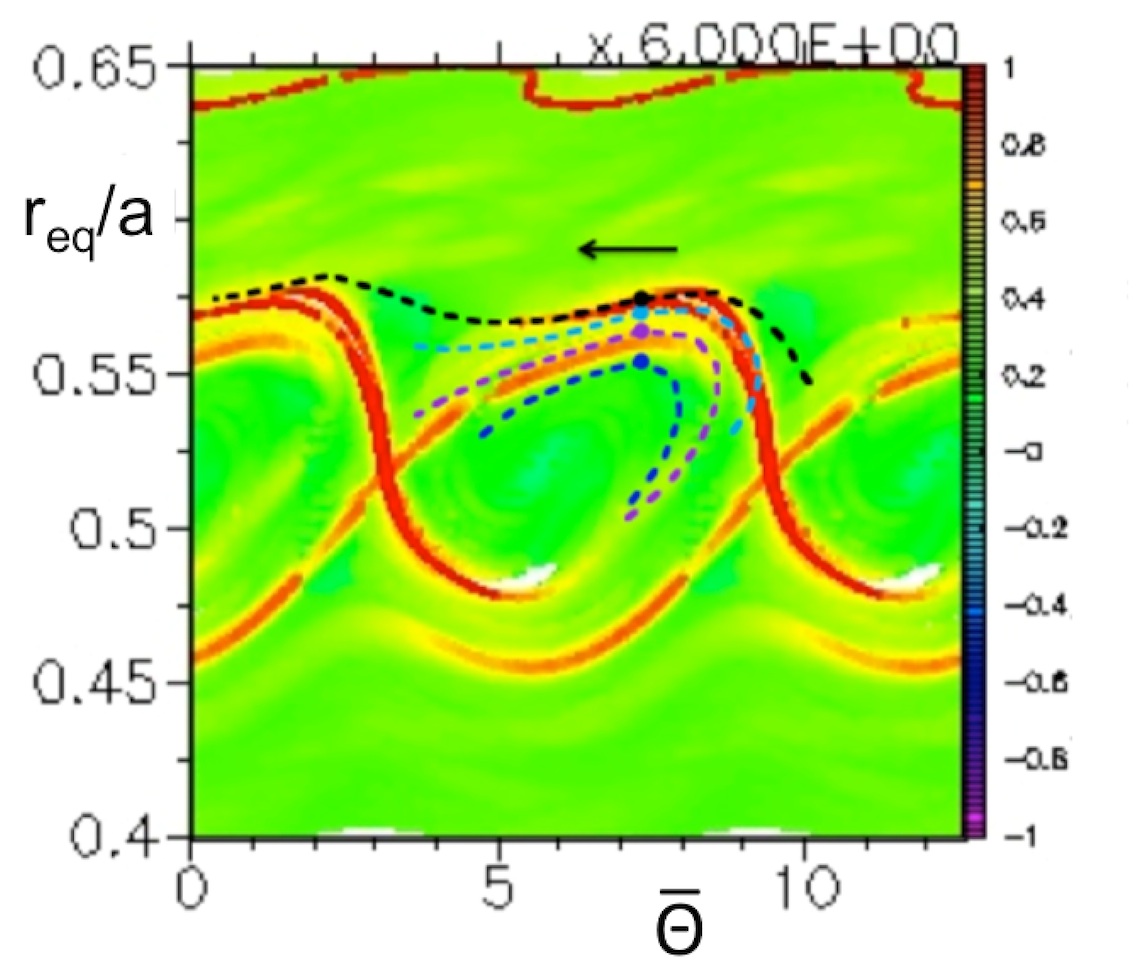}
    \caption{Analogous to Fig.~\ref{fig:lyapunov_trapping}, but with particles, at the reference time, located just below (the blue big dot), within (violet and light blue big dots) and just above (the black big dot) the upper channel. The arrow indicates the verse of motion. Particles that enter the upper channel get de-trapped.}
    \label{fig:lyapunov_detrapping} 
\end{figure}

The continuous trapping/de-trapping phenomenon can also be observed by resorting to the Lagrangian Coherent Structure (LCS) technique~\cite{haller2015lagrangian,falessi2015lagrangian,pegoraro2019,di2018coherent,di2018coherent2}. It is well known that these structures generalise dynamical patterns observed in autonomous and periodic systems to temporally aperiodic flows like the one that we are analysing. In particular, as already shown in Refs.~\cite{carlevaro2015nonlinear,carlevaro_montani_falessi_2020}, we can describe the shape of the structures enclosing instantaneously trapped particles and, thus, partition the phase space into different sub-domains where particle motion is qualitatively different, i.e. trapped/unbounded trajectories. 
The application of this technique yields the results shown in Fig.~\ref{fig:lyapunov}. For each marker, the forward Lyapunov exponent, $\lambda_{f}$ and the backward one, $\lambda_{b}$, are defined respectively as $\lambda_{f} \equiv \log \left(d_{t+\Delta t} / d_{t}\right)$ and $\lambda_{b} \equiv \log \left(d_{t-\Delta t} / d_{t}\right)$, where $d_{t+\tau}$ is the distance in the plane $\left(\bar{\Theta}, r_{\mathrm{eq}}\right)$, at time $t+\tau$ ($\tau = 0,\pm \Delta t$, between a marker and its nearest neighbour at time $t$. In this Figure, $t=504 \tau_{A0}$ and $\Delta t=200 \tau_{A0}$. Markers are coloured, respectively, according to the values of $\lambda_{f}$ (Fig.~\ref{fig:lyapunov}-left), $\lambda_{b}$ (Fig.~\ref{fig:lyapunov}-centre) or the largest of the two (Fig.~\ref{fig:lyapunov}-right).
Large values of the forward exponent identify the so-called repulsive lines (shown in Fig.~\ref{fig:lyapunov}-left): particles on opposite sides of such lines tend to diverge with increasing time. Large values of the backward exponent correspond to attractive lines (Fig.~\ref{fig:lyapunov}-center): particles on opposite sides of these lines have converged to close positions coming from distant ones. The superposition of the two kind of lines (Fig.~\ref{fig:lyapunov}-right) would correspond, in the case of constant mode frequency and amplitude, to the separatrix delimiting the island. In the present case (both amplitude and frequency depending on time), we observe that the attractive and repulsive lines do not match at the top and bottom of the island. The channels left open between the attractive and repulsive lines allow for an incoming flow of particles, destined to become trapped, into the island from the lower edge and an outgoing flow of particles, destined to become de-trapped, from the upper edge.
This fact can be better appreciated from Fig.~\ref{fig:lyapunov_trapping}, showing the LCSs at $t=558 \tau_{A0}$, with the same choice of $\Delta t$ as in Fig.~\ref{fig:lyapunov}. The trajectories of some particles are also shown. Particles have been chosen in such a way that they are located, at $t=558 \tau_{A0}$, just below the lower channel (the blue big dot), within the channel (violet and light blue big dots) and just above the channel (the black bid dot). We observe that at the reference time, the first and second particles lay on the opposite sides of a repulsive line; the second and third particles are not separated neither by repulsive nor by attractive lines; the third and fourth particles lay on opposite sides of an attractive line. Coherent with such observation, the trajectories of the different particles show that the first and second particles come from relatively close positions in the unbounded-orbit region, and tend to diverge from each other, as the first particle maintains an unbounded orbit, while the second gets trapped. The second and third particles tend to move together (both come from unbounded-orbit region; both get trapped). The third and fourth particles come from well separated positions (the fourth particle comes from the region of trapped particles) and converge towards close positions (both in the trapped-particle region). This results, that is substantially independent on the specific particles examined, show that particles that enter the lower channel become trapped.
Figure~\ref{fig:lyapunov_detrapping} shows the same LCSs, along with the trajectories of four different particles, located, at the reference time, just below (the first particle), within (second and third particles) and just above (the fourth particle) the upper channel. It can be seen that the second and third particles become de-trapped, while the first particle remains trapped and the fourth one maintains an unbounded orbit. These findings can be easily justified repeating, \textit{mutatis mutandis}, the above considerations concerning the relative positions of each couple with respect to repulsive and attractive lines. We then arrive at the specular conclusion: particles that enter the upper channel get de-trapped.

\begin{figure}
\includegraphics[width=0.60\textwidth]{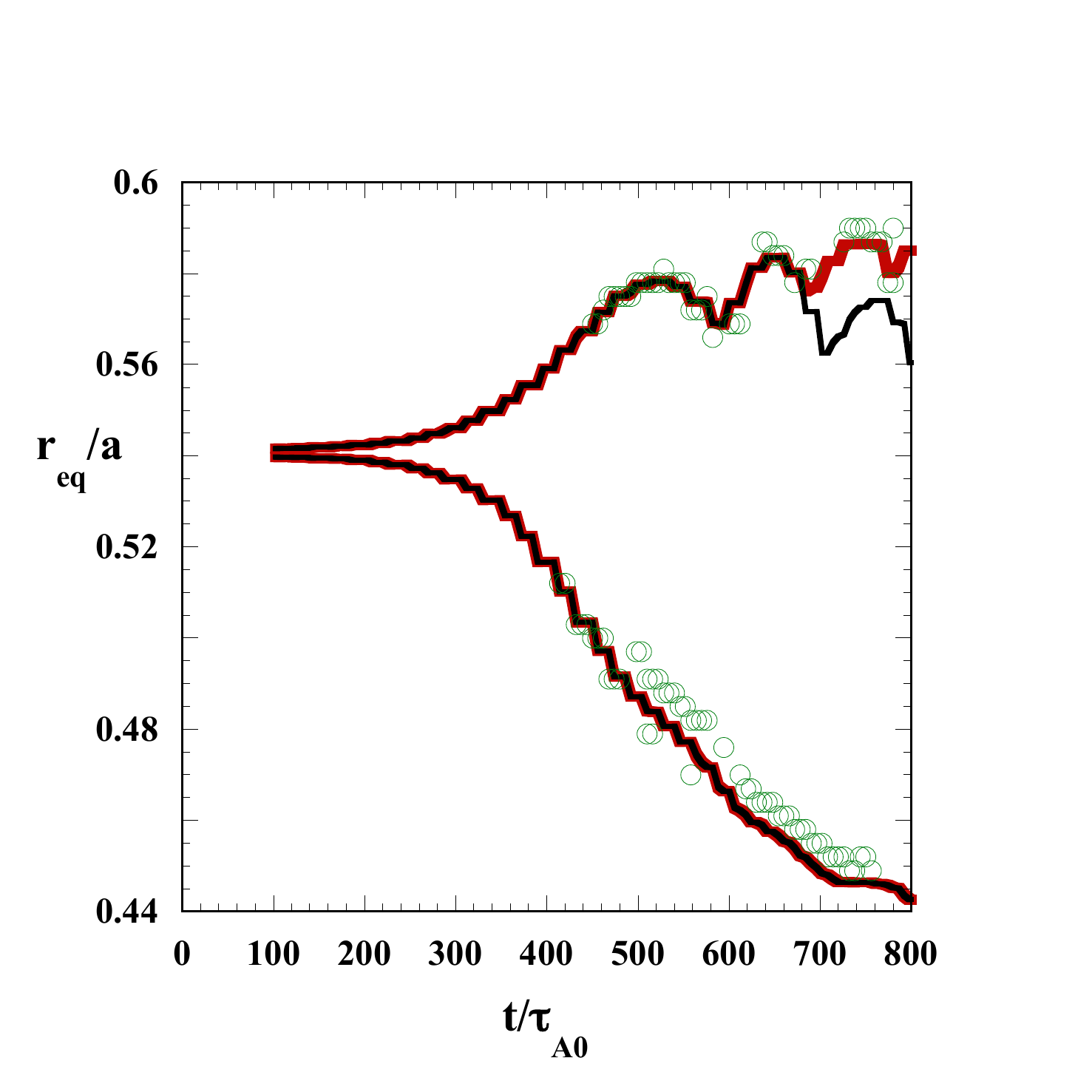}
    \caption{Radial boundaries of the already-trapped region, for the \textit{linear} slice (thick red line), compared with those of the actually-trapped region (black lines). As trapping of new particles takes place on the inner side of the island structure, and de-trapping on the outer side, the inner boundary is the same for the two regions; the outer boundaries are instead different. The radial positions of the maximum negative density gradients (green circles) delimiting the density-flattening region are also shown. The boundaries of the flattened-density region are then adequately represented by those of the region including the already-trapped particles; as far as the inner boundary is concerned, however, it is well represented by that of the actually-trapped region as well; that is, by the inner boundary of the island.}
    \label{fig:er_alr_act_trapp_grad_13} 
\end{figure}

The possibility of distinguishing actually-trapped particles from already-trapped ones (Fig.~\ref{fig:trapping_detrapping}) allows us to evidence a subtle distinction between the island structure and the density-flattening region. Indeed, the island only includes, by definition, actually-trapped particles. However, particles that get de-trapped continue to take part to the density flattening, as they contribute to the mixing phenomenon around the resonance radius. Then, the density flattening region extends over the whole region including the set of already-trapped particles (both actually-trapped and de-trapped particles). This can be seen from Fig.~\ref{fig:er_alr_act_trapp_grad_13}, which shows the time evolution of the inner and outer radial boundaries of the regions including the set of actually-trapped particles and, respectively, that of the already-trapped particles (both actually-trapped and de-trapped particles). The radial position of the maximum negative density gradients, delimiting the density-flattening region, are also shown. The latter are reported only after a certain time, and their calculation appears to be affected by noise; the reason is that computing them requires a discretisation of the radial domain, with a trade off between the need of accuracy and that of containing noise. It can be seen that identifying the boundaries of the already-trapped set with the radial position of the gradients is a fair approximation. Then, in the following, we will assume that such boundaries adequately represent the large-gradient positions and, then, the boundaries of the flattened-density region. On the other side, we have observed that, due to the inward island drift (consequent to the downward frequency chirping), trapping of new particles takes place mainly on the inner side of the island structure\footnote{Particle trapping also occurs, to a smaller extent, on the outer side of the island, because of the mode-amplitude growth.}; particle de-trapping, on the outer side only. Therefore, the inner boundary is the same for the region including the actually trapped particles (the island) and that including the already trapped ones (the flattening region); outer boundaries differ instead from each other. In this respect, as far as the inner boundary of the density-flattening region is concerned, we can identify its radial position (and that of the corresponding density gradient) with the inner boundary of the island as well.

As last remark, we want to emphasize that this trapping and detrapping process that accompanies frequency chirping fluctuations has been recently pointed out by hybrid simulations~\cite{tao2021} and corresponding theoretical framework~\cite{zonca2021,zonca2021_2} of ``chorus emission" in the Earth's magnetosphere, suggesting the universal nature of the underlying nonlinear dynamics.

\section{Frequency chirping and resonance evolution}
\label{sec:chirping}

We want to investigate how the collective power transfer is related to the the fulfilment of the resonance condition and the evolution of the free-energy source (namely, the radial density gradient). Let us start from considering the \textit{linear} test-particle set.

We have already observed that, during the linear stage (Fig.~\ref{fig:an_er_phase_1}-left), the maximum power transfer is yielded by resonant particles, around $r_{\textrm{eq}}\simeq 0.54 a$. In the early nonlinear stage, the island formation and growth and the corresponding flattening of the density profile around the resonance radius is accompanied by the splitting of the power transfer maximum into two separate maxima, each of them following one of the two large negative density gradients delimiting the flattening region and moving, respectively, inward and outward (Fig.~\ref{fig:an_er_phase_1} right).

\begin{figure}
\includegraphics[width=0.60\textwidth]{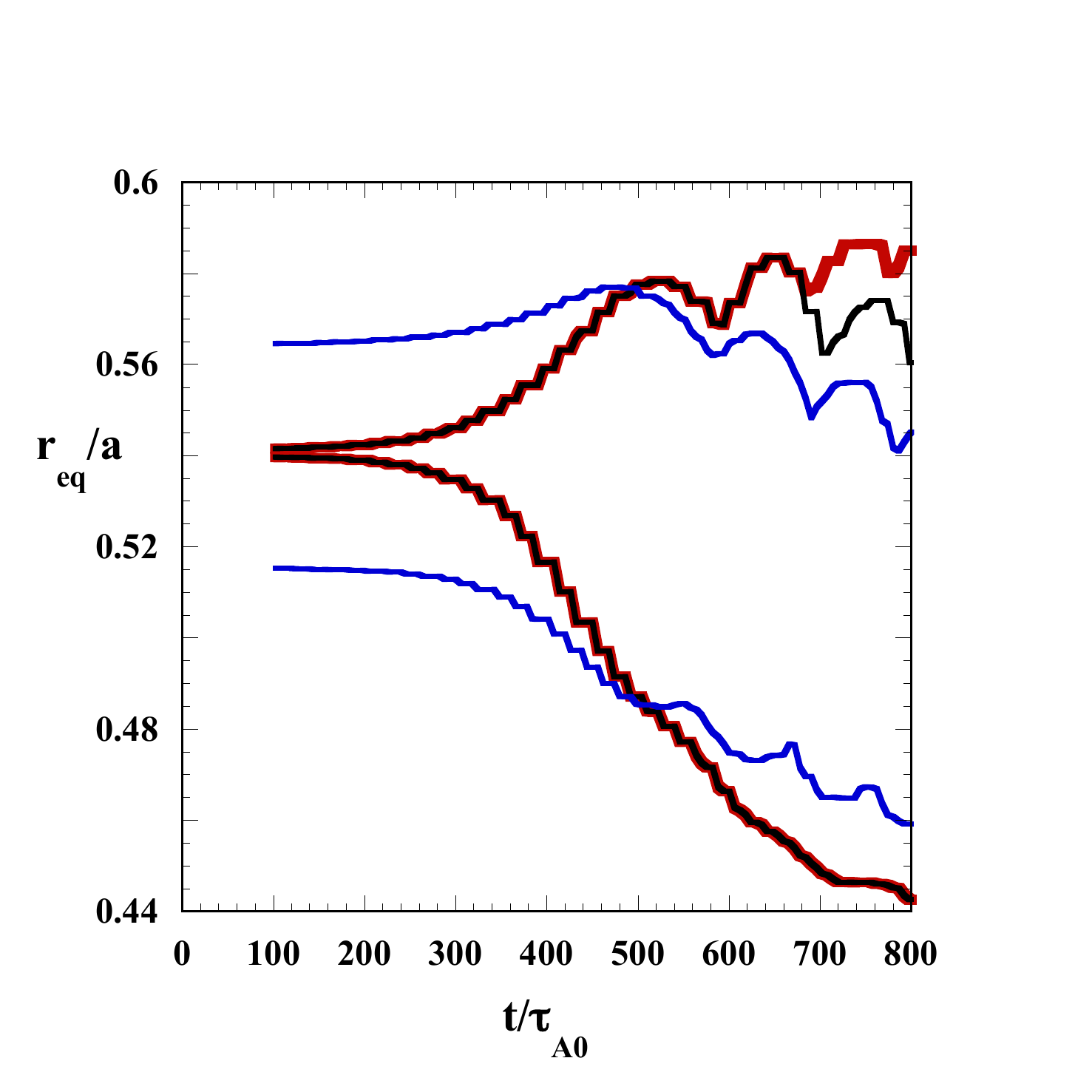}
    \caption{Time evolution of the radial boundaries of density-flattening region (thick red lines), island (black lines) and resonance region (blue lines). The latter is conventionally defined by the condition 
$|\Delta\overline{\Theta}_j| \leq 2\gamma_L (t_j-t_{j-1}) \sim  4\pi  \gamma_L q R_0/U_{\textrm{eq0}}  \sim 0.153 \,\,\pi$.}
    \label{fig:er_alr_act_trapp_res_13} 
\end{figure}

Figure~\ref{fig:er_alr_act_trapp_res_13} compares the radial boundaries of the island and the density-flattening regions (shown in Fig.~\ref{fig:er_alr_act_trapp_grad_13}) with those of the resonance region, conventionally defined as the region containing particles that satisfy the condition $|\Delta \overline{\Theta_j}| \leq 2\gamma_L (t_j-t_{j-1}) \sim   4\pi \gamma_L q R_0/U_{\textrm{eq0}}  \sim 0.153 \,\,\pi$, with $\gamma_L$ being the linear growth rate, and $\Delta \overline{\Theta_j}$, $t_j$ and $t_{j-1}$ being defined in Eq.~\ref{eq:resonance_condition}; this choice corresponds to include in the resonance width up to one fifth of the resonance peak.
Some broadening of the resonance region is observed. Such broadening keeps the drifting gradients in the resonance region for a slightly longer time than the ``linear" resonance width would be able to do, allowing for further driving the mode.
This would however go quite soon to an end if the mode were not able to vary its frequency. As soon as the density-flattening region completely covered the broadened resonance region (whose width, in the present case, is smaller than the mode width), the drive would be exhausted.
In the present case, however, the frequency can chirp down. This moves the resonance radius and the entire resonance region inward. The symmetry of the two large density gradients is, thus, broken: the inner one can contribute to resonant drive much more efficiently than the outer one. 
On the other side, wave-trapped trajectories (Fig.~\ref{fig:traces}-right and Fig.~\ref{fig:an_er_phase_0_2}) and the wave-trapping process (Fig.~\ref{fig:trapping}) will move their center around the new resonance radius, along with the island structure (cf. the black lines in Fig.~\ref{fig:er_alr_act_trapp_grad_13}), and the inner gradient will further drift inward, requesting further frequency variation.

\begin{figure}
       \includegraphics[width=0.80\textwidth]{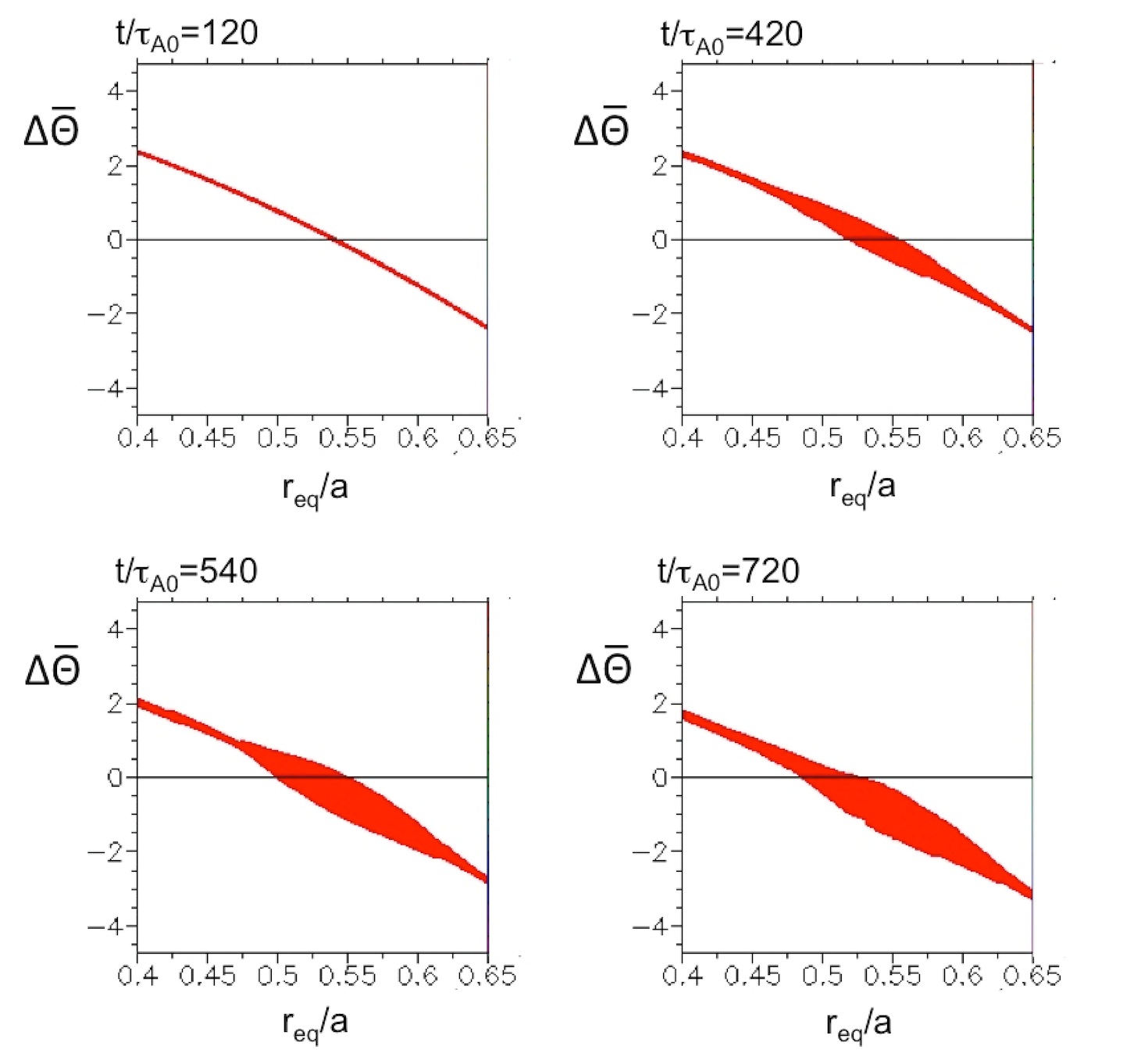}
 \caption{Test-particle markers in the plane $(r_{\textrm{eq}},\Delta\overline{\Theta}_j)$. Four times are considered: $t=120 \tau_{A0}$, $t=420 \tau_{A0}$, $t=540 \tau_{A0}$ and $t=720 \tau_{A0}$. The resonance broadening can be considered and the effect of two competing factors: the variation of the shape of the resonant structure and its migration towards the left-bottom corner of the plot; resonance broadening is favoured by the former, contrasted by the latter.}
        \label{fig:resonance_region}
 \end{figure}

\begin{figure}
      \includegraphics[width=0.5\textwidth]{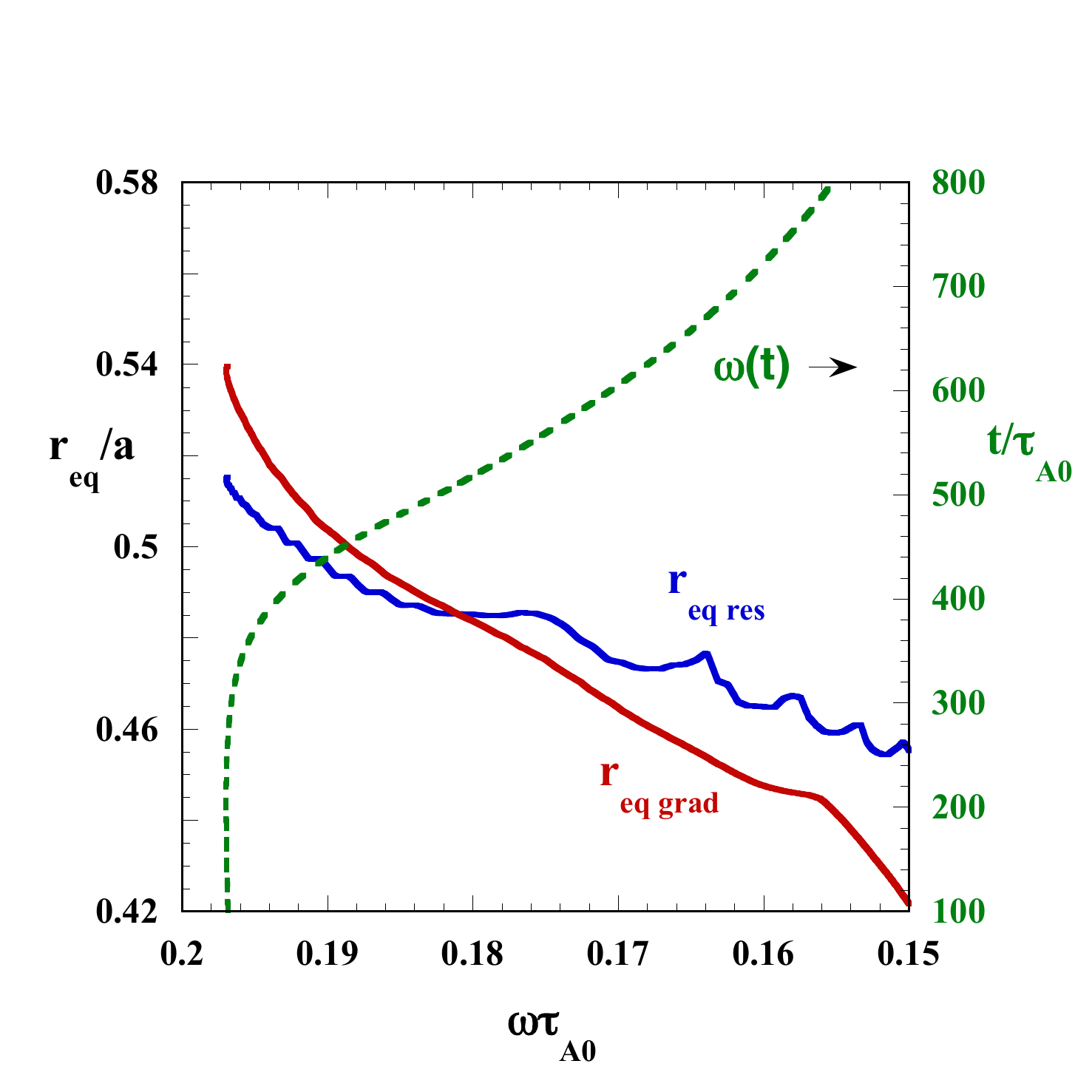}  
 \caption{Trajectories, in the space $(\omega,r_{\mathrm{eq}})$ of the inner boundary of the density-flattening region (red) and that of the resonance region (blue). The time evolution of the mode frequency is also reported (green dashed line) to make it clear how these trajectories are travelled over time. As the frequency falls below values of the order $\omega \tau_{A0} \simeq 0.18$, at $t\simeq 500\tau_{A0}$, the inner gradient remains irretrievably outside the resonance region.}
        \label{fig:omega_r_res_r_grad_13}
 \end{figure}

\begin{figure}
\includegraphics[width=0.60\textwidth]{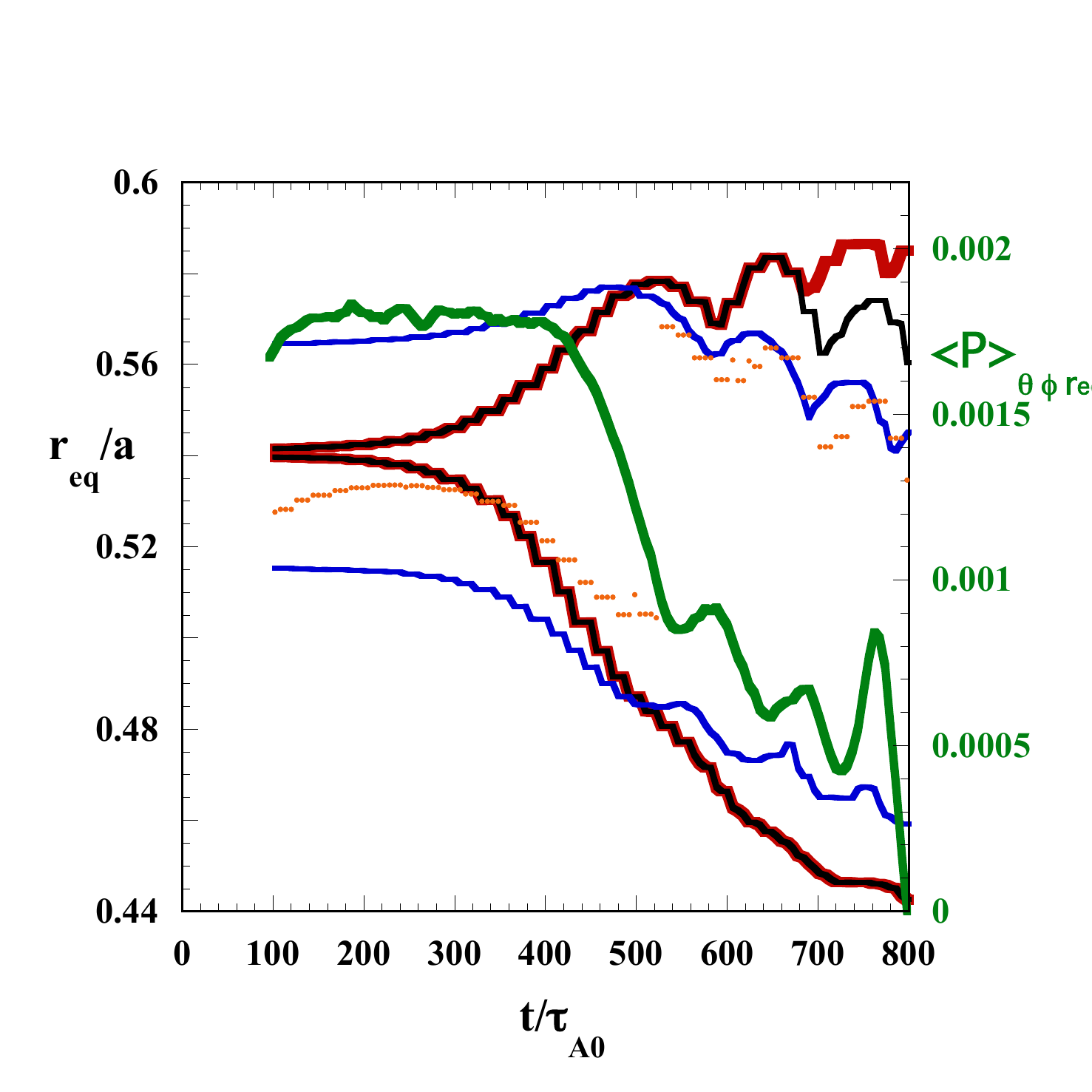}
    \caption{Same boundaries as in Fig.~\ref{fig:er_alr_act_trapp_res_13}, along with the radial coordinates of the time dependent maximum of the power transfer (orange dots) and the time evolution of the slice-integrated power (green line; cf Fig.~\ref{fig:relevant_slices}).}
    \label{fig:er_alr_act_trapp_res_13_2} 
\end{figure}

We can expect that this process continues until the frequency change becomes either detrimental to the growth of the mode from the point of view of the drive/damping balance, or unable to cause a significant inward shift of the resonance radius and/or the resonance region.
The latter fact effectively occurs, as seen from Fig.~\ref{fig:resonance_region}, in which all markers are reported, at four different times, in the plane $(r_{\mathrm{eq}},\Delta\overline{\Theta}_j)$, with $j$ referring, for each marker, at the last poloidal orbit completed. 
We observe that the resonance broadening (corresponding to the broadening of the red region around $\Delta\overline{\Theta}=0$) can be considered as the effect of two competing factors: the spindle-shaping of the resonant structure and its migration towards the left-bottom corner of the plot (due to the downward frequency chirping); resonance broadening is favoured by the former, contrasted by the latter, as it moves the resonance condition towards the spindle end. At a certain time the spindle-end factor prevails and the inward drift of the resonance region suddenly slows down. This is clearly shown in Fig.~\ref{fig:omega_r_res_r_grad_13}, which compares the trajectories, in the space $(\omega,r_{\mathrm{eq}})$ of the inner boundary of the density-flattening region and that of the resonance region. The time evolution of the mode frequency is also reported to make it clear how these trajectories are travelled over time.
It is possible to see that as the frequency falls below values of the order $\omega \tau_{A0} \simeq 0.18$ (that is, at $t\simeq 500\tau_{A0}$), the inner gradient remains irretrievably outside the resonance region and any further frequency decrease is not able to prolong the relevance of the destabilising contribution yielded by the linear slice. 
Correspondingly, the inner power-transfer maximum loses its prevalence, as shown in Fig.~\ref{fig:er_alr_act_trapp_res_13_2}, which plots the same boundaries as in Fig.~\ref{fig:er_alr_act_trapp_res_13}, along with the radial coordinates of the time dependent maximum of the power transfer and the time evolution of the slice-integrated power.
No further frequency decrease is then able to prolong the relevance of the destabilising contribution yielded by the linear slice, and the mode has to tap to a different slice in order to extract more power from the energetic particles. In the following, we examine the behaviour of the nonlinear slice; that is, the slice that takes the role of most destabilising slice after the linear one has exhausted its drive capability.

\begin{figure}
\includegraphics[width=0.90\textwidth]{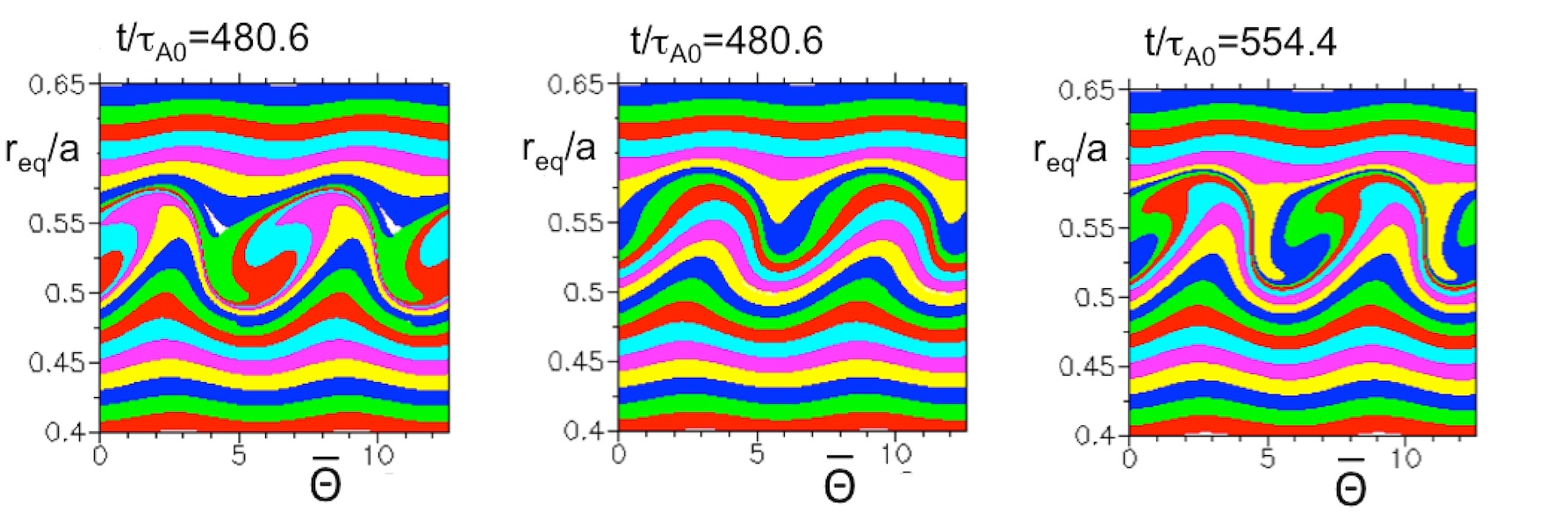}
    \caption{Test-particle markers in the $(\overline{\Theta},r_{\textrm{eq}})$ at $t=480.6 \tau_{A0}$, for the linear slice (left) and for the nonlinear one (center). The plot relative to the nonlinear slice at a later time ($t=554.4 \tau_{A0}$) is also shown (right). Marker colours follow the recipe adopted in Fig.~\ref{fig:an_er_phase_0}. The delay in the island formation for the nonlinear slice is evident.}
    \label{fig:an_er_phase_0_nl} 
\end{figure}

\begin{figure}
\includegraphics[width=0.50\textwidth]{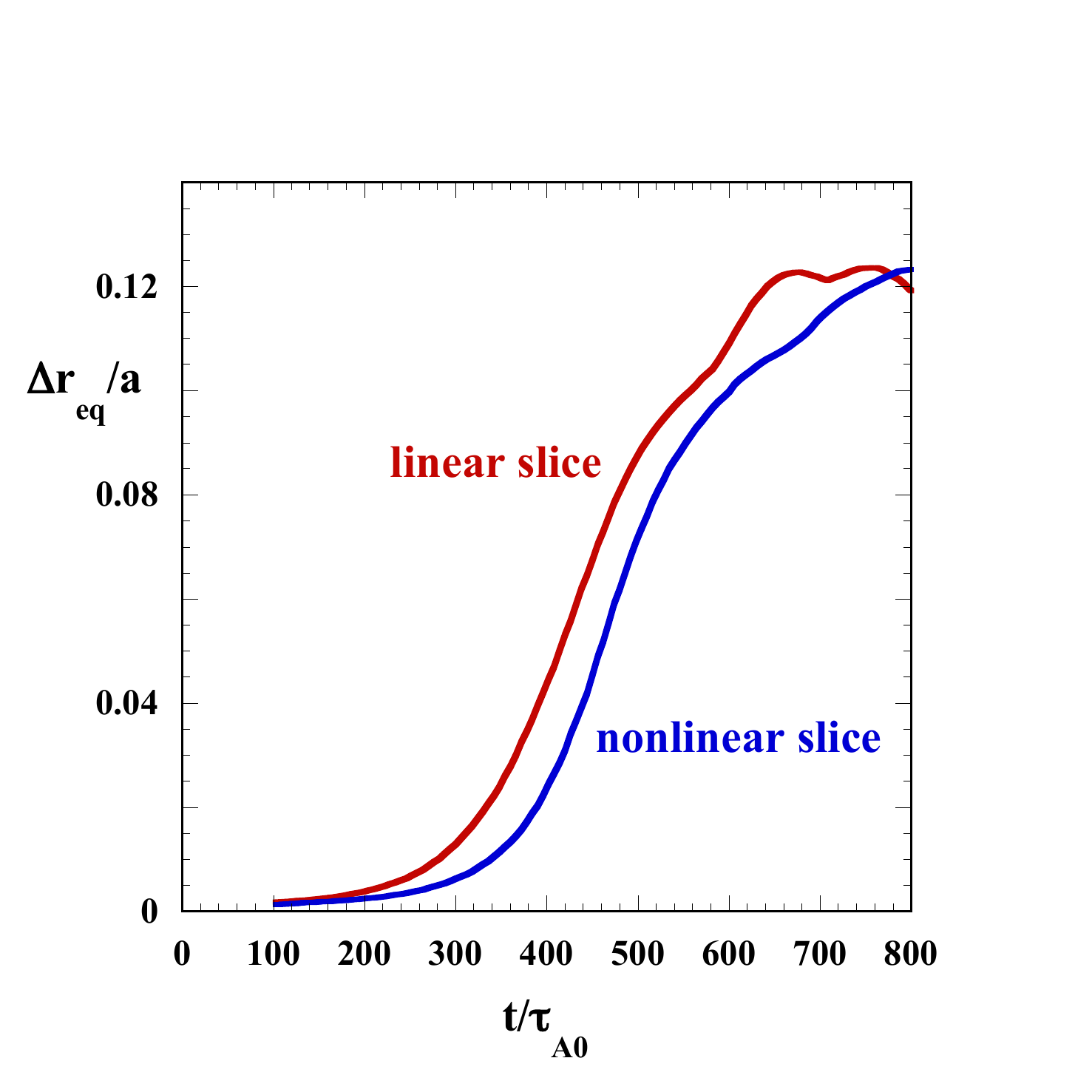}
    \caption{Time evolution of the island radial width for the two slices.}
    \label{fig:island_width_13_11} 
\end{figure}

\begin{figure}
      \includegraphics[width=0.5\textwidth]{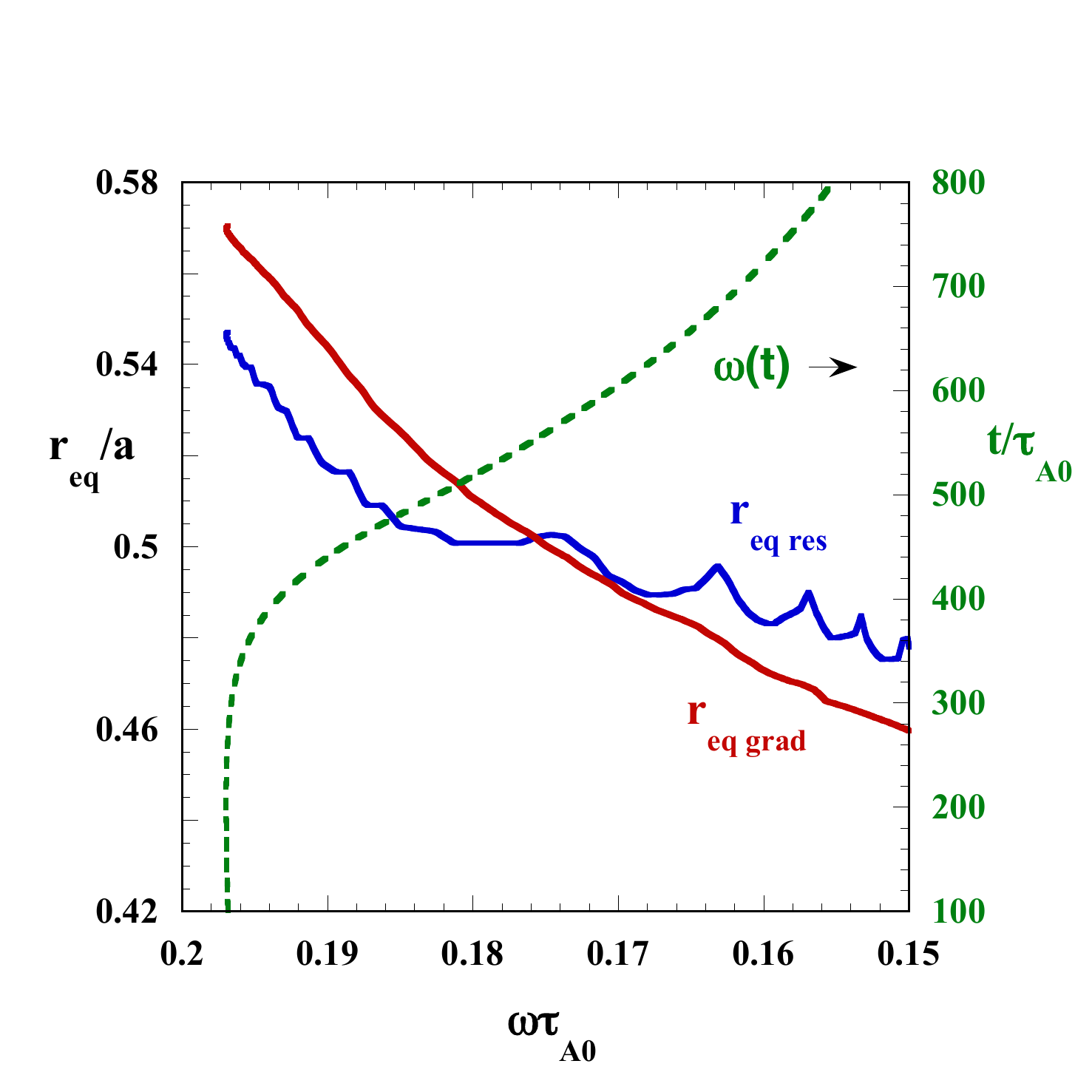} 
 \caption{Same as Fig.~\ref{fig:omega_r_res_r_grad_13}, but for the nonlinear slice. In this case, the decoupling between density-flattening and resonance-region boundaries is both delayed and less pronounced than in the linear-slice case.}
        \label{fig:omega_r_res_r_grad_11}
 \end{figure}

\begin{figure}
\includegraphics[width=0.60\textwidth]{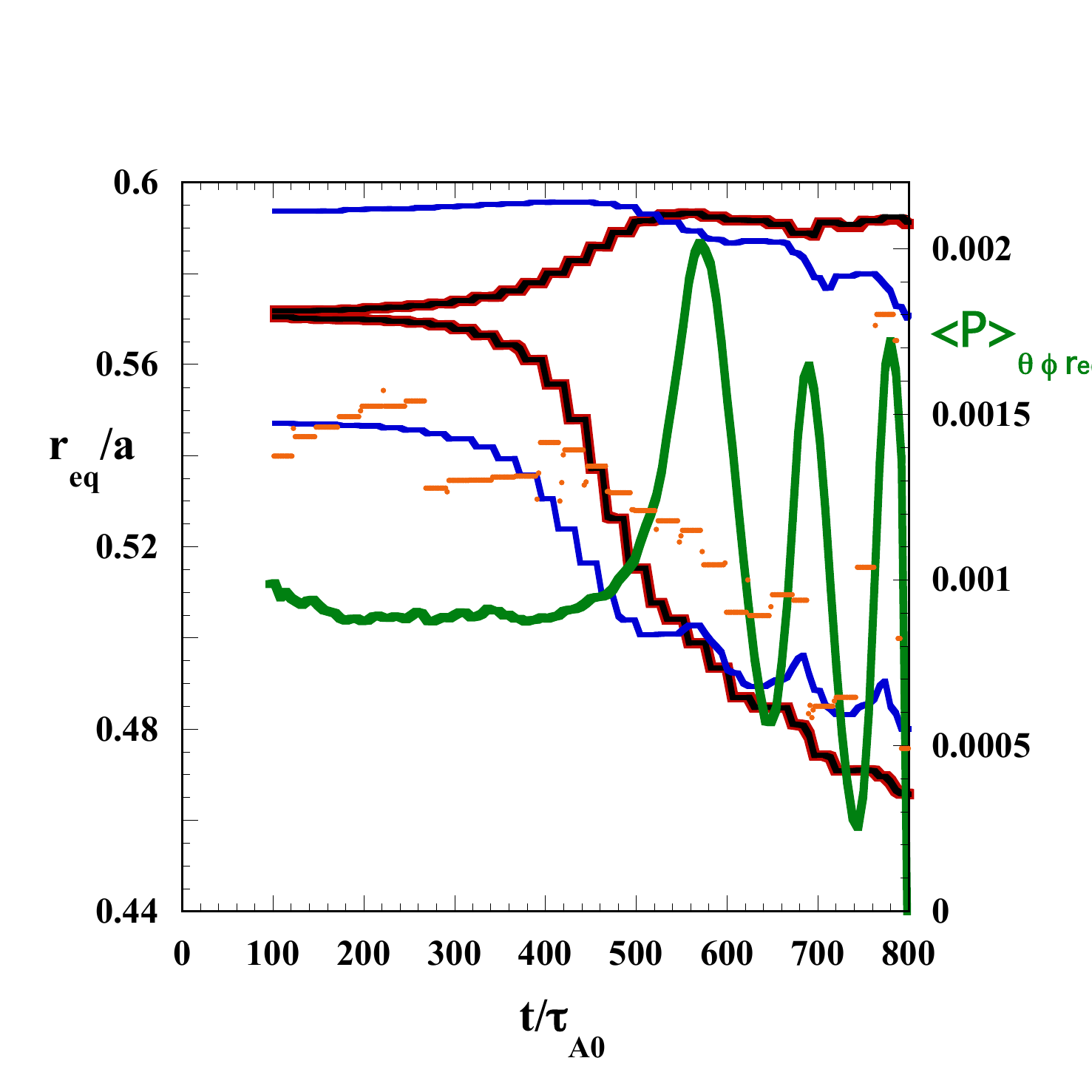}
    \caption{Same as Fig.~\ref{fig:er_alr_act_trapp_res_13_2} for the nonlinear slice.}
    \label{fig:er_alr_act_trapp_res_11_2} 
\end{figure}

The first fact we observe is that, for the nonlinear slice, the island formation is delayed when compared to the linear one. This can be seen from Fig.~\ref{fig:an_er_phase_0_nl}, where the plot of Fig.~\ref{fig:an_er_phase_0}-right, relative to the linear slice, is compared with the plots obtained, for the nonlinear slice, at the same time ($t=480.6 \tau_{A0}$) and at a later time ($t=554.4 \tau_{A0}$).
The delay in the island formation can also be observed from Fig. ~\ref {fig:island_width_13_11}, which compares the time evolution of the island radial width for the two slices.
The second observation (Fig. ~\ref {fig:er_alr_act_trapp_res_11_2}) is that during the linear stage the maximum power transfer of the nonlinear slice neither occurs around the resonance radius, nor is directly related to the density gradient: it is influenced by the fluctuating field localisation, mainly determined by the interaction with the linear slice. 
Only later, because of resonance broadening and frequency chirping, the maximum power transfer matches a full resonance condition and appears dominated by the inner density gradient. 
The third relevant element shown by Fig.~\ref{fig:er_alr_act_trapp_res_11_2} is that the de-trapping is negligible, in the considered time interval, for such slice, so that even the outer actually-trapped particle boundary and the already-trapped particle one essentially coincide. Finally, and more important, the decoupling between density-flattening and resonance-region boundaries is not only delayed, but also less pronounced than in the linear-slice case, as shown in Fig.~\ref{fig:omega_r_res_r_grad_11}. 
This is consistent with the fact that the inner power-transfer maximum maintains its prevalence, as shown in Fig.~\ref{fig:er_alr_act_trapp_res_11_2}, for a longer time and that the the power transfer yielded by the slice, after reaching its maximum, falls down at $t\simeq 600.0 \tau_{A0}$.

\section{Summary and discussion}
\label{sec:conclusions}

The dynamics of a single-$n$ chirping Alfv\'{e}n mode has been investigated.

We have discussed the relevance of adopting a coordinate system containing global invariants of motion in order to unambiguously identify resonant particles destabilising/stabilising the mode. In particular, this allows us to distinguish whether, in the non-linear evolution of the mode, it is driven by the same set of particles, with phase-space coordinates modified by the interaction with the fluctuating fields, or by a succession of different sets. Moreover, it makes us able to cut the phase space into slices characterised by fixed values of the global invariants; as there is no particle flux from one slice to another, 
the only gradients that can play the role of free-energy source for the mode destabilisation are those along the coordinates that vary inside the slice, while the gradients orthogonal to the slice do not play any role. The analysis of the nonlinear dynamics and, in particular, the saturation mechanism can be simplified by investigating separately the most relevant slices (i.e., the most effective in exchanging power with the mode).

In the gyrokinetic collisionless limit, for Alfv\'{e}n modes characterised by a single toroidal mode number and a constant frequency, both the magnetic moment $M$ and the linear combination of energy and angular momentum, defined as $C=\omega P_{\phi} - n E$, are global invariants of motion. In the general case and, in particular, for chirping modes (i.e., modes with varying frequency), $C$ is no longer a constant. It is, however, possible to choose a coordinate system including other conserved quantities, with the same advantages offered by global invariants of motion.
Here, in the view of a numerical approach to the investigation of mode-particle dynamics, we have proposed to adopt as constant-of-motion coordinates $M$ and the initial values of the parallel velocity (more precisely, the initial values of the corresponding \textit{equatorial} coordinate; that is, the value assumed when the particle crosses the equatorial plane at $\theta=0$). These coordinates are readily connected with the actual particle coordinates within a numerical approach by simply storing, in addition to the actual particle coordinates, their initial values.
Such choice allows us to analyse the nonlinear evolution of the system in terms of isolated resonances.

The evolution of the particle-mode power transfer shows that a succession of different resonance takes place in driving the mode. We have identified two dominant driving phase-space slices during the linear and nonlinear phases. The analysis of the dynamics of these slices has been performed by sampling each of them by sets of test particles evolving in the fluctuating fields computed by a full-population, self-consistent simulation.
We have shown that,
within each set, the same bunch of particles drives or damps the mode, depending on their inward or, respectively, outward radial drift during their bounce in the potential well of the wave. Moreover, efficient power transfer to or from the mode always corresponds to the instantaneous fulfilment of the resonance condition.

The nonlinear motion of particles gives rise to two different kind of trajectories in the space  $(\overline{\Theta},r_{\mathrm{eq}})$; namely, bounded orbits for particles instantaneously trapped in the wave, and unbounded ones for streaming particles. The former yield the formation of a relatively closed island-like structure around the resonance radius, characterised by mixing of particles originating from the different sides of that radius. A density-flattening region is then produced, delimited by large (negative) density gradients.
 In the absence of other effects, the mode would saturate as the flattening region covers the whole resonant-interaction region; that is, the intersection between the resonance region and the mode structure. In the considered case, because of the relatively small value of the growth rate and the relatively large mode structure, saturation would occur as the density gradients reach the limit of the resonance region. Frequency decreasing causes, however, the resonance region to drift inward and keeps the inner gradient within the resonant-interaction region, helped by a significant resonance broadening. 
The island now grows around the new resonance radius, by trapping new particles on its inner boundary, while other particles get de-trapped from the outer one. De-trapped particles still continue to contribute to the density flattening, as they have previously taken part it the mixing phenomenon. Then, while the inner density gradient is still associated to the inner boundary of the island, the position of the outer gradient is mainly related to the outer boundary of the larger region including particles that have been wave-trapped, even if they are no longer contained in the island.
The inward drift of the island and its further growth cause the inner gradient to drift inward as well; to further maintain the drive capability of the slice, further frequency decreasing is needed. The process continues as long as the frequency change is effective in causing a inward shift of the resonance region able to recapture the density gradient. 

At a certain time, the combined effect of frequency chirping and resonance broadening weakens the sensitivity of the inner boundary of the resonance region to the frequency decreasing; this boundary is no longer able to lock to the inner density gradient, and the process goes to an end.
After that time, the power transfer of the considered slice decreases.
To further grow, the mode has to resort to the interaction with a different resonant structure (the ``nonlinear slice"), using, if needed, additional frequency variations. 
This slice is poorly resonant during the linear phase. It becomes fully resonant because of frequency chirping and resonance broadening. The corresponding island formation is delayed with respect to the linear slice. Moreover, its resonance broadening is more effective than that of the linear set, and it is able to prolong the permanence of the large density gradient in the resonance region. This gives the mode further drive and makes it reaching larger amplitudes than that obtained because of the linear-slice drive.

In this paper the analysis has been focused on the drive evolution during the ``early" nonlinear stage, where by early we mean the stage that brings to the first fall of the destabilising contribution of the linearly dominant resonant structure (the linear slice) and to the growth and successive fall of the nonlinear-slice contribution. We have not addressed the subsequent rich and complicate evolution. Moreover, we have especially looked at the drive evolution, without investigating the nonlinear damping mechanisms. In this respect, our comprehension of the nonlinear dynamics of a chirping mode is still far from being complete, although we hope to have successfully enlightened some relevant aspects.

We have also to stress that including a constant of motion (in our case, the initial value of the equatorial parallel velocity) in the coordinate system, though useful for treating a case characterised by varying frequency (and, then, by non conservation of the global invariant $C$), does not represent a unique recipe. In particular, there is no certainty, a priori, of being able to describe the nonlinear dynamics in terms of the contribution of a single slice, characterised by unique values of $M$ and the chosen constant. In principle, it is possible that a particular constant of motion (global invariant or not) exists, able to yield such a result, but there is no known way to identify \textit{a priori} such a constant. Thus, only \textit{a posteriori} we can discern whether the particular choice made is suited enough for our scopes: that is, whether it allows the relevant dynamics to be described in terms of only a few slices.

Finally, we observe that, as stated above, the rate of variation of $C$ is quite small in the case here examined. In this respect, neglecting such rate and treating $C$ as an invariant produces, in fact, very similar results; this means that the fluxes along $C$ do not play a relevant role in the nonlinear dynamics of the mode. We get a completely different conclusion in the case of Alfv\'{e}n spectra characterised by multiple toroidal mode numbers. This case will be the subject of future investigations.

\section{Acknowledgements}

This work has been carried out within the framework of the EUROfusion Consortium [Enabling Research Projects: MET (CfP-AWP19-ENR-01-ENEA-05) and ATEP
12 (CfP-FSD-AWP21-ENR-03)] and has received funding from the Euratom research and training programme 2014-2018 and 2019-2020 under grant agreement No 633053. The views and opinions expressed herein do not necessarily reflect those of the European Commission.
The computing resources and the related technical support used for
this work have been provided by CRESCO/ENEAGRID High Performance
Computing infrastructure and its staff ~\cite{cresco}. CRESCO/ENEAGRID High
Performance Computing infrastructure is funded by ENEA, the Italian
National Agency for New Technologies, Energy and Sustainable Economic
Development and by Italian and European research programmes, see
http://www.cresco.enea.it/english for information.

\end{document}